\documentclass[12pt]{article}  

\usepackage{graphicx}

\usepackage{a41}
\usepackage[rflt]{floatflt}
\usepackage{color}
\usepackage{cite}
\usepackage{float}
\usepackage{amsmath}
\usepackage{amssymb}
\usepackage{amsxtra}
\usepackage{graphicx}
\usepackage{slashed}

\newcommand{\LtwoWpWqNStwo}{L_{2,q}^{W^{+}+W^{-},\text{NS},(2)}}
\newcommand{\HtwoWpWqNStwo}{H_{2,q}^{W^{+}+W^{-},\text{NS},(2)}}
\newcommand{\HtwoWmWqNStwo}{H_{2,q}^{W^{+}-W^{-},\text{NS},(2)}}
\newcommand{\HtwoWqPStwo}{H_{2,q}^{W,\text{PS},(2)}}
\newcommand{\LtwoWgtwo}{L_{2,g}^{W,(2)}}
\newcommand{\HtwoWgtwo}{H_{2,g}^{W,(2)}}
\newcommand{\LthreeWpWqNStwo}{L_{3,q}^{W^{+}+W^{-},\text{NS},(2)}}
\newcommand{\HthreeWpWqNStwo}{H_{3,q}^{W^{+}+W^{-},\text{NS},(2)}}
\newcommand{\HthreeWmWqNStwo}{H_{3,q}^{W^{+}-W^{-},\text{NS},(2)}}
\newcommand{\HthreeWqPStwo}{H_{3,q}^{W,\text{PS},(2)}}
\newcommand{\HthreeWgtwo}{H_{3,g}^{W,(2)}}
\newcommand{\eph}{\frac{\varepsilon}{2}}
\newcommand{\Ftwon}[3]{{}_2F_1\left[\begin{array}{c}#1\\#2\end{array};#3\right]}
\newcommand\MSbar{$\overline{\mbox{MS}}$}

\newcommand{\N}{\nonumber}
\newcommand{\CKM}[1]{\vert V_{#1}\vert^2}
\newcommand{\period}{\,.}
\newcommand{\comma}{\,,}

\newcommand{\HSums}{{\sf HarmonicSums}}

\newcommand{\citeANCONT}{\cite{Blumlein:2000hw, Blumlein:2005jg, Blumlein:2009ta,
Blumlein:2009fz}}
\newcommand{\citeHSums}{\cite{Ablinger:2010kw, Ablinger:2013hcp, Ablinger:2011te, Ablinger:2013cf}}

\usepackage[implicit=true]{hyperref}

\usepackage{array}

\usepackage[english]{babel}
\usepackage[latin1]{inputenc}
\usepackage[T1]{fontenc}
\usepackage{ae}

\usepackage{url}

\usepackage{amsmath, amsthm, amssymb}
\newtheorem{thm}{Theorem}[section]

\newtheorem{definition}[thm]{Definition}

\newcommand{\Li}{{\rm Li}}

\newcommand{\ep}{\varepsilon}

\usepackage{rotating}

\usepackage{algorithm}
\usepackage{algorithmicx}
\usepackage{algpseudocode}
\usepackage{graphicx}

\newcounter{mmacnt}
\def\restartmma{\setcounter{mmacnt}{0}}
\restartmma \catcode`|=\active
\def|#1|{\mathrm{#1}}
\catcode`|=12
\newenvironment{mma}{
 \par\smallskip
 \catcode`|=\active
 \parskip=0pt\parindent=0pt 
 \small
 \def\In##1\\{%
   \def\linebreak{\hfill\break\null\qquad}%
   \refstepcounter{mmacnt}
   \hangindent=2.5em\hangafter=0
   \leavevmode
   \llap{\tiny\sffamily In[\arabic{mmacnt}]:=\kern.5em}%
   \mathversion{bold}\footnotesize$\displaystyle##1$\normalsize
   \mathversion{normal}\par
 }%
 \def\Print##1\\{%
   \def\linebreak{\hfill\break}%
   \hangindent=2.5em\hangafter=0
   \leavevmode ##1\par}%
 \def\Out##1\\{%
   \def\linebreak{$\hfill\break\null\hfill$}%
   \kern\abovedisplayskip\par
   \hangindent=2.5em\hangafter=0
   \leavevmode
   \llap{\tiny\sffamily Out[\arabic{mmacnt}]=\kern.5em}
   \footnotesize$\displaystyle##1$\normalsize\hfill\null\par
   \kern\belowdisplayskip
 }%
 \def\Warning##1##2\\{%
   \def\linebreak{\hfill\break}%
   \hangindent=2.5em\hangafter=0
   \leavevmode
   {\scriptsize##1 : ##2}\par}%
}{%
 \par\smallskip
}

\usepackage{color}

\newenvironment{fshaded}{%
\MakeFramed {\FrameRestore}
}%
{\endMakeFramed}

\usepackage{tikz}
\usetikzlibrary{matrix}

\allowdisplaybreaks[4]

\begin{document}
\setlength{\baselineskip}{0.515cm}
\sloppy
\thispagestyle{empty}
\begin{flushleft}
DESY 13--247
\hfill {\tt arXiv:1401.4352 [hep-ph]}
\\
DO--TH 13/34\\
SFB/CPP-14-005\\
LPN 14-006\\
Higgstools 14--2\\
TUM-HEP-921/13\\
December 2013\\
\end{flushleft}

\mbox{}
\vspace*{\fill}
\begin{center}

{\LARGE\bf The \boldmath{$O(\alpha_s^2)$} Heavy Quark Corrections to}

\vspace*{2mm}
{\LARGE\bf Charged Current Deep-Inelastic Scattering}

\vspace*{2mm}
{\LARGE\bf at large Virtualities}

\vspace{4cm}
\large
Johannes Bl\"umlein$^a$, Alexander Hasselhuhn$^{a,b}$, and Torsten Pfoh$^a$\footnote{Present 
address: Physik Department T31, James-Franck-Strasse 1,
Technische Universitaet Muenchen, 85748 Garching, Germany.}

\vspace{1.5cm}
\normalsize
{\it  $^a$ Deutsches Elektronen--Synchrotron, DESY,}\\
{\it  Platanenallee 6, D-15738 Zeuthen, Germany}
\\
{\it $^b$~Research Institute for Symbolic Computation (RISC),\\
                          Johannes Kepler University, Altenbergerstra\ss{}e 69,
                          A--4040, Linz, Austria}\\

\vspace*{3mm}

\end{center}
\normalsize
\vspace{\fill}
\begin{abstract}
\noindent 
We calculate the $O(\alpha_s^2)$ heavy flavor corrections to charged current deep-inelastic 
scattering at large scales $Q^2 \gg m^2$. The contributing Wilson coefficients are given as
convolutions between massive operator matrix elements and massless Wilson coefficients. 
Foregoing results in the literature are extended and corrected. Numerical results are 
presented for the kinematic region of the HERA data.
\end{abstract}

\vspace*{\fill}
\noindent
\numberwithin{equation}{section}
\newpage

\section{Introduction}

\vspace*{1mm}
\noindent
The heavy flavor corrections to deep-inelastic scattering obey different scaling violations
both in neutral and charged current scattering if compared to the massless contributions
\cite{Blumlein:1996sc}. Furthermore, for charged current reactions these contributions 
constitute out of flavor excitation on the one hand, e.g. $s \rightarrow c$ transitions, and also 
heavy quark pair production at higher orders in the coupling constant $\alpha_s(M_Z^2)$. In the 
charged current case most of the data are situated at higher values of $Q^2$, 
cf.~\cite{Blumlein:1987xk,Blumlein:1989pd}. Therefore the representation of the heavy flavor Wilson 
coefficients in the region $Q^2 \gg m^2$ can be obtained using the factorization \cite{Buza:1995ie}
into massive operator matrix elements (OMEs) and the massless Wilson coefficients 
\cite{Furmanski:1981cw,vanNeerven:1991nn,Zijlstra:1991qc,Zijlstra:1992kj,Zijlstra:1992qd,Moch:1999eb}.
In the past a series of analytic results has been calculated for neutral current reactions in 
this way \cite{Buza:1996xr, Buza:1996wv,Bierenbaum:2007dm,Blumlein:2006mh,Bierenbaum:2007qe,
Bierenbaum:2008yu,Bierenbaum:2009zt,Bierenbaum:2009mv,Ablinger:2010ty,Ablinger:2012qm,
Blumlein:2012vq,Behring:2013dga}. 

In the present paper we calculate the $O(\alpha_s^2)$ corrections in the charged current case,
extending and correcting Ref.~\cite{Buza:1997mg}. The $O(\alpha_s)$ corrections were computed in
\cite{Gottschalk:1980rv,Gluck:1996ve,Blumlein:2011zu} before. In \cite{Buza:1997mg} the heavy 
flavor Wilson coefficients $H_{2,g}^{W,(2)}$, $H_{2,q}^{\text{PS},(2)}$, $H_{3,g}^{W,\text{PS},(2)}$ 
and $H_{3,q}^{\text{PS},(2)}$ were calculated. Since this was not the complete set we also calculate
the remaining Wilson coefficients and compare the present results with the previous ones.
The heavy flavor Wilson coefficients to $O(\alpha_s^2)$ will allow to refine QCD fits w.r.t. the
extraction of the individual sea quarks, in particular also the strange quark distribution,
cf.~\cite{Alekhin:2013nda,Alekhin:2012ig}.

The paper is organized as follows. We give first a summary of the charged current 
structure functions with emphasis on the heavy flavor contributions and present the
general structure of the different heavy flavor Wilson coefficients in the limit $Q^2 \gg m^2$. 
Here combinations which are invariant under current crossing are important to allow for proper 
renormalization. In Section~3 the Wilson coefficients are presented in Mellin-$N$ space to 
$O(\alpha_s^2)$. Numerical results are given in Section~4 and Section~5 contains the conclusions. 
In the Appendices technical aspects are dealt with and we also present the Wilson coefficients 
in $x$-space there.
\section{The Structure Functions}

\vspace*{1mm}
\noindent
The scattering cross sections for charged current deep-inelastic lepton-nucleon
scattering are parameterized by the three structure functions $F_1$, $F_2$,
$F_3$\,:
\begin{align}
\label{eq:XSa}
 \frac{d\sigma^{\nu(\bar{\nu})}}{dx dy}
 ={}&
 \frac{G_F^2 s}{4 \pi}
\frac{M_W^4}{(M_W^2 + Q^2)^2}
 \left\{
      (1+(1-y)^2) F_2^{W^{\pm}}
  - y^2 F_L^{W^{\pm}}
  \pm (1-(1-y)^2) x F_3^{W^{\pm}}
 \right\}
\comma
\\
\label{eq:XS}
 \frac{d\sigma^{e^-(e^+)}}{dx dy}
 ={}&
 \frac{G_F^2 s}{4 \pi} 
\frac{M_W^4}{(M_W^2 + Q^2)^2}
 \left\{
      (1+(1-y)^2) F_2^{W^{\mp}}
  - y^2 F_L^{W^{\mp}}
  \pm (1-(1-y)^2) x F_3^{W^{\mp}}
 \right\}
\comma
\end{align}
where
\begin{eqnarray}
F_L = F_2 - 2x F_1~.
\end{eqnarray}
Here $x = Q^2/(sy)$ and $y$ denote the Bjorken variables, $Q^2$ is the virtuality of the
exchanged electro-weak gauge boson, $s$ the cms-energy squared, $M_W$ the mass of the 
$W^\pm$-bosons, and $G_F$ Fermi's constant.

At Born level the structure functions are given by the following combinations
of parton distribution functions (PDFs) $q \equiv q(x,Q^2)$, cf.~\cite{Schmitz1997}~:
\begin{align}
\label{eq:F2plBorn}
 F_2^{W^+} ={}& 2x[ (\CKM{ud} + \CKM{cd}) d       
                 +(\CKM{us} + \CKM{cs}) s
                 +(\CKM{ud} + \CKM{us}) \bar{u}
                ]
\comma
\\
\label{eq:F2miBorn}
 F_2^{W^-} ={}& 2x[ (\CKM{ud} + \CKM{cd}) \bar{d} 
                 +(\CKM{us} + \CKM{cs}) \bar{s} 
                 +(\CKM{ud} + \CKM{us}) u
                ]
\comma
\\
\label{eq:F3plBorn}
 xF_3^{W^+} ={}& 2x[ (\CKM{ud} + \CKM{cd}) d       
                 +(\CKM{us} + \CKM{cs}) s
                 -(\CKM{ud} + \CKM{us}) \bar{u}
                ]
\comma
\\
\label{eq:F3miBorn}
 xF_3^{W^-} ={}& 2x[-(\CKM{ud} + \CKM{cd}) \bar{d} 
                 -(\CKM{us} + \CKM{cs}) \bar{s} 
                 +(\CKM{ud} + \CKM{us}) u
                ]
\comma
\\
 F_L^{W^+} ={}& F_L^{W^-} = 0
\comma
\end{align}
where $V_{ij}$ denote the Cabibbo-Kobayashi-Maskawa matrix elements 
\cite{Cabibbo:1963,Kobayashi:1973fv}. In the following we refer to the four-quark picture.

It is worthwhile to study combinations of cross sections
\begin{align}
 \frac{d\sigma^{\nu}}{dx dy}
 +
 \frac{d\sigma^{\bar{\nu}} }{dx dy}
 =:&
 \frac{G_F^2 s}{4 \pi}
 \left\{
    (1+(1-y)^2) F_2^{W^++W^-}
  - y^2 F_L^{W^++W^-}
  + (1-(1-y)^2) x F_3^{W^++W^-}
 \right\}
\comma
\\
 \frac{d\sigma^{\nu}}{dx dy}
 -
 \frac{d\sigma^{\bar{\nu}} }{dx dy}
 =:&
 \frac{G_F^2 s}{4 \pi}
 \left\{
    (1+(1-y)^2) F_2^{W^+-W^-}
  - y^2 F_L^{W^+-W^-}
  + (1-(1-y)^2) x F_3^{W^+-W^-}
 \right\}
\comma
\end{align}
which are symmetric/antisymmetric under crossing, respectively.
The following partonic quantities are introduced\, :
\begin{align}
  \mathcal{F}_{2}^{W^\pm} := \frac{1}{2x} F_{2}^{W\pm}
\comma
\quad
  \mathcal{F}_{3}^{W^\pm} := \frac{1}{2} F_{3}^{W\pm}
\period
\end{align}
The Mellin transforms of the structure functions read
\begin{align}
 F_2^{W^\pm}(N) 
 :={}& \int_0^1 dx x^{N-2} F_2^{W^\pm}(x) 
 = 2\int_0^1 dx x^{N-1} \mathcal{F}_2^{W^\pm}(x)
 =: 2 \mathcal{F}_2^{W^\pm}(N)
\comma
\N\\
 F_3^{W^\pm}(N) 
 :={}& \int_0^1 dx x^{N-1} F_3^{W^\pm}(x) 
 = 2\int_0^1 dx x^{N-1} \mathcal{F}_3^{W^\pm}(x)
 =: 2 \mathcal{F}_3^{W^\pm}(N)
\period
\end{align}
In the following formulae, we will work in Mellin space and drop the
argument $N$ for brevity.

\begin{figure}[htbp]
  \centering
  \parbox[t]{.3\textwidth}{\centering
  \includegraphics[width=.15\textwidth]{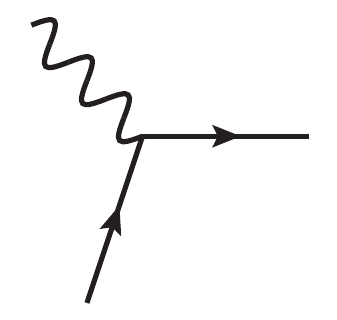}\\
  (a)
  }
  \parbox[t]{.3\textwidth}{\centering
  \includegraphics[width=.15\textwidth]{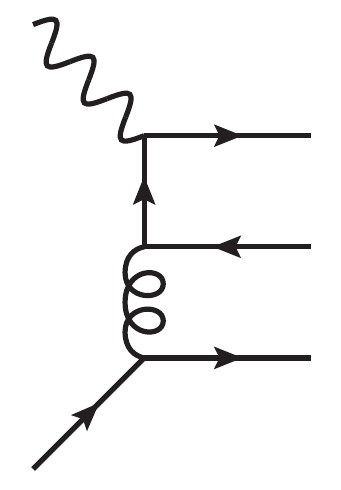}\\
  (b)
  }
  \caption{Born diagrams for main processes contributing to
  $W$-boson exchange. The weavy lines denote $W$ bosons, the curly lines gluons, and the arrow-lines
quarks.}
  \label{fig:IncVertCombs}
\end{figure}

There are diagrams in which the incoming fermion line runs through the
$W$-boson-quark vertex, and others where these two fermion lines are
separated. Examples are given in Figure~\ref{fig:IncVertCombs}.

It is useful to separate the corresponding terms in the Wilson
coefficients into ``valence'' and ``sea'' contributions, respectively.
The valence parts are flavor-diagonal while the sea parts do not
distinguish different flavors. However, differences in the quark
masses are detected. Hence the $c$-quark is treated differently from
$u,d,s$.  

\begin{figure}[htpb]
  \centering
  \includegraphics[width=.32\textwidth]{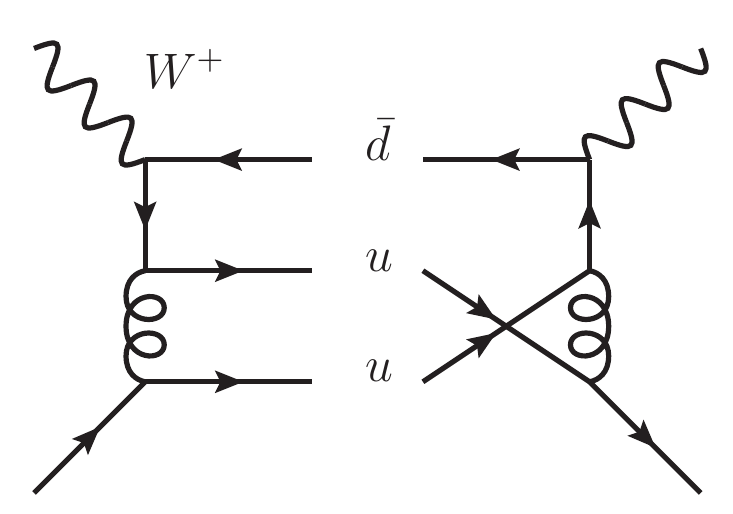}
  \hspace{2em}
  \includegraphics[width=.32\textwidth]{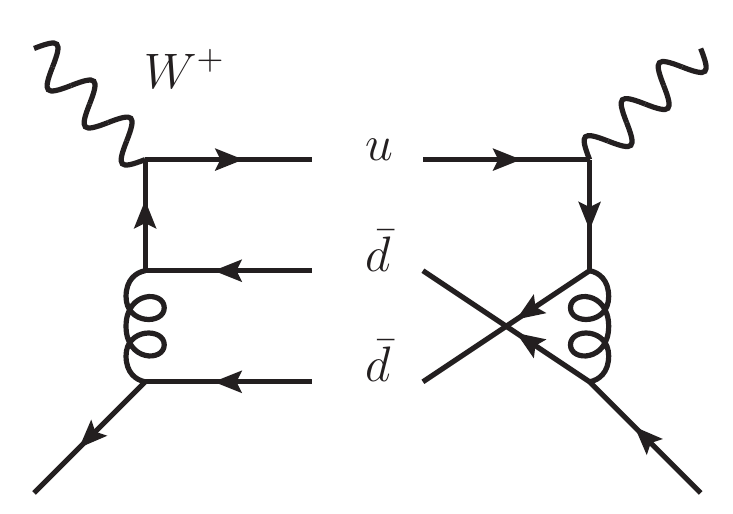}
  \caption{Valence like interference terms for $W^+u$-scattering and
    $W^+\bar{d}$-scattering, which have no counter parts on tree
    level.}
  \label{fig:CCinterferenceDiags}
\end{figure}

Obviously, all terms built from graphs like
Figure~\ref{fig:IncVertCombs}(a) and their QCD corrections form valence
contributions, and all sea contributions are built from graphs like
Figure~\ref{fig:IncVertCombs}(b).  However, there are interference
contributions from the latter class of graphs, see e.g.\ 
Figure~\ref{fig:CCinterferenceDiags}, which clearly form valence
terms.

Note that minus signs derive from the charge conjugation antisymmetry
of the fermion line to which the $W$-boson is attached.  This
antisymmetry is due to the presence of a single $\gamma_5$-matrix and hence
only occurs in contributions to $F_3$.  The emergence of these minus
signs is shortly illustrated in the following.  In these
considerations, factors of $i$ or $(-1)$ stemming from the Feynman
rules are not of relevance, since expressions with the same number of
vertices and propagators are compared.

\begin{figure}
 \centering
 \parbox[t]{.3\textwidth}{\centering 
 \includegraphics[width=.3\textwidth]{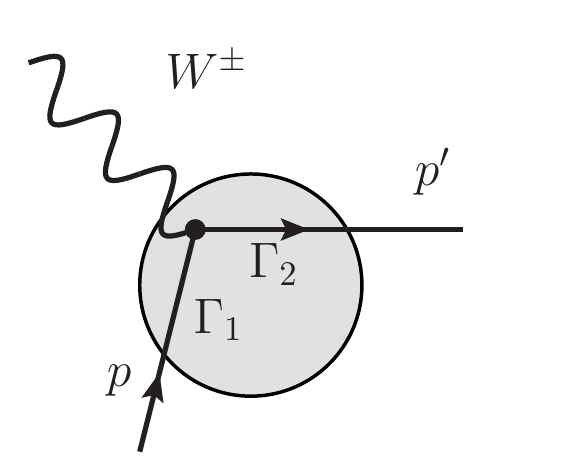}\\
  (a)}
 \parbox[t]{.3\textwidth}{\centering
 \includegraphics[width=.3\textwidth]{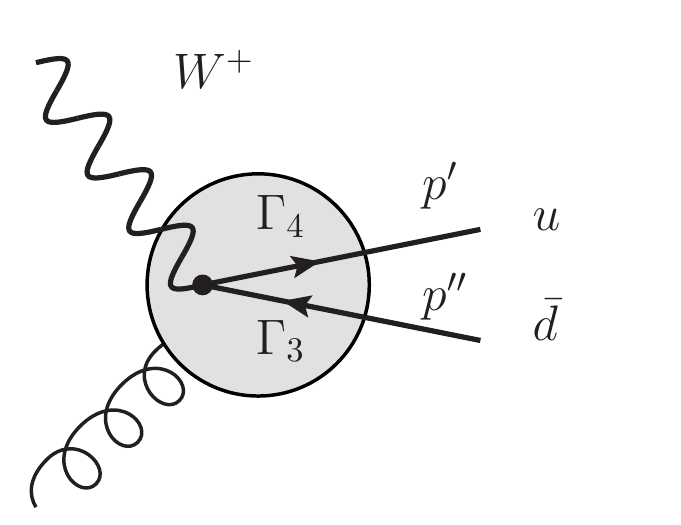}
  (b)}
 \parbox[t]{.3\textwidth}{\centering
 \includegraphics[width=.3\textwidth]{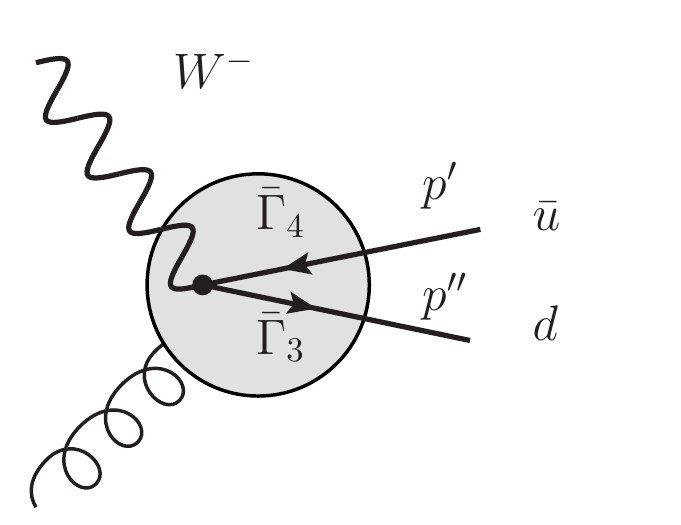}
  (c)}
 \caption{The QCD corrections, denoted by the gray area, are connected
          to the scattered quark line through gluon exchange.}
 \label{fig:NSscatter}
\end{figure}
Figure~\ref{fig:NSscatter} schematically shows the structure of the
graphs in which (a) the fermion line coupling to the $W^\pm q q$-vertex is
incoming or in which (b,c) this fermion is pair produced. The incoming
gluon line in the latter could also be replaced by an incoming fermion
line, that passes through to the final state. The gray area denotes
any QCD correction, that couples to the fermion line via the gluon
field.
The fermion trace of a diagram depicted by Figure~\ref{fig:NSscatter}(a)
can be written as
\begin{align}
  T^{W^+,V}_{q}
  =\;
  \sum_{\text{Spins}}
  \bar{u}(p') \Gamma_2 \frac{1+\gamma_5}{2}\gamma_\mu \Gamma_1 u(p) 
  \bar{u}(p) \bar{\Gamma}_1 \frac{1+\gamma_5}{2}\gamma_\nu
\bar{\Gamma}_2 u(p')
\comma
\end{align}
where the $\Gamma_i$ denote products of Dirac matrices, multiplied by
real numbers which also include the denominators of the propagators,
and $\bar{\Gamma}_i = \gamma_0 \Gamma_i^\dagger \gamma_0$ is just
$\Gamma_i$ with inverted order of the factors.  Due to the charge
conjugation properties of the Dirac matrices and Dirac bispinors, there
is a bijection onto diagrams. Assuming an
antifermion in the initial state, the same trace has the form 
\begin{align}
 T_{\bar{q}}^{W^+,V}
 =\;&
 \sum_{\text{Spins}}
 \bar{v}(p) \bar{\Gamma}_1 \frac{1+\gamma_5}{2} \gamma_\mu
 \bar{\Gamma}_2 v(p')
 \bar{v}(p') \Gamma_2  \frac{1+\gamma_5}{2} \gamma_\nu \Gamma_1 v(p)
\N\\
 =\;&
 \sum_{\text{Spins}}
 \bar{v}(p') \Gamma_2 \frac{1+\gamma_5}{2} \gamma_\nu \Gamma_1 v(p)
 \bar{v}(p) \bar{\Gamma}_1  \frac{1+\gamma_5}{2} \gamma_\mu \bar{\Gamma}_2 v(p')
\period
\end{align}
Since the difference between fermion and antifermion bispinors in the
trace only affects the part $\propto m^2$, it contributes to the power
corrections only, and due to antisymmetry of the $\gamma_5$ part,
$T_{\bar{q}}^{W^+}$ and $T_{q}^{W^+}$ only differ by a minus sign in
front of $\gamma_5$.  This leads to the minus signs in the Eqs.
(\ref{eq:3flEvenComb3}, \ref{eq:3flOddComb3}) below when compared to
(\ref{eq:3flEvenComb2}, \ref{eq:3flOddComb2}).

In case of the ``sea''-contributions depicted in Figure
\ref{fig:NSscatter}(b) and (c) the traces read\,:
\begin{align}
\label{eq:TraceWpm}
  T^{W^+,S} 
  =\;&
 \sum_{\text{Spins}}
  \bar{u}_u(p') 
  \Gamma_4 
  \frac{1+\gamma_5}{2} \gamma_\mu 
  \Gamma_3
  v_d(p'')
  \bar{v}_d(p'')
  \bar{\Gamma}_3
  \frac{1+\gamma_5}{2} \gamma_\nu
  \bar{\Gamma}_4
  u_u(p')
\comma
\\
  T^{W^-,S} 
  =\;&
 \sum_{\text{Spins}}
  \bar{u}_d(p'')
  \bar{\Gamma}_3
  \frac{1+\gamma_5}{2} \gamma_\mu
  \bar{\Gamma}_4
  v_u(p')
  \bar{v}_{u}(p')
  \Gamma_4 
  \frac{1+\gamma_5}{2} \gamma_\nu 
  \Gamma_3
  u_d(p'')
\N\\
  =\;&
 \sum_{\text{Spins}}
  \bar{v}_{u}(p')
  \Gamma_4 
  \frac{1+\gamma_5}{2} \gamma_\nu 
  \Gamma_3
  u_d(p'')
  \bar{u}_d(p'')
  \bar{\Gamma}_3
  \frac{1+\gamma_5}{2} \gamma_\mu
  \bar{\Gamma}_4
  v_u(p')
\period
\end{align}
Here, bispinors of down-type (anti)quarks are marked by the subscript
$d$ and the ones of up-type (anti)quarks are marked by $u$.  
In the complete contribution one will find a corresponding diagram with $\Gamma_4$ and 
$\bar{\Gamma}_3$ interchanged, if the down- and up-type lines have the same mass or are both massless. 
This leads to the symmetry between $T^{W^+,\text{S}}$ and $T^{W^-,\text{S}}$ when summed over all 
diagrams.
The combination of this symmetry and the antisymmetry
from above leads to the relations\cite{Buza:1997mg}\,:
\begin{align}
\label{eq:C3PSL3PS00}
  C_{3,q}^{W,\text{PS}} = C_{3,g}^{W} = 0
  ~\text{and}~
  L_{3,q}^{W,\text{PS}} = L_{3,g}^{W} = 0
\period
\end{align}
Another source for negative signs are the valence like interference terms. While the coupling of $W^+$ 
with the down-type quarks in the nonsinglet channel is described by (QCD corrections to) diagrams like 
Fig. 1(a), the nonsinglet coupling of $W^-$ to a down-type quark is only possible through the 
interference terms of Fig.2. Calling these different contributions $C_d^{W^+,\text{NS}}$ and 
$C_d^{W^-,\text{NS}}$, the Wilson coefficients of the combinations of $W^+$ and $W^-$ take the form:
\begin{align}
  C_d^{W^-+W^-,\text{NS}}
 = 
  C_d^{W^+,\text{NS}}
  +
  C_d^{W^-,\text{NS}}
\comma\quad
  C_d^{W^--W^-,\text{NS}}
 = 
  C_d^{W^+,\text{NS}}
  -
  C_d^{W^-,\text{NS}}
\period
\end{align}
As a consequence these combinations are even/odd under crossing, but the individual 
$W^\pm$-contributions are not (cf. [32]). Interestingly, in Mellin space
this lack of symmetry is reflected in an oscillating
behaviour characterized by a factor $(-1)^N$ which, by Carlson's
theorem \cite{Carlson1914, Titchmarsh39, Ablinger:2013cf}, prevents the Wilson
coefficients from being Mellin invertible.

For these reasons it is useful to study combinations of structure
functions, which have a crossing symmetry by construction. The
factorization of these combinations reads\,:
\begin{align}
\label{eq:3flEvenComb2}
  \mathcal{F}_2^{W^{+}} 
  +
  \mathcal{F}_2^{W^{-}} 
  = &
  \phantom{+}
  \left(
  \vert V_{du} \vert^2
  (d + \bar{d})
  + 
  \vert V_{su} \vert^2
  (s + \bar{s})
  +
  V_u
  (u + \bar{u})
  \right)
  (
    C_{2,q}^{W^++W^-,\text{NS}} 
   +L_{2,q}^{W^++W^-,\text{NS}}
  )
\N\\ &
  + 
  \left(
  \vert V_{dc} \vert^2
  (d + \bar{d})
  + 
  \vert V_{sc} \vert^2
  (s + \bar{s})
  \right)
  H_{2,q}^{W^++W^-,\text{NS}}
\N\\ &
  +
  2V_u
  \left[
  (
   C_{2,q}^{W,\text{PS}}
  +L_{2,q}^{W,\text{PS}}
  )
  \Sigma
  +
  (
   C_{2,g}^{W}
  +L_{2,g}^{W}
  )
  G
  \right]
\N\\ &
  +
  2V_c
  \left[
  H_{2,q}^{W,\text{PS}}
  \Sigma
  +
  H_{2,g}^{W}
  G
  \right]
\comma
\end{align}
\begin{align}
\label{eq:3flOddComb2}
  \mathcal{F}_2^{W^{+}} 
  -
  \mathcal{F}_2^{W^{-}} 
  = &
  \phantom{+}
  \left(
  \vert V_{du} \vert^2
  (d - \bar{d})
  + 
  \vert V_{su} \vert^2
  (s - \bar{s})
  -
  V_u
  (u - \bar{u})
  \right)
  (
    C_{2,q}^{W^+-W^-,\text{NS}} 
   +L_{2,q}^{W^+-W^-,\text{NS}}
  )
\N\\ &
  + 
  \left(
  \vert V_{dc} \vert^2
  (d - \bar{d})
  + 
  \vert V_{sc} \vert^2
  (s - \bar{s})
  \right)
  H_{2,q}^{W^+-W^-,\text{NS}}
\comma
\end{align}
\begin{align}
\label{eq:3flEvenComb3}
  \mathcal{F}_3^{W^{+}} 
  +
  \mathcal{F}_3^{W^{-}} 
  = &
  \phantom{+}
  \left(
  \vert V_{du} \vert^2
  (d - \bar{d})
  + 
  \vert V_{su} \vert^2
  (s - \bar{s})
  +
  V_u
  (u - \bar{u})
  \right)
  (
   C_{3,q}^{W^+-W^-,\text{NS}} 
  +L_{3,q}^{W^+-W^-,\text{NS}}
  )
\N\\ &
  + 
  \left(
  \vert V_{dc} \vert^2
  (d - \bar{d})
  + 
  \vert V_{sc} \vert^2
  (s - \bar{s})
  \right)
  H_{3,q}^{W^+-W^-,\text{NS}}
\comma
\end{align}
\begin{align}
\label{eq:3flOddComb3}
  \mathcal{F}_3^{W^{+}} 
  -
  \mathcal{F}_3^{W^{-}} 
  = &
  \phantom{+}
  \left(
  \vert V_{du} \vert^2
  (d + \bar{d})
  + 
  \vert V_{su} \vert^2
  (s + \bar{s})
  -
  V_u
  (u + \bar{u})
  \right)
  (
   C_{3,q}^{W^++W^-,\text{NS}} 
  +L_{3,q}^{W^++W^-,\text{NS}}
  )
\N\\ &
  + 
  \left(
  \vert V_{dc} \vert^2
  (d + \bar{d})
  + 
  \vert V_{sc} \vert^2
  (s + \bar{s})
  \right)
  H_{3,q}^{W^++W^-,\text{NS}}
\N\\ &
  +
  2V_c
  \left[
  H_{3,q}^{W,\text{PS}}
  \Sigma
  +
  H_{3,g}^{W}
  G
  \right]
\comma
\end{align}
with
\begin{eqnarray}
V_u &=& |V_{du}|^2 + |V_{su}|^2
\\
V_d &=& |V_{du}|^2 + |V_{dc}|^2
\\
V_s &=& |V_{su}|^2 + |V_{sc}|^2
\\
V_c &=& |V_{dc}|^2 + |V_{sc}|^2~.
\end{eqnarray}

As a result, either even or odd moments contribute to the combinations
in Mellin space.  One finds \cite{Buras:1979yt, Politzer:1974fr, Blumlein:1996tp,
Blumlein:1998nv}\,:
\begin{align}
  F_2^{W^{+}} + F_2^{W^{-}}\,:
  &~\text{even $N$}
\comma
\\
  F_2^{W^{+}} - F_2^{W^{-}}\,:
  &~\text{odd $N$}
\comma
\\
  F_3^{W^{+}} + F_3^{W^{-}}\,:
  &~\text{odd $N$}
\comma
\\
  F_3^{W^{+}} - F_3^{W^{-}}\,:
  &~\text{even $N$}
\period
\end{align}
These sequences of even or odd moments then have well defined
$x$-space counter parts.

In order to derive factorization formulae, we choose to take a safe
detour via the relations of parton distributions in the variable
flavor number scheme ($q',\bar{q}'$) \cite{Buza:1996wv, Bierenbaum:2009mv}\,:
\begin{align}
  q' + \bar{q}' 
  =\;& 
   A_{qq,Q}^{\text{NS}} (q+\bar{q}) 
  +\tilde{A}_{qq,Q}^{\text{PS}} \Sigma
  +\tilde{A}_{qg,Q} G
\comma
\N\\
  c' + \bar{c}' 
  =\;& 
   A_{Qq}^{\text{PS}} \Sigma
  +A_{Qg} G
\comma
\N\\
  \Sigma'
  =\;&
   ( n_f \tilde{A}_{qq,Q}^{\text{PS}}
    +A_{Qq}^{\text{PS}}
    +A_{qq,Q}^{\text{NS}}
   ) \Sigma
  +( n_f \tilde{A}_{qg,Q}
    +A_{Qg}
   ) G
\comma
\N\\
  G'
  =\;&
   A_{gq,Q} \Sigma
  +A_{gg,Q} G
\comma
\N\\
\Delta'_q =\;& q' + \bar{q}' - \frac{1}{n_f +1} \Sigma' 
\period
\end{align}
Here the following notation was used\,:
\begin{align}
\label{eq:DefAtilde}
  \tilde{A}_{ij}(n_f+1) \equiv \frac{1}{n_f} A_{ij}(n_f+1)
\period
\end{align}

From this point on, the number of light flavors contributing to the
light flavor Wilson coefficients is written explicitly as an argument.
The four-flavor expressions read\,:
\begin{align}
 \label{eq:4flEvenCombF2}
 \mathcal{F}_{2}^{W^+}
 +
 \mathcal{F}_{2}^{W^-}
 =\;&
  (V_d \Delta_d' + V_s \Delta_s' + V_u \Delta_u' + V_c \Delta_c') 
  C_{2,q}^{W^++ W^-,\text{NS}}(n_f+1)
\N\\&
 +\frac{V_u+V_c}{n_f+1}
  \left[
    C_{2,q,(n_f+1)}^{W^++ W^-,\text{S}} \Sigma'
   +2(n_f+1) C_{2,g}^{W}(n_f+1) G'
  \right]
\comma
\\
 \label{eq:4flOddCombF2}
  \mathcal{F}_2^{W^{+}} 
  -
  \mathcal{F}_2^{W^{-}} 
  = &
  \left(
  V_d
  (d' - \bar{d}')
  + 
  V_s
  (s' - \bar{s}')
  -
  V_u
  (u' - \bar{u}')
  -
  V_c
  (c' - \bar{c'})
  \right)
  C_{2,q}^{W^+-W^-,\text{NS}} (n_f+1)
\comma
\\
\label{eq:4flEvenCombF3}
 \mathcal{F}_{3}^{W^+}
 -
 \mathcal{F}_{3}^{W^-}
 =\;&
  (V_d \Delta_d' + V_s \Delta_s' - V_u \Delta_u' - V_c \Delta_c') 
  C_{3,q}^{W^++ W^-,\text{NS}}(n_f+1)
\comma
\\
\label{eq:4flOddCombF3}
  \mathcal{F}_3^{W^{+}} 
  +
  \mathcal{F}_3^{W^{-}} 
  = &
  \left(
  V_d
  (d' - \bar{d}')
  + 
  V_s
  (s' - \bar{s}')
  +
  V_u
  (u' - \bar{u}')
  +
  V_c
  (c' - \bar{c}')
  \right)
  C_{3,q}^{W^+-W^-,\text{NS}}(n_f+1)
\period
\end{align}
Comparing the coefficients of $\Sigma, G, \Delta_q, V_u, V_c$ in
these relations with the 3-flavor representation (\ref{eq:3flEvenComb2}) one finds
\begin{align}
   C_{2,q}^{W^+\pm W^-,\text{NS}}(n_f)
  +L_{2,q}^{W^+\pm W^-,\text{NS}}
  =\;&
   A_{qq,Q}^{\text{NS}}
   C_{2,q}^{W^+\pm W^-,\text{NS}}(n_f+1)
\comma
\N\\
   H_{2,q}^{W^+\pm W^-,\text{NS}}
  =\;&
   A_{qq,Q}^{\text{NS}}
   C_{2,q}^{W^+\pm W^-,\text{NS}}(n_f+1)
\comma
\N\\
   C_{2,q}^{W,\text{PS}}(n_f)
  +L_{2,q}^{W,\text{PS}}
  =\;&
   \tilde{A}_{qq,Q}^{\text{PS}}
   C_{2,q}^{W^++ W^-,\text{NS}}(n_f+1)
\N\\&
  +C_{2,q}^{W,\text{PS}}(n_f+1)
   \left(
     n_f 
     \tilde{A}_{qq,Q}^{\text{PS}}
    +A_{Qq}^{\text{PS}}
    +A_{qq,Q}^{\text{NS}}
   \right)
\N\\&
  +A_{gq,Q}
   C_{2,g}^{W}(n_f+1)
\comma
\N\\
  H_{2,q}^{W,\text{PS}}
  =\;&
  \frac{1}{2}
  \left(
    \tilde{A}_{qq,Q}^{\text{PS}}
   + 
    A_{Qq}^{\text{PS}}
  \right)
  C_{2,q}^{W^++ W^-,\text{NS}}(n_f+1)
\N\\&
  +
  \left(
    n_f 
    \tilde{A}_{qq,Q}^{\text{PS}}
   +A_{Qq}^{\text{PS}}
   +A_{qq,Q}^{\text{NS}}
  \right)
  C_{2,q}^{W,\text{PS}}(n_f+1)
\N\\&
  + 
  A_{gq,Q}
  C_{2,g}^{W}(n_f+1)
\comma
\N\\
   C_{2,g}^{W}(n_f)
  +L_{2,g}^{W}
  =\;&
  \tilde{A}_{qg,Q}
  C_{2,q}^{W^++ W^-,\text{NS}}(n_f+1)
\N\\&
  + \left( 
    n_f
    \tilde{A}_{qg,Q}
   +A_{Qg}
  \right)
  C_{2,q}^{W,\text{PS}}(n_f+1)
  + 
  A_{gg,Q}
  C_{2,g}^{W}(n_f+1)
\comma
\N\\
   H_{2,g}^{W}
  =\;&
  \frac{1}{2}
  \left(
    \tilde{A}_{qg,Q}
   +A_{Qg}
  \right)
  C_{2,q}^{W^++ W^-,\text{NS}}(n_f+1)
\N\\&
  +
  \left(
    n_f
    \tilde{A}_{qg,Q}
   +A_{Qg}
  \right)
  C_{2,q}^{W,\text{PS}}(n_f+1)
  +
  A_{gg,Q}
  C_{2,g}^{W}(n_f+1)
\comma
\end{align}
where the odd-$N$ combinations are included in analogy to the even-$N$
ones.  From Eqs.~(\ref{eq:4flEvenCombF3}) and (2.22)
one can deduce similarly
\begin{align}
  L_{3,q}^{W^+\pm W^-,\text{NS}}
  =\;&
   A_{qq,Q}^{\text{NS}}
   C_{3,q}^{W^+\pm W^-,\text{NS}}(n_f+1)
   -C_{3,q}^{W^+\pm W^-,\text{NS}}(n_f)
\comma
\N\\
   H_{3,q}^{W^+\pm W^-,\text{NS}}
  =\;&
   A_{qq,Q}^{\text{NS}}
   C_{3,q}^{W^+\pm W^-,\text{NS}}(n_f+1)
\comma
\N\\
   H_{3,q}^{W,\text{PS}}
  =\;&
   \frac{1}{2}
   (
    \tilde{A}_{qq,Q}^{\text{PS}}
   -
    A_{Qq}^{\text{PS}}
   )
   C_{3,q}^{W^++ W^-,\text{NS}}(n_f+1)
\comma
\N\\
   H_{3,g}^{W}
  =\;&
   \frac{1}{2}
   (
    \tilde{A}_{qg,Q}
   -
    A_{Qg}
   )
   C_{3,q}^{W^++ W^-,\text{NS}}(n_f+1)
\period
\label{eq237}
\end{align}
By inserting the odd-$N$ factorization relations into
(\ref{eq:3flOddComb2}) and (2.21), and comparing with
(\ref{eq:4flOddCombF2}) and (\ref{eq:4flOddCombF3}), respectively,
one finds\,:
\begin{align}
  q' - \bar{q}' 
  =\;& 
   A_{qq,Q}^{\text{NS}} (q-\bar{q}) 
\comma
\N\\
  c'-\bar{c}' =\;& 0
\period  
\end{align}
Expanding the above relations up to order $a_s^2$, 
\begin{align}
  f(a_s) = \sum_{l=0}^{\infty} a_s^l f^{(l)}
\comma
\end{align}
one finds the asymptotic representations. The relations for the
longitudinal structure function $F_L$ are almost complete analogs to
the ones for $F_2$, so they are included using the index $i=2/L$,
where the only structural difference, denoted by Kronecker symbols
$\delta_{i,2}$, derives from the fact, that the coefficients
$C_L^{\text{NS}}$ do not have a Born contribution. On the Born level,
one obviously has
\begin{align}
 H_{i,q}^{W^+\pm W^-,\text{NS},(0)} =\;& \delta_{i,2}
\comma
\N\\
 H_{3,q}^{W^+\pm W^-,\text{NS},(0)} =\;& 1
\period
\end{align}
At 1-loop level, one obtains
\begin{align}
  H_{i,q}^{W^+\pm W^-,\text{NS},(1)} 
  =\;& 
  C_{i,q}^{W^+\pm W^-,\text{NS},(1)}(n_f+1)
\comma
\N\\
  H_{i,g}^{W,(1)} 
  =\;& 
  \frac{1}{2} \delta_{i,2} A_{Qg}^{(1)} 
  + C_{i,g}^{W,(1)}(n_f+1)
\comma
\N\\
  H_{3,q}^{W^+\pm W^-,\text{NS},(1)} 
  =\;&
  C_{3,q}^{W^+\pm W^-,\text{NS},(1)}(n_f+1)
\comma
\N\\
  H_{3,g}^{W,(1)}
  =\;&
  -\frac{1}{2} A_{Qg}^{(1)}
\comma
\end{align}
in accordance with the asymptotic expressions derived in
Ref.~\cite{Blumlein:2011zu}.  At 2-loop
order, the asymptotic formulae take the form\,:
\begin{align}
\label{eq:asympt2loop}
  L_{i,q}^{W^+\pm W^-,\text{NS},(2)}
  =\;&
  \delta_{i,2}
  A_{qq,Q}^{\text{NS},(2)}
  + C_{i,q}^{W^+\pm W^-,\text{NS},(2)}(n_f+1)
  - C_{i,q}^{W^+\pm W^-,\text{NS},(2)} (n_f)
\comma
\N\\
  H_{i,q}^{W^+\pm W^-,\text{NS},(2)}
  =\;&
  \delta_{i,2} 
  A_{qq,Q}^{\text{NS},(2)}
  +C_{i,q}^{W^+\pm W^-,\text{NS},(2)}(n_f+1)
\comma
\N\\ 
  L_{i,q}^{W,\text{PS},(2)}
  =\;&
    C_{i,q}^{W,\text{PS},(2)}(n_f+1)
  - C_{i,q}^{W,\text{PS},(2)}(n_f)
  =0
\comma
\N\\
  H_{i,q}^{W,\text{PS},(2)}
  =\;&
  \frac{1}{2}
  \delta_{i,2} 
  A_{Qq}^{\text{PS},(2)}
  +C_{i,q}^{W,\text{PS},(2)}(n_f+1)
\comma
\N\\
  L_{i,g}^{W,(2)}
  =\;&
  A_{gg,Q}^{(1)}
  C_{i,g}^{W,(1)}(n_f+1)
  +C_{i,g}^{W,(2)}(n_f+1)
  -C_{i,g}^{W,(2)}(n_f)
\comma
\N\\
  H_{i,g}^{W,(2)}
  =\;&
  A_{gg,Q}^{(1)}
  C_{i,g}^{W,(1)}(n_f+1)
  +C_{i,g}^{W,(2)}(n_f+1)
\N\\&
  +\frac{1}{2} 
  \left(
    \delta_{i,2}
    A_{Qg}^{(2)}
   +A_{Qg}^{(1)}
    C_{i,q}^{W^++ W^-,\text{NS},(1)}(n_f+1)
  \right)
\comma
\N\\
  L_{3,q}^{W^+\pm W^-,\text{NS},(2)}
  =\;&
  A_{qq,Q}^{\text{NS},(2)}
  +C_{3,q}^{W^+\pm W^-,\text{NS},(2)}(n_f+1)
  -C_{3,q}^{W^+\pm W^-,\text{NS},(2)}(n_f)
\comma
\N\\
  H_{3,q}^{W^+\pm W^-,\text{NS},(2)}
  =\;&
  A_{qq,Q}^{\text{NS},(2)}
  +C_{3,q}^{W^+\pm W^-,\text{NS},(2)}(n_f+1)
\comma
\N\\
  H_{3,q}^{W,\text{PS},(2)}
  =\;&
  -\frac{1}{2} A_{Qq}^{\text{PS},(2)}
\comma
\N\\
  H_{3,g}^{W,(2)}
  =\;&
  \frac{1}{2} 
  \left(
    -A_{Qg}^{(2)}
    -A_{Qg}^{(1)}
     C_{3,q}^{W^++ W^-,\text{NS},(1)}(n_f+1)
   \right)
\period
\end{align}
Comparing with results given in \cite{Buza:1997mg}, one finds that the
above relations agree for $H_{2,g}^{W,(1)}$ and
$H_{2,q}^{W,\text{PS},(2)}$.  They further correct $H_{2,g}^{W,(2)}$
with regard to heavy quark loop contributions on external lines, cf.
\cite{Bierenbaum:2009mv}, and correct signs in $H_{3,g}^{W,(1)}$,
$H_{3,g}^{W,(2)}$ and $H_{3,g}^{W,\text{PS},(2)}$. The correctness of these signs was checked in two 
ways: First by a careful independent recalculation of the exact 1-loop gluon-boson fusion 
contributions to $H_{3,g}^{W}$. Then we checked the signs of $H_{3,g}^{W,(1)}$, 
$H_{3,q}^{W,\text{PS}}$ by calculating their leading $\ln(m^2/Q^2)$-contributions. Details of these 
calculations are given in Appendix A.

\section{The Wilson Coefficients}

\vspace*{1mm}
\noindent
In the following, we present the $N$-space expressions for the Wilson coefficients having been
derived in the previous Section. The non-singlet light flavor Wilson coefficients
$c_{i,q}^{(i),\text{ns},\pm}$ defined in Eq.~(94) of \cite{Moch:1999eb}
are related to the ones used above via
\begin{align}
 C_{i,q}^{W^+\pm W^-,\text{NS},(i)} 
 ={}&\;
 c_{i,q}^{(i),\text{ns},+}
 \pm
 c_{i,q}^{(i),\text{ns},-},
\quad
 i=2,3
\comma
\end{align}
where the $\pm$-signs correspond to each other on the left and right
hand sides.  The splitting denoted by superscripts $+$ or $-$ is the
same as in Eq.~(14) in \cite{Zijlstra:1992kj}.  The gluonic and pure singlet
Wilson coefficients can be taken over from the electromagnetic case.
Using $c_{i,\text{ps}}^{(i)}$ and $c_{i,g}^{(i)}$ from \cite{Vermaseren:2005qc},
one finds
\begin{align}
  C_{i,q}^{W,\text{PS},(i)}(n_f)
  =
  \frac{1}{n_f}
  c_{i,\text{ps}}^{(i)},
\quad
  C_{i,g}^{W,\text{PS},(i)}(n_f)
  =
  \frac{1}{n_f}
  c_{i,g}^{(i)},
\quad i=2,L
\period
\end{align}
The contributions to the non-singlet Wilson coefficients of the
structure functions $F_{2,3}$ were given in \cite{Zijlstra:1991qc,
vanNeerven:1991nn, Zijlstra:1992qd, Zijlstra:1992kj}, and confirmed in \cite{Moch:1999eb}\footnote{ 
See also \cite{Moch:2007rq}, where also the even-odd-$N$ difference for the Wilson
coefficient of $F_L$ is published.}.

Very often the massless Wilson coefficients given in the literature for general values of $N$ 
are understood to be valid either for even or odd values only. The representations for general 
values of $N$ can, however, be obtained by a Mellin transform of the $x$-space expressions, 
cf. e.g.~\cite{Vermaseren:2005qc,Blumlein:2009tj}, resp. for odd moments \cite{Moch:2007rq} and 
the even-/odd-$N$ combinations from \cite{vanNeerven:1991nn}.

The heavy flavor Wilson coefficients in Mellin $N$-space are constructed as described
above and are given in terms of harmonic sums \cite{Vermaseren:1998uu,Blumlein:1998if}
\begin{eqnarray}
S_{b,\vec{a}}(N) = \sum_{k=1}^N \frac{({\rm sign}(b))^k}{k^{|b|}} S_{\vec{a}}(k),~~~~S_\emptyset = 
1,~~~N \in \mathbb{N} \backslash \{0\}~.
\end{eqnarray}
For brevity we use the notation $S_{\vec{a}}(N) \equiv S_{\vec{a}}$.
The individual Wilson coefficients read~:
\begin{eqnarray}
\LtwoWpWqNStwo &=&
  C_F T_F \Biggl\{
    \frac{16}{3} S_{2,1}
   -\frac{2 (29 N^2+29 N-6)}{9 N (N+1)} S_1{}^2
   +\frac{2 (35 N^2+35 N-2)}{3 N (N+1)} S_2
\nonumber\\ &&
   -\frac{2 (359 N^4+844 N^3+443 N^2+66 N+72)}{27 N^2 (N+1)^2} S_1
   +\frac{P_{1}}{27 N^3 (N+1)^3}
   -\frac{8}{9} S_1{}^3
\nonumber\\ &&
   +\frac{8}{3} S_2 S_1
   -\frac{112 S_3}{9}
  \Biggr\}
 +C_F T_F \ln^2\Biggl(\frac{m^2}{Q^2}\Biggr) \Biggl\{
    \frac{2 (3 N^2+3 N+2)}{3 N (N+1)}
   -\frac{8 S_1}{3}
  \Biggr\}
\nonumber\\ &&
 +C_F T_F \ln\Biggl(\frac{m^2}{Q^2}\Biggr) \Biggl\{
    \frac{2 (3 N^4+6 N^3+47 N^2+20 N-12)}{9 N^2 (N+1)^2}
   -\frac{80 S_1}{9}
   +\frac{16 S_2}{3}
  \Biggr\},
\nonumber\\
\end{eqnarray}
with
\begin{equation}
P_{1} = 795 N^6+1587 N^5+1295 N^4+397 N^3+50 N^2+300 N+216,
\end{equation}
\begin{eqnarray}
\lefteqn{\HtwoWpWqNStwo =} \nonumber\\ &&
   C_F^2 \Biggl\{
      8 (-1)^N \Biggl[
        5 S_{-4}
       -4 S_{-3,1}
       -2 S_{-2,2}
       +2 S_{-3} S_1
       +4 S_{-2,1} S_1
       +2 S_{-2} S_2
\nonumber\\ &&
       -4 S_{-2} S_1{}^2
       -3 \zeta_2 S_{-2}
       +  \zeta_2 S_2
       -2 \zeta_2 S_1{}^2
       +4 \zeta_3 S_1
       -\frac{8}{5} \zeta_2^2
      \Biggr]
     +12 S_4
     +40 S_{3,1}
     -24 S_{2,1,1}
\N\\ &&
     -20 S_2 S_1{}^2
     +2 S_1{}^4
     -24 S_3 S_1
     +16 S_{2,1} S_1
     +24 S_{-2}{}^2
     +6 S_2{}^2
     -8 \zeta_2 S_2
     +24 \zeta_2 S_{-2}
\N\\ &&
     +16 \zeta_2 S_1{}^2
     +16 \zeta_3 S_1
     +\frac{64}{5} \zeta_2^2
     +4 (-1)^N \Biggl[
        -\frac{4 (4 N-3)}{N (N+1)} S_{-2} S_1
        -\frac{2 (4 N-3)}{N (N+1)} \zeta_2 S_1
\N\\ &&
        -\frac{8}{N+1} S_{-3}
        +\frac{  (4 N-5)}{N (N+1)} \zeta_3
        +\frac{8 (2 N-1)}{N (N+1)} S_{-2,1}
\N\\ &&
        -\frac{2 P_{2}}{(N-2) N^2 (N+1)^2 (N+3)} S_{-2}
        -\frac{  P_{2}}{(N-2) N^2 (N+1)^2 (N+3)} \zeta_2
      \Biggr]
\N\\ &&
     +\frac{2 (3 N^2+3 N-2)}{N (N+1)} S_1{}^3
     -\frac{2 (9 N^2+9 N-10)}{N (N+1)} S_2 S_1
\N\\ &&
     -\frac{27 N^4+26 N^3-9 N^2-40 N-24}{2 N^2 (N+1)^2} S_1{}^2
     -\frac{P_{3}}{2 N^3 (N+1)^3} S_1
     +\frac{8 (4 N-3)}{N (N+1)} \zeta_2 S_1
\N\\ &&
     +\frac{4 P_{2}}{(N-2) N^2 (N+1)^2 (N+3)} \zeta_2
     -\frac{4 (18 N^2-2 N+7)}{N (N+1)} \zeta_3
\N\\ &&
     +\frac{95 N^4+162 N^3+35 N^2-32 N-16}{2 N^2 (N+1)^2} S_2
     -\frac{2 (9 N^2+25 N-10)}{N (N+1)} S_3
\N\\ &&
     +\frac{4 (3 N^2+3 N-2)}{N (N+1)} S_{2,1}
     +\frac{P_{4}}{8 (N-2) N^4 (N+1)^4 (N+3)}
   \Biggr\}
\N\\ &&
  +T_F C_F \ln^2\Biggl(\frac{m^2}{Q^2}\Biggr) \Biggl\{
      \frac{2 (3 N^2+3 N+2)}{3 N (N+1)}
     -\frac{8 S_1}{3}
   \Biggr\}
\N\\ &&
  +T_F C_F\ln\Biggl(\frac{m^2}{Q^2}\Biggr) \Biggl\{
      \frac{2 (3 N^4+6 N^3+47 N^2+20 N-12)}{9 N^2 (N+1)^2}
     -\frac{80 S_1}{9}
     +\frac{16 S_2}{3}
   \Biggr\}
\N\\ &&
  +T_F C_F \Biggl\{
     -\frac{8}{9} S_1{}^3
     -\frac{2 (29 N^2+29 N-6)}{9 N (N+1)} S_1{}^2
     +\frac{2 (35 N^2+35 N-2)}{3 N (N+1)} S_2
\N\\ &&
     -\frac{2 (359 N^4+844 N^3+443 N^2+66 N+72)}{27 N^2 (N+1)^2} S_1
     +\frac{8}{3} S_2 S_1
     +\frac{P_{5}}{27 N^3 (N+1)^3}
\N\\ &&
     -\frac{112}{9} S_3
     +\frac{16}{3} S_{2,1}
   \Biggr\}
  +n_f T_F C_F \Biggl\{
     -\frac{8}{9} S_1{}^3
     -\frac{2 (29 N^2+29 N-6) S_1{}^2}{9 N (N+1)}
\N\\ &&
     -\frac{2 (247 N^4+620 N^3+331 N^2+66 N+72) S_1}{27 N^2 (N+1)^2}
     +\frac{8}{3} S_2 S_1
     +\frac{P_{6}}{54 N^3 (N+1)^3}
\N\\ &&
     +\frac{2 (85 N^2+85 N-6) S_2}{9 N (N+1)}
     -\frac{88 S_3}{9}
     +\frac{16}{3} S_{2,1}
   \Biggr\}
\N\\ &&
  +C_A C_F \Biggl\{
      4 (-1)^N \Biggl[
        -5 S_{-4}
        +4 S_{-3,1}
        +2 S_{-2,2}
        -2 S_{-3} S_1
        -4 S_{-2,1} S_1
        -2 S_{-2} S_2
\N\\ &&
        +4 S_{-2} S_1{}^2
        +3 \zeta_2 S_{-2}
        -  \zeta_2 S_2
        +2 \zeta_2 S_1{}^2
        -4 \zeta_3 S_1
        +\frac{8}{5} \zeta_2^2
      \Biggr]
     +\frac{22}{9} S_1{}^3
     -8 \zeta_2 S_1{}^2
\N\\ &&
     +4 S_2 S_1{}^2
     -32 \zeta_3 S_1
     +24 S_3 S_1
     -16 S_{2,1} S_1
     -12 S_{-2}{}^2
     -4 S_2{}^2
     -12 \zeta_2 S_{-2}
     +4 \zeta_2 S_2
\N\\ &&
     -8 S_4
     -24 S_{3,1}
     +24 S_{2,1,1}
     -\frac{32 \zeta_2^2}{5}
     +\frac{367 N^2+367 N-66}{18 N (N+1)} S_1{}^2
\N\\ &&
     +\frac{P_{7}}{54 N^2 (N+1)^3} S_1
     -\frac{4 (4 N-3)}{N (N+1)} \zeta_2 S_1
     -\frac{2 (11 N^2+11 N+6)}{3 N (N+1)} S_2 S_1
\N\\ &&
     -\frac{2 P_{2}}{(N-2) N^2 (N+1)^2 (N+3)} \zeta_2
     +\frac{2 (27 N^2+7 N+13)}{N (N+1)} \zeta_3
\N\\ &&
     -\frac{1067 N^3+2134 N^2+929 N-66}{18 N (N+1)^2} S_2
     +\frac{2 (121 N^2+193 N-72)}{9 N (N+1)} S_3
\N\\ &&
     -\frac{4 (11 N^2+11 N-6)}{3 N (N+1)} S_{2,1}
     -\frac{P_{8}}{216 (N-2) N^3 (N+1)^3 (N+3)}
\N\\ &&
     +2 (-1)^N \Biggl[
        -\frac{8 (2 N-1)}{N (N+1)} S_{-2,1}
        +\frac{2 (4 N-3)}{N (N+1)} \zeta_2 S_1
        +\frac{4 (4 N-3)}{N (N+1)} S_{-2} S_1
\N\\ &&
        +\frac{  P_{2}}{(N-2) N^2 (N+1)^2 (N+3)} \zeta_2
        -\frac{  (4 N-5)}{N (N+1)} \zeta_3
        +\frac{8}{N+1} S_{-3}
\N\\ &&
        +\frac{2 P_{2}}{(N-2) N^2 (N+1)^2 (N+3)} S_{-2}
      \Biggr]
   \Biggr\},
\end{eqnarray}
with
\begin{eqnarray}
 P_{2} &=& 2 N^6-2 N^5-3 N^4+26 N^3-45 N^2-34 N-48\\
 P_{3} &=& 51 N^6+203 N^5+207 N^4+33 N^3+106 N^2+160 N+48\\
 P_{4} &=& 331 N^{10}+1179 N^9-848 N^8-4754 N^7-2157 N^6+4247 N^5+3474 N^4-2528 N^3
\N\\&&
     -4976 N^2-2704 N-480\\
 P_{5} &=& 795 N^6+1587 N^5+1295 N^4+397 N^3+50 N^2+300 N+216\\
 P_{6} &=& 1371 N^6+2517 N^5+1397 N^4+31 N^3+140 N^2+648 N+360\\
 P_{7} &=& 3155 N^5+11607 N^4+12279 N^3+3329 N^2+510 N+792\\
 P_{8} &=& 16395 N^8+47520 N^7-51416 N^6-162042 N^5-99843 N^4+7930 N^3+21432 N^2
\N\\&&
-25848 N-23760,
\end{eqnarray}
\begin{eqnarray}
\HtwoWmWqNStwo &=&
  \HtwoWpWqNStwo
\N\\ &&
 +C_F (C_F - C_A/2) \Biggl\{
     64 (-1)^N S_{-3,1}
    -\frac{64 (-1)^N (2 N-1)}{N (N+1)} S_{-2,1}
\N\\&&
    -64 (-1)^N S_1 S_{-2,1}
    +32 (-1)^N S_{-2,2}
    +\frac{16 (2 N^2+2 N+1)}{N^3 (N+1)^3} S_1
\N\\&&
    -\frac{16 P_{9}}{(N-2) N^2 (N+1)^2 (N+2) (N+3)} \zeta_2
\N\\&&
    +\frac{16 (-1)^N P_{10}}{(N-2) N^2 (N+1)^2 (N+2) (N+3)} S_{-2}
\N\\&&
    +\frac{8 (-1)^N P_{10}}{(N-2) N^2 (N+1)^2 (N+2) (N+3)} \zeta_2
\N\\&&
    -\frac{4 P_{11}}{(N-2) N^4 (N+1)^4 (N+2) (N+3)}
    +32 (-1)^N \zeta_2 S_1{}^2
\N\\&&
    +48 (-1)^N \zeta_2 S_{-2}
    +\frac{16 (-1)^N (4 N-3)}{N (N+1)} S_1 \zeta_2
    -16 (-1)^N \zeta_2 S_2
\N\\&&
    -64 (-1)^N \zeta_3 S_1
    +64 (-1)^N S_{-2} S_1{}^2
    -80 (-1)^N S_{-4}
    +\frac{64 (-1)^N}{N+1} S_{-3}
\N\\&&
    -32 (-1)^N S_{-3} S_1
    +\frac{32 (-1)^N (4 N-3)}{N (N+1)} S_1 S_{-2}
    -32 (-1)^N S_{-2} S_2
\N\\&&
    +\frac{128}{5} (-1)^N \zeta_2^2
    -\frac{8 (-1)^N (4 N-5)}{N (N+1)} \zeta_3
  \Biggr\}
\comma
\end{eqnarray}
with
\begin{eqnarray}
P_{9} &=& 2 N^5+6 N^4-3 N^3-33 N^2-26 N-24
\comma
\\
P_{10} &=& 2 N^7+2 N^6-11 N^5+8 N^4+13 N^3-58 N^2-64 N-48
\comma
\\
P_{11} &=& N^9+6 N^8-3 N^7+75 N^6+278 N^5+239 N^4-186 N^3-386 N^2-264 N-72,
\nonumber\\
\end{eqnarray}
\begin{align}
  \HtwoWqPStwo
  ={}&
  -C_F T_F \ln^2\Biggl(\frac{m^2}{Q^2}\Biggr) \frac{2 (N^2+N+2)^2}{(N-1) N^2 (N+1)^2 (N+2)}
\N\\&
  -C_F T_F \ln\Biggl(\frac{m^2}{Q^2}\Biggr) \frac{4 (5 N^5+32 N^4+49 N^3+38 N^2+28 N+8)}{(N-1) N^3 (N+1)^3 (N+2)^2}
\N\\&
  +C_F T_F \Biggl\{
      (-1)^N \frac{32}{(N-1) N (N+1) (N+2)} (2S_{-2}+\zeta_2)
\N\\&
     +\frac{4 (N^2+N+2)^2}{(N-1) N^2 (N+1)^2 (N+2)} S_1{}^2
     -\frac{8 (N^2+N+2)^2}{(N-1) N^2 (N+1)^2 (N+2)} S_2
\N\\&
     +\frac{8 P_{12}}{(N-1) N^3 (N+1)^3 (N+2)^2} S_1
     +\frac{2 P_{13}}{(N-1) N^4 (N+1)^4 (N+2)^3}
\N\\&
     -\frac{32}{(N-1) N (N+1) (N+2)} \zeta_2
   \Biggr\},
\end{align}
with
\begin{align}
 P_{12} ={}& N^7-15 N^5-58 N^4-92 N^3-76 N^2-48 N-16\\
 P_{13} ={}& 7 N^{10}+36 N^9+95 N^8+207 N^7+583 N^6+1567 N^5+2585 N^4+2464 N^3
\N\\&
+1512 N^2+656 N+144,\\
\end{align}
\begin{align}
\LtwoWgtwo
  ={}&
  T_F^2 \ln\Biggl(\frac{m^2}{Q^2}\Biggr) \Biggl\{
    -\frac{16 (N^2+N+2) S_1}{3 N (N+1) (N+2)}
    -\frac{16 (N^3-4 N^2-N-2)}{3 N^2 (N+1) (N+2)}
  \Biggr\}
\end{align}

\begin{align}
  \HtwoWgtwo
  ={}&
   \ln^2\Biggl(\frac{m^2}{Q^2}\Biggr) \Biggl\{
     -T_F^2 \frac{8 (N^2+N+2)}{3 N (N+1) (N+2)}
\N\\&
     +C_F T_F \Biggl[
        \frac{3 N^4+6 N^3+11 N^2+8 N+4}{N^2 (N+1)^2 (N+2)}
       -\frac{4 (N^2+N+2)}{N (N+1) (N+2)} S_1
      \Biggr]
\N\\&
     +C_A T_F \Biggl[
        \frac{4 (N^2+N+2)}{N (N+1) (N+2)} S_1
       -\frac{8 (N^4+2 N^3+4 N^2+3 N+2)}{(N-1) N^2 (N+1)^2 (N+2)^2}
      \Biggr]
   \Biggr\}
\N\\&
  +\ln\Biggl(\frac{m^2}{Q^2}\Biggr) \Biggl\{
      T_F^2 \Biggl[
       -\frac{16 (N^3-4 N^2-N-2)}{3 N^2 (N+1) (N+2)}
       -\frac{16 (N^2+N+2) S_1}{3 N (N+1) (N+2)}
      \Biggr]
\N\\&
     +C_A T_F \Biggl[
        \frac{4 (-1)^N (N^2+N+2)}{N (N+1) (N+2)} (2S_{-2}+\zeta_2)
       +\frac{4 (N^2+N+2)}{N (N+1) (N+2)} S_1{}^2
\N\\&
       -\frac{16 (2 N+3)}{(N+1)^2 (N+2)^2} S_1
       -\frac{8 (N^2+N+2)}{N^3 (N+1) (N+2)}
       -\frac{4 P_{14}}{(N-1) N^3 (N+1)^3 (N+2)^3}
\N\\&
       -\frac{4 (N^2+N+2)}{N (N+1) (N+2)} \zeta_2
       +\frac{4 (N^2+N+2)}{N (N+1) (N+2)} S_2
      \Biggr]
     +C_F T_F \Biggl[
       -\frac{8 (N^2+N+2)}{N (N+1) (N+2)} S_1{}^2
\N\\&
       -\frac{2 (3 N^4+2 N^3-9 N^2-16 N-12)}{N^2 (N+1)^2 (N+2)} S_1
       +\frac{2 P_{15}}{N^3 (N+1)^3 (N+2)}
\N\\&
       +\frac{8 (N^2+N+2)}{N (N+1) (N+2)} S_2
      \Biggr]
   \Biggr\}
  +C_A T_F \Biggl[
     -\frac{2 (N^2+N+2)}{3 N (N+1) (N+2)} S_1{}^3
\N\\&
     -\frac{2 P_{16}}{(N-1) N (N+1)^2 (N+2)^2} S_1{}^2
     -\frac{2 P_{17}}{(N-1) N^3 (N+1)^3 (N+2)^3} S_1
\N\\&
     -\frac{4 (3 N^2+3 N-2)}{N (N+1) (N+2)} \zeta_2 S_1
     +\frac{26 (N^2+N+2)}{N (N+1) (N+2)} S_2 S_1
     -\frac{2 P_{18}}{(N-1) N^4 (N+1)^4 (N+2)^4}
\N\\&
     -\frac{4 P_{19}}{(N-1) N (N+1)^2 (N+2)^2} \zeta_2
     -\frac{10 (N^2+N-6)}{N (N+1) (N+2)} \zeta_3
\N\\&
     +\frac{2 P_{20}}{(N-1) N^2 (N+1)^2 (N+2)^2} S_2
     +\frac{8 (7 N^2+7 N+2)}{3 N (N+1) (N+2)} S_3
     -\frac{16 (N^2+N+2)}{N (N+1) (N+2)} S_{2,1}
\N\\&
     +2 (-1)^N \Biggl[
        +\frac{2 (3 N^2+3 N-2)}{N (N+1) (N+2)} \zeta_2 S_1
        +\frac{4 (3 N^2+3 N-2)}{N (N+1) (N+2)} S_{-2} S_1
\N\\&
        +\frac{2 P_{19}}{(N-1) N (N+1)^2 (N+2)^2} \zeta_2
        -\frac{  (7 N^2+7 N+6)}{N (N+1) (N+2)} \zeta_3
\N\\&
        -\frac{2 (3 N^2+3 N+14)}{N (N+1) (N+2)} S_{-3}
        +\frac{4 P_{19}}{(N-1) N (N+1)^2 (N+2)^2} S_{-2}
\N\\&
        -\frac{4 (N^2+N-6)}{N (N+1) (N+2)} S_{-2,1}
      \Biggr]
   \Biggr]
  +C_F T_F \Biggl[
     -\frac{22 (N^2+N+2)}{3 N (N+1) (N+2)} S_1{}^3
\N\\&
     -\frac{2 (9 N^4+9 N^3-7 N^2-21 N-18)}{N^2 (N+1)^2 (N+2)} S_1{}^2
     +\frac{2 P_{21}}{N^3 (N+1)^3 (N+2)} S_1
\N\\&
     -\frac{64}{N (N+1) (N+2)} \zeta_2 S_1
     +\frac{10 (N^2+N+2)}{N (N+1) (N+2)} S_2 S_1
\N\\&
     +\frac{P_{22}}{(N-2) N^4 (N+1)^4 (N+2) (N+3)}
     -\frac{8 P_{23}}{(N-2) N^2 (N+1)^2 (N+2) (N+3)} \zeta_2
\N\\&
     +\frac{16 (3 N^2+3 N-4)}{N (N+1) (N+2)} \zeta_3
     +\frac{2 (10 N^3+15 N^2+11 N-10)}{N^2 (N+1) (N+2)} S_2
     -\frac{8 (7 N^2+7 N-10)}{3 N (N+1) (N+2)} S_3
\N\\&
     +\frac{16 (N^2+N+2)}{N (N+1) (N+2)} S_{2,1}
     +8 (-1)^N \Biggl[
        +\frac{8 }{N (N+1) (N+2)} \zeta_2 S_1
\N\\&
        +\frac{16}{N (N+1) (N+2)} S_{-2} S_1
        +\frac{P_{23}}{(N-2) N^2 (N+1)^2 (N+2) (N+3)} \zeta_2
\N\\&
        -\frac{4 }{N (N+1) (N+2)} \zeta_3
        +\frac{8 }{N (N+1) (N+2)} S_{-3}
        -\frac{16}{N (N+1) (N+2)} S_{-2,1}
\N\\&
        +\frac{2 P_{23}}{(N-2) N^2 (N+1)^2 (N+2) (N+3)} S_{-2}
     \Biggr]
   \Biggr],
\end{align}
with
\begin{align}
 P_{14} ={}& N^9+6 N^8+13 N^7+13 N^6+8 N^5+53 N^4+118 N^3+132 N^2+104 N+32\\
 P_{15} ={}& 4 N^6+5 N^5-10 N^4-39 N^3-40 N^2-24 N-8\\
 P_{16} ={}& 4 N^5-7 N^4-17 N^3-9 N^2-57 N-10\\
 P_{17} ={}& 15 N^9+17 N^8-71 N^7+81 N^6+632 N^5+974 N^4+984 N^3+664 N^2+288 N+64\\
 P_{18} ={}& 6 N^{12}+48 N^{11}+114 N^{10}+40 N^9-361 N^8-1273 N^7-3057 N^6-5691 N^5-7482 N^4
\N\\&
-6456 N^3-3712 N^2-1456 N-288\\
 P_{19} ={}& 2 N^5-N^4-12 N^3+3 N^2+32 N+24\\
 P_{20} ={}& 4 N^6-7 N^5-61 N^4-49 N^3-37 N^2-26 N-16\\
 P_{21} ={}& 3 N^6-14 N^5-27 N^4-40 N^3-74 N^2-84 N-32\\
 P_{22} ={}& 8 N^{10}+28 N^9-84 N^8-160 N^7+465 N^6+1091 N^5-163 N^4-1671 N^3-1646 N^2
\N\\&
-852 N-216\\
 P_{23} ={}& N^6+7 N^5-7 N^4-39 N^3+14 N^2+40 N+48,\\
\end{align}
\begin{align}
  \LthreeWpWqNStwo
  =\LtwoWpWqNStwo
  +C_F T_F \Biggl[
     \frac{8 (2 N+1)}{3 N (N+1)} S_1
    +\frac{4 (38 N^3+27 N^2-17 N-12)}{9 N^2 (N+1)^2}
   \Biggr],
\end{align}

\begin{align}
\HthreeWpWqNStwo
  ={}&
  \HtwoWpWqNStwo
 +C_F^2 \Biggl\{
     \frac{128}{5} (-1)^N \zeta_2^2
    +32 (-1)^N S_1{}^2 \zeta_2
\N\\&
    +\frac{32 P_{24}}{(N-2) (N-1) N^2 (N+1)^2 (N+2) (N+3)} \zeta_2
\N\\&
    +\frac{8 (-1)^N P_{25}}{(N-2) (N-1) N^2 (N+1)^2 (N+2) (N+3)} \zeta_2
    +48 (-1)^N S_{-2} \zeta_2
\N\\&
    -\frac{16 (2 N-1)}{N (N+1)} \zeta_2 S_1
    +\frac{32 (-1)^N (N-1)}{N (N+1)} \zeta_2 S_1
    -16 (-1)^N S_2 \zeta_2
\N\\&
    -\frac{4 (2 N+1)}{N (N+1)} S_1{}^2
    +64 (-1)^N S_{-2} S_1{}^2
\N\\&
    +\frac{P_{26} }{(N-2) (N-1) N^4 (N+1)^4 (N+2) (N+3)}
    -\frac{40 (2 N-1)}{N (N+1)} \zeta_3
\N\\&
    -\frac{16 (-1)^N (N-2)}{N (N+1)} \zeta_3
    -80 (-1)^N S_{-4}
    +\frac{16 (-1)^N (2 N+1)}{N (N+1)} S_{-3}
\N\\&
    +\frac{16 (-1)^N P_{25} }{(N-2) (N-1) N^2 (N+1)^2 (N+2) (N+3)} S_{-2}
    +\frac{2 P_{27}}{N^3 (N+1)^3} S_1
\N\\&
    -64 (-1)^N \zeta_3 S_1
    -32 (-1)^N S_{-3} S_1
    +\frac{64 (-1)^N (N-1)}{N (N+1)} S_1 S_{-2}
    +\frac{4 (2 N+1)}{N (N+1)} S_2
\N\\&
    -32 (-1)^N S_{-2} S_2
    +\frac{16 (2 N-1)}{N (N+1)} S_3
    +64 (-1)^N S_{-3,1}
    -\frac{32 (-1)^N (2 N-1)}{N (N+1)} S_{-2,1}
\N\\&
    -64 (-1)^N S_1 S_{-2,1}
    +32 (-1)^N S_{-2,2}
  \Biggr\}
 +n_f C_F T_F \Biggl\{
     \frac{4 (38 N^3+27 N^2-17 N-12)}{9 N^2 (N+1)^2}
\N\\&
    +\frac{8 (2 N+1)}{3 N (N+1)} S_1
  \Biggr\}
 +C_F T_F \Biggl\{
     \frac{4 (38 N^3+27 N^2-17 N-12)}{9 N^2 (N+1)^2}
    +\frac{8 (2 N+1)}{3 N (N+1)} S_1
  \Biggr\}
\N\\&
 +C_A C_F \Biggl\{
    -\frac{64}{5} (-1)^N \zeta_2^2
    -16 (-1)^N S_1{}^2 \zeta_2
\N\\&
    -\frac{16 P_{24}}{(N-2) (N-1) N^2 (N+1)^2 (N+2) (N+3)} \zeta_2
\N\\&
    -\frac{4 (-1)^N P_{25}}{(N-2) (N-1) N^2 (N+1)^2 (N+2) (N+3)} \zeta_2
    -24 (-1)^N S_{-2} \zeta_2
\N\\&
    +\frac{8 (2 N-1)}{N (N+1)} \zeta_2 S_1
    -\frac{16 (-1)^N (N-1)}{N (N+1)} \zeta_2 S_1
    +8 (-1)^N S_2 \zeta_2
    -32 (-1)^N S_{-2} S_1{}^2
\N\\&
    +\frac{P_{28} }{9 (N-2) (N-1) N^4 (N+1)^4 (N+2) (N+3)}
    +\frac{20 (2 N-1)}{N (N+1)} \zeta_3
\N\\&
    +\frac{8 (-1)^N (N-2)}{N (N+1)} \zeta_3
    +40 (-1)^N S_{-4}
    -\frac{8 (-1)^N (2 N+1)}{N (N+1)} S_{-3}
\N\\&
    -\frac{8 (-1)^N P_{25}}{(N-2) (N-1) N^2 (N+1)^2 (N+2) (N+3)} S_{-2}
\N\\&
    -\frac{2 (46 N^5+67 N^4-4 N^3-N^2+24 N+12)}{3 N^3 (N+1)^3} S_1
    +32 (-1)^N \zeta_3 S_1
\N\\&
    +16 (-1)^N S_{-3} S_1
    -\frac{32 (-1)^N (N-1)}{N (N+1)} S_1 S_{-2}
    +16 (-1)^N S_{-2} S_2
    -\frac{8 (2 N-1)}{N (N+1)} S_3
\N\\&
    -32 (-1)^N S_{-3,1}
    +\frac{16 (-1)^N (2 N-1)}{N (N+1)} S_{-2,1}
    +32 (-1)^N S_1 S_{-2,1}
\N\\&
    -16 (-1)^N S_{-2,2}
  \Biggr\}
\comma
\end{align}
with
\begin{align}
P_{24} ={}& N^7+N^6-7 N^5-N^4+16 N^3-6 N^2-4 N-12
\comma
\\
P_{25} ={}& 2 N^8+4 N^7-5 N^6-N^5-17 N^4-67 N^3-16 N^2+4 N+48
\comma
\\
P_{26} ={}& 34 N^{11}+161 N^{10}-135 N^9-1238 N^8-832 N^7+1573 N^6+2113 N^5
\N\\&
+1352 N^4+884 N^3+120 N^2-672 N-288
\comma
\\
P_{27} ={}& 18 N^5+23 N^4-4 N^3+13 N^2+22 N+8
\comma
\\
P_{28} ={}& -430 N^{11}-2089 N^{10}+159 N^9+11688 N^8+11736 N^7-9189 N^6-16613 N^5-8006 N^4
\N\\&
-3708 N^3-1260 N^2+2592 N+1296
\comma
\end{align}

\begin{align}
\HthreeWmWqNStwo
  ={}&
  \HthreeWpWqNStwo
 +C_F (C_F - C_A/2) \Biggl\{
    -64 (-1)^N S_{-3,1}
    +64 (-1)^N S_1 S_{-2,1}
\N\\&
    -32 (-1)^N S_{-2,2}
    -\frac{16 (2 N^2+2 N+1)}{N^3 (N+1)^3} S_1
\N\\&
    -\frac{16 (-1)^N (2 N^4+2 N^3+N^2+2 N-4)}{(N-1) N^2 (N+2)} S_{-2}
\N\\&
    +\frac{16 (N^4+2 N^3-3 N^2-4 N-2)}{(N-1) N^2 (N+1)^2 (N+2)} \zeta_2
\N\\&
    -\frac{8 (-1)^N (2 N^4+2 N^3+N^2+2 N-4)}{(N-1) N^2 (N+2)} \zeta_2
    +\frac{4 P_{29} }{(N-1) N^4 (N+1)^4 (N+2)}
\N\\&
    -32 (-1)^N \zeta_2 S_1{}^2
    -48 (-1)^N \zeta_2 S_{-2}
    +\frac{16 (-1)^N}{N (N+1)} S_1 \zeta_2
    +16 (-1)^N \zeta_2 S_2
\N\\&
    +64 (-1)^N \zeta_3 S_1
    -64 (-1)^N S_{-2} S_1{}^2
    +80 (-1)^N S_{-4}
    -\frac{32 (-1)^N}{N (N+1)} S_{-3}
\N\\&
    +32 (-1)^N S_{-3} S_1
    +\frac{32 (-1)^N}{N (N+1)} S_1 S_{-2}
    +32 (-1)^N S_{-2} S_2
\N\\&
    -\frac{128}{5} (-1)^N \zeta_2^2
    -\frac{24 (-1)^N}{N (N+1)} \zeta_3
  \Biggr\}
\comma
\end{align}
with
\begin{align}
P_{29} ={}& 9 N^8+36 N^7+41 N^6+13 N^5+44 N^4+67 N^3+20 N^2-26 N-12
\comma
\end{align}

\begin{align}
\HthreeWqPStwo
  ={}&
  C_F T_F \ln^2\Biggl(\frac{m^2}{Q^2}\Biggr) 
  \frac{2 (N^2+N+2)^2}{(N-1) N^2 (N+1)^2 (N+2)}
\N\\&
 +C_F T_F \ln\Biggl(\frac{m^2}{Q^2}\Biggr) 
  \frac{4 P_{30}}{(N-1) N^3 (N+1)^3 (N+2)^2}
\N\\&
 +C_F T_F \Biggl\{
     \frac{4 (N^2+N+2)^2}{(N-1) N^2 (N+1)^2 (N+2)} S_2
    -\frac{2 P_{31}}{(N-1) N^4 (N+1)^4 (N+2)^3}
  \Biggr\}
\comma
\end{align}
with
\begin{align}
P_{30} ={}& 5 N^5+32 N^4+49 N^3+38 N^2+28 N+8
\comma
\\
P_{31} ={}& N^{10}+8 N^9+29 N^8+49 N^7-11 N^6-131 N^5-161 N^4-160 N^3
\N\\&
-168 N^2-80 N-16
\comma
\end{align}
and
\begin{align}
\HthreeWgtwo
  ={}&
  \ln^2\Biggl(\frac{m^2}{Q^2}\Biggr) \Biggl\{
     T_F^2 \frac{8 (N^2+N+2)}{3 N (N+1) (N+2)}
    +C_A T_F \Biggl(
        \frac{8 (N^4+2 N^3+4 N^2+3 N+2)}{(N-1) N^2 (N+1)^2 (N+2)^2}
\N\\&
       -\frac{4 (N^2+N+2)}{N (N+1) (N+2)} S_1
     \Biggr)
    +C_F T_F \Biggl(
       \frac{4 (N^2+N+2)}{N (N+1) (N+2)} S_1
\N\\&
      -\frac{3 N^4+6 N^3+11 N^2+8 N+4}{N^2 (N+1)^2 (N+2)}
     \Biggr)
  \Biggr\}
 +\ln\Biggl(\frac{m^2}{Q^2}\Biggr) \Biggl\{
     C_F T_F \Biggl(
        \frac{8 (N^2+N+2) S_1{}^2}{N (N+1) (N+2)}
\N\\&
       +\frac{2 (3 N^4+2 N^3-9 N^2-16 N-12) S_1}{N^2 (N+1)^2 (N+2)}
       -\frac{2 P_{32}}{N^3 (N+1)^3 (N+2)}
\N\\&
       -\frac{8 (N^2+N+2)}{N (N+1) (N+2)} S_2
     \Biggr)
    +C_A T_F \Biggl(
       -\frac{4 (N^2+N+2)}{N (N+1) (N+2)} S_1{}^2
       +\frac{16 (2 N+3)}{(N+1)^2 (N+2)^2} S_1
\N\\&
       +\frac{4 P_{33}}{(N-1) N^3 (N+1)^3 (N+2)^3}
       -\frac{4 (-1)^N (N^2+N+2)}{N (N+1) (N+2)} \zeta_2
       +\frac{4 (N^2+N+2)}{N (N+1) (N+2)} \zeta_2
\N\\&
       -\frac{8 (-1)^N (N^2+N+2)}{N (N+1) (N+2)} S_{-2}
       -\frac{4 (N^2+N+2)}{N (N+1) (N+2)} S_2
     \Biggr)
  \Biggr\}
\N\\&
 +C_F T_F \Biggl\{
     \frac{2 (N^2+N+2)}{3 N (N+1) (N+2)} S_1{}^3
    -\frac{2 (3 N+2)}{N^2 (N+2)} S_1{}^2
\N\\&
    -\frac{2 (N^4-N^3-20 N^2-10 N-4)}{N^2 (N+1)^2 (N+2)} S_1
    +\frac{2 (N^2+N+2)}{N (N+1) (N+2)} S_1 S_2
    -\frac{P_{34} }{N^4 (N+1)^4 (N+2)}
\N\\&
    -\frac{2 (N^4+17 N^3+17 N^2-5 N-2)}{N^2 (N+1)^2 (N+2)} S_2
    -\frac{8 (N^2+N+2)}{3 N (N+1) (N+2)} S_3
  \Biggr\}
\N\\&
 +C_A T_F \Biggl\{
    -\frac{2 (N^2+N+2)}{3 N (N+1) (N+2)} S_1{}^3
    +\frac{2 (N^3+8 N^2+11 N+2)}{N (N+1)^2 (N+2)^2} S_1{}^2
\N\\&
    +\frac{2 P_{35}}{N (N+1)^3 (N+2)^3} S_1
    -\frac{4 (-1)^N (N^2+N+2)}{N (N+1) (N+2)} S_1 \zeta_2
    +\frac{4 (N^2+N+2)}{N (N+1) (N+2)} S_1 \zeta_2
\N\\&
    -\frac{8 (-1)^N (N^2+N+2)}{N (N+1) (N+2)} S_1 S_{-2}
    -\frac{6 (N^2+N+2)}{N (N+1) (N+2)} S_1 S_2
\N\\&
    -\frac{2 P_{36}}{(N-1) N^4 (N+1)^4 (N+2)^4}
    -\frac{4 (-1)^N (N^2-N-4)}{(N+1)^2 (N+2)^2} \zeta_2
    +\frac{4 (N^2-N-4)}{(N+1)^2 (N+2)^2} \zeta_2
\N\\&
    +\frac{2 (-1)^N (N^2+N+2)}{N (N+1) (N+2)} \zeta_3
    -\frac{2 (N^2+N+2)}{N (N+1) (N+2)} \zeta_3
    -\frac{4 (-1)^N (N^2+N+2)}{N (N+1) (N+2)} S_{-3}
\N\\&
    -\frac{8 (-1)^N (N^2-N-4)}{(N+1)^2 (N+2)^2} S_{-2}
    +\frac{2 (7 N^5+21 N^4+13 N^3+21 N^2+18 N+16)}{(N-1) N^2 (N+1)^2 (N+2)^2} S_2
\N\\&
    -\frac{16 (N^2+N+2)}{3 N (N+1) (N+2)} S_3
    +\frac{8 (-1)^N (N^2+N+2)}{N (N+1) (N+2)} S_{-2,1}
  \Biggr\}
\comma
\end{align}
with
\begin{align}
P_{32} ={}& {4 N^6+9 N^5-23 N^3-26 N^2-20 N-8}
\comma
\\
P_{33} ={}& N^9+6 N^8+15 N^7+25 N^6+36 N^5+85 N^4+128 N^3+104 N^2+64 N+16
\comma
\\
P_{34} ={}& 12 N^8+52 N^7+132 N^6+216 N^5+191 N^4+54 N^3-25 N^2-20 N-4
\comma
\\
P_{35} ={}& N^6+8 N^5+23 N^4+54 N^3+94 N^2+72 N+8
\comma
\\
P_{36} ={}& 2 N^{12}+20 N^{11}+86 N^{10}+192 N^9+199 N^8-N^7-297 N^6-495 N^5
\N\\&
-514 N^4-488 N^3-416 N^2-176 N-32
\period
\end{align}

The harmonic sums appearing are reduced to the
following basis\,:
\begin{align}
  \{&
   S_1(N),
   S_2(N),
   S_{-2}(N),
   S_3(N),
   S_{-3}(N),
   S_{2,1}(N),
   S_{-2,1}(N),
\N\\&
   S_4(N),
   S_{-4}(N),
   S_{3,1}(N),
   S_{-3,1}(N),
   S_{-2,2}(N),
   S_{2,1,1}(N)
  \}
\period
\end{align}
As the harmonic sums, the different Wilson coefficients obey recursion relations
for $N \rightarrow {N-1}, N \in \mathbb{C}$, which may be used in their analytic
continuation. In this way one may shift a value $N \in \mathbb{C}$ to a complex number
with large negative real part for which the asymptotic representation of the corresponding
Wilson coefficient holds in the analyticity region $N \neq -k, k \in \mathbb{N}$.
As examples the asymptotic representation for two Wilson coefficients is given in 
Appendix~B. 

\begin{figure}[htbp]
\begin{center}
  \includegraphics[width=.48\textwidth]{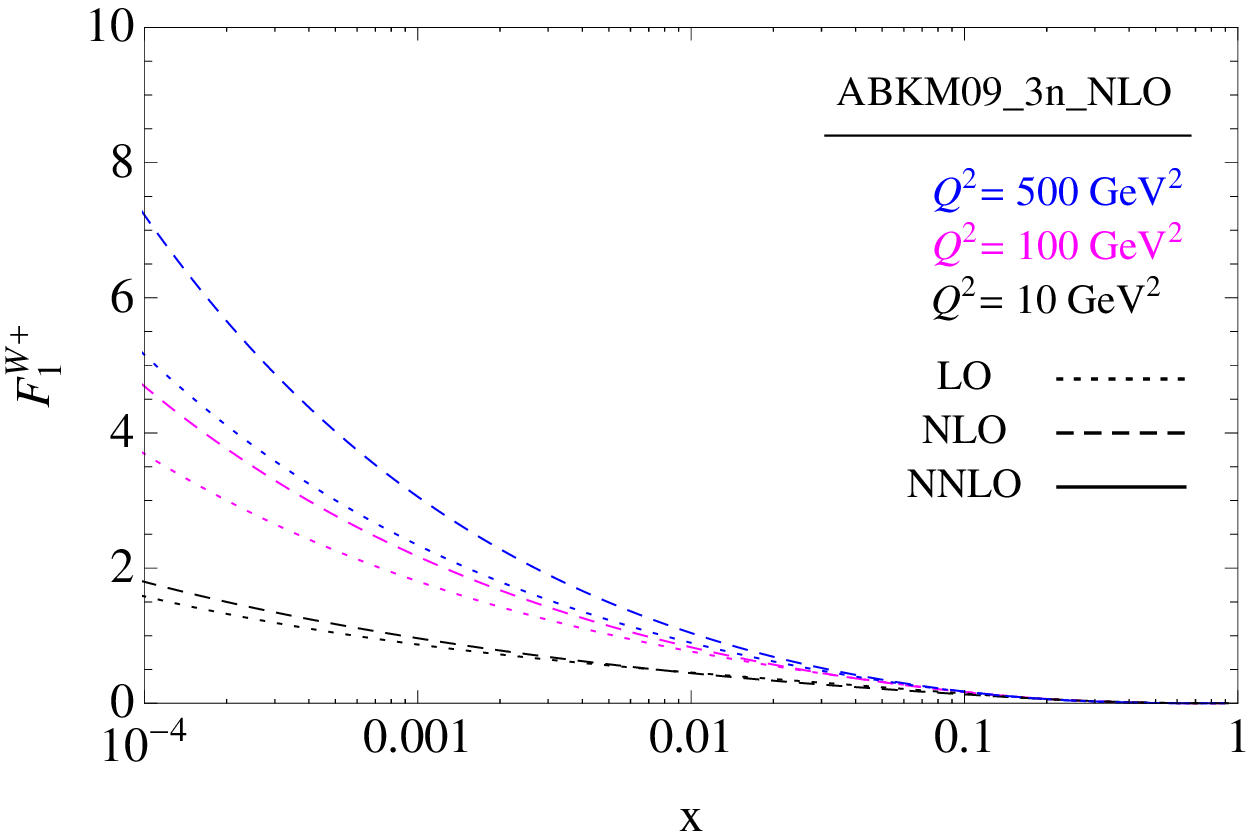}
  \includegraphics[width=.48\textwidth]{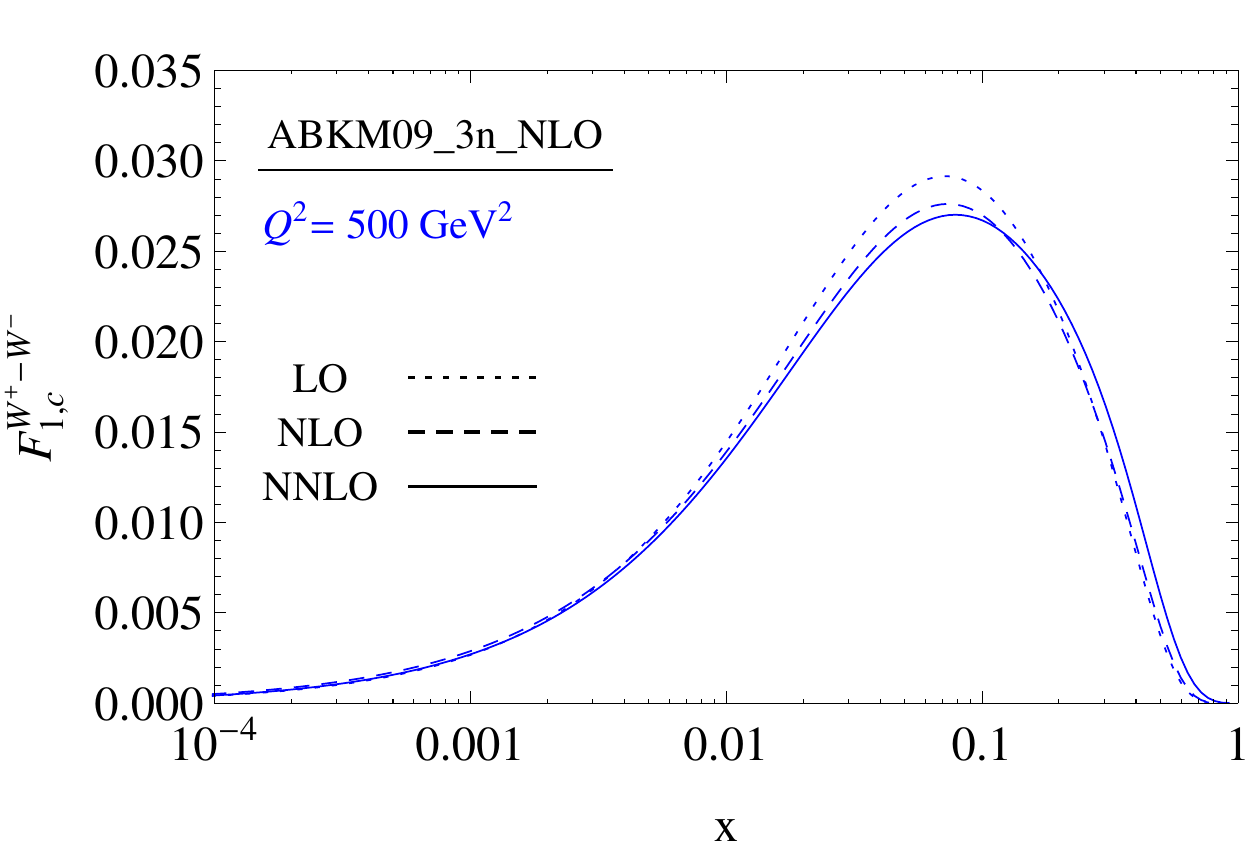}\\
  \includegraphics[width=.48\textwidth]{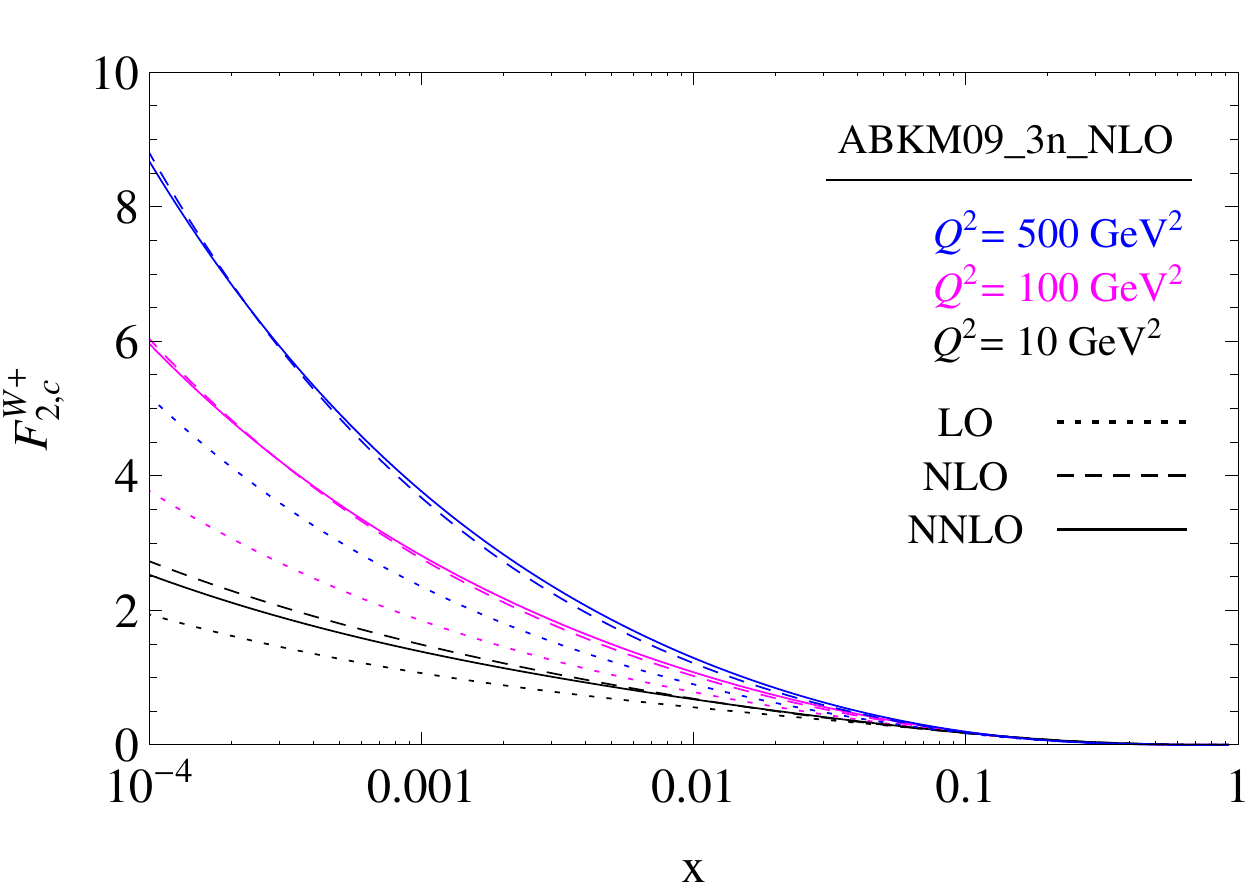}
  \includegraphics[width=.48\textwidth]{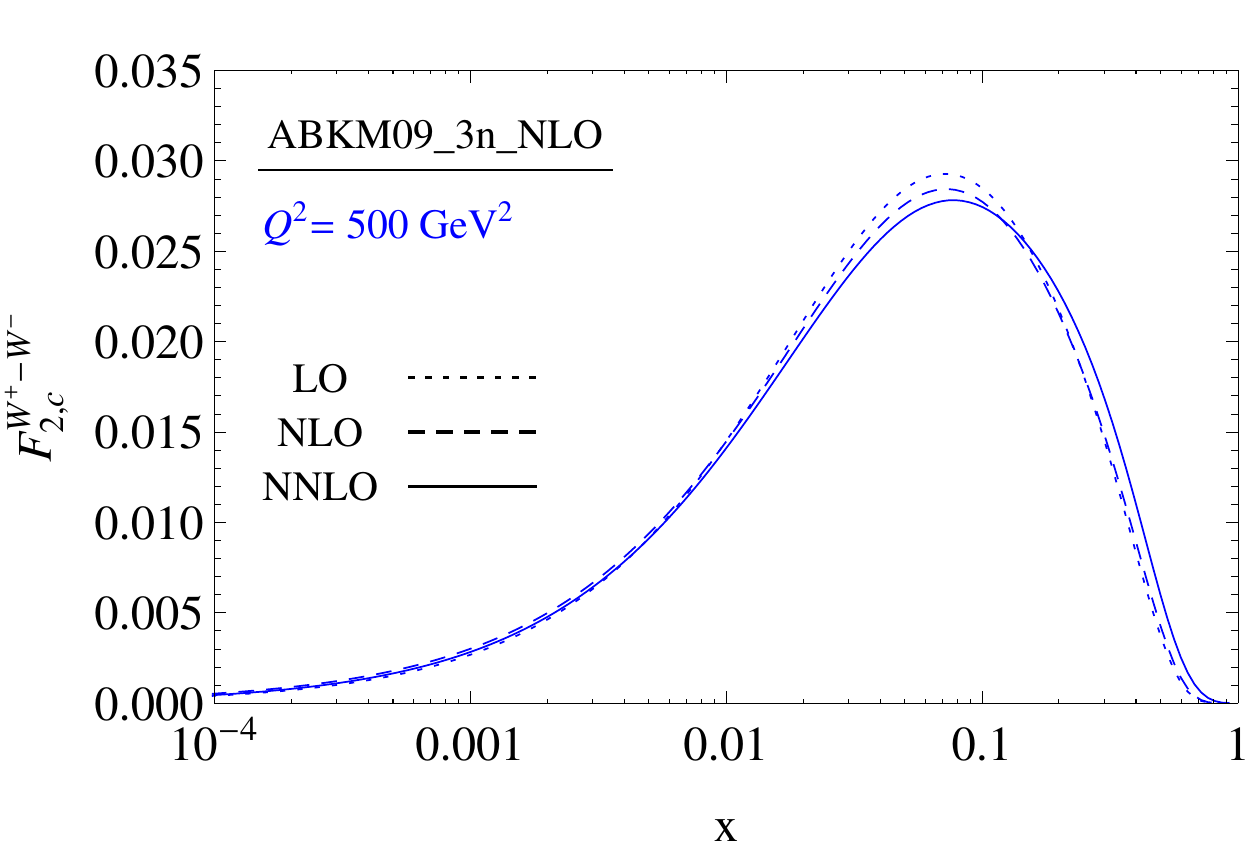}\\
  \includegraphics[width=.48\textwidth]{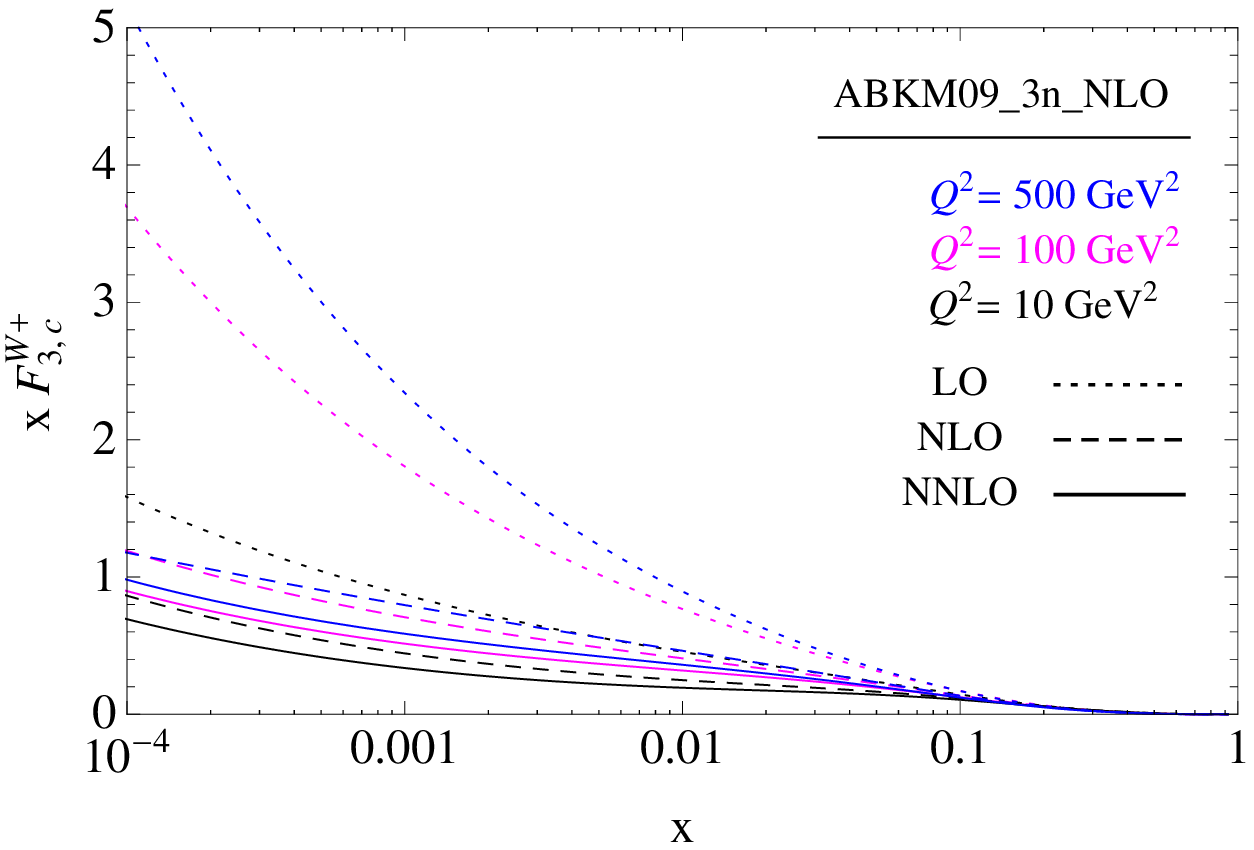}
  \includegraphics[width=.48\textwidth]{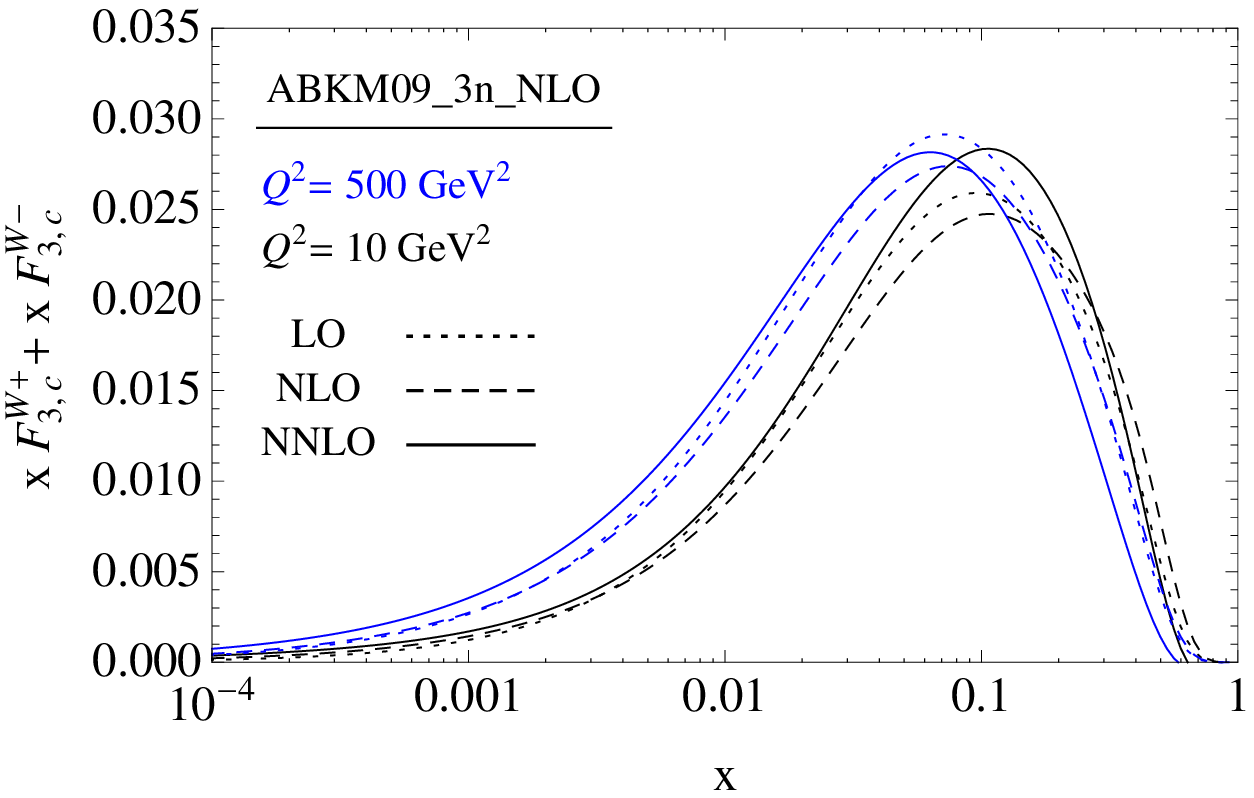}
\end{center}
  \caption{The charm contributions to the structure functions
           $F_{1}$, $F_{2}$, $F_{3}$ at different scales $Q^2$ and
           with increasing precision\,: LO, NLO, NNLO using the ABKM09 parameterization 
\cite{Alekhin:2009ni}.
\label{fig:strfunc}
}
\end{figure}

Since QCD analyses of deep-inelastic scattering data are often being
performed in $x$-space, we present the heavy flavor Wilson coefficients 
also in this space in Appendix~C. The corresponding Mellin inversions were performed
using the package \HSums{} \citeHSums. Here harmonic polylogarithms \cite{Remiddi:1999ew}
occur, which are reduced to the following basis set\,:
\begin{align}
  \{&
   H_0(x),
   H_1(x),
   H_{-1}(x),
   H_{0,1}(x),
   H_{0,-1}(x),
   H_{0,0,1}(x),
   H_{0,0,-1}(x),
\N\\&
   H_{0,1,1}(x),
   H_{0,1,-1}(x),
   H_{0,-1,1}(x),
   H_{0,-1,-1}(x)
  \}
\period
\end{align}
These functions have the following representations in terms of Nielsen integrals \cite{NIELSEN}~:
\begin{eqnarray}
S_{n,p}(x) &=& \frac{(-1)^{n+p-1}}{(n-1)! p!} \int_0^1 \frac{dz}{z} \ln^{n-1}(z) \ln^p(1-zx)~, \\
\Li_n(x)    &=& S_{n-1,1}(x)
\end{eqnarray}
and are given by
\begin{eqnarray}
H_0(x) &=& \ln(x) 
\\
H_1(x) &=& -\ln(1-x) 
\\
H_{-1}(x) &=& \ln(1+x) 
\\
H_{0,1}(x) &=& \Li_2(x)
\\
H_{0,-1}(x) &=& \Li_2(-x)
\\
H_{0,0,1}(x) &=& \Li_3(x)
\\
H_{0,0,-1}(x) &=& \Li_3(-x)
\\
H_{0,1,1}(x)  &=& S_{1,2}(x)
\\
H_{0,-1,1}(x) &=& 
 \frac{1}{2} \zeta_2 \ln(2)
-\frac{1}{8} \zeta_3
-\frac{1}{6} \ln^3(2)
+\frac{1}{2} \ln(1-x) \zeta_2
+\ln(1-x) \Li_2(-x)
-2 \ln(1+x) \zeta_2
\nonumber\\ &&
+\frac{1}{2} \ln(1+x) \ln^2(2)
-\frac{1}{2} \ln^2(1+x) \ln(2)
+\frac{1}{6} \ln^3(1+x)
\nonumber\\ &&
+\ln(x) \ln(1-x) \ln(1+x)
-\frac{1}{2} \ln(x) \ln^2(1+x)
+ \Li_3\left(\frac{1+x}{2}\right)
+\Li_3(1-x)
\nonumber\\ &&
-\Li_3(-x)
+\Li_3\left(-\frac{1-x}{1+x}\right)
-\Li_3\left(\frac{1-x}{1+x}\right)
-\Li_3\left(\frac{2x}{1+x}\right)
+\Li_3(x)
\\
H_{0,-1,1}(x) &=& 
-\frac{1}{2} \zeta_2 \ln(2)
-\frac{1}{8} \zeta_3
+\frac{1}{6} \ln^3(2)
+\frac{3}{2} \ln(1+x) \zeta_2
-\frac{1}{2} \ln(1+x) \ln^2(2)
\nonumber\\ &&
+\frac{1}{2} \ln^2(1+x) \ln(2)
-\frac{1}{3} \ln^3(1+x)
+\ln(1+x) \Li_2(x)
+\frac{1}{2} \ln(x) \ln^2(1+x)
\nonumber\\ &&
-\Li_3\left(\frac{1+x}{2}\right)
+\Li_3(-x)
+\Li_3\left(\frac{1}{1+x}\right)
+\Li_3\left(\frac{2x}{1+x}\right)
-\Li_3(x)
\\
H_{0,-1,-1}(x) &=& S_{1,2}(-x),
\end{eqnarray}
see also \cite{Moch:1999eb}. Fast numerical implementations for the functions
$\Li_{2,3}(x)$ and $S_{1,2}(x)$ are provided in the code {\tt ANCONT} 
\cite{Blumlein:2000hw}.
\section{Numerical Results}

\vspace{1mm}
\noindent
In the following we illustrate the effect of the heavy flavor Wilson coefficients up to 
$O(\alpha_s^2)$ on the structure functions $F_i(x,Q^2),~i = 1,2,3$ for $W^+$-exchange 
and the respective differences 
$F_i^{W^+}(x,Q^2) - F_i^{W^-}(x,Q^2)$ 
within the kinematic
range of HERA referring to the PDFs of Ref.~\cite{Alekhin:2009ni}. In Figure~\ref{fig:strfunc} 
these distributions are given at different values  of $Q^2$ comparing the contributions at LO, NLO, 
and NNLO. While the difference between the LO and NLO terms are generally large for the individual 
structure functions due to the newly contributing gluonic term at NLO, the effect is less pronounced 
in the differences $F_i^{W^+}(x,Q^2) - F_i^{W^-}(x,Q^2)$, showing the typical valence-type shape.
In general the NNLO corrections are close to the NLO ones over a wide range in $Q^2$ displaying the
scale evolution of the charged current structure functions. Both the functions $F_i^{W^+}(x,Q^2)$
and $F_i^{W^-}(x,Q^2)$ grow with rising $Q^2$ and towards small values of $x$.

\begin{figure}[htbp]
  \includegraphics[width=.5\textwidth]{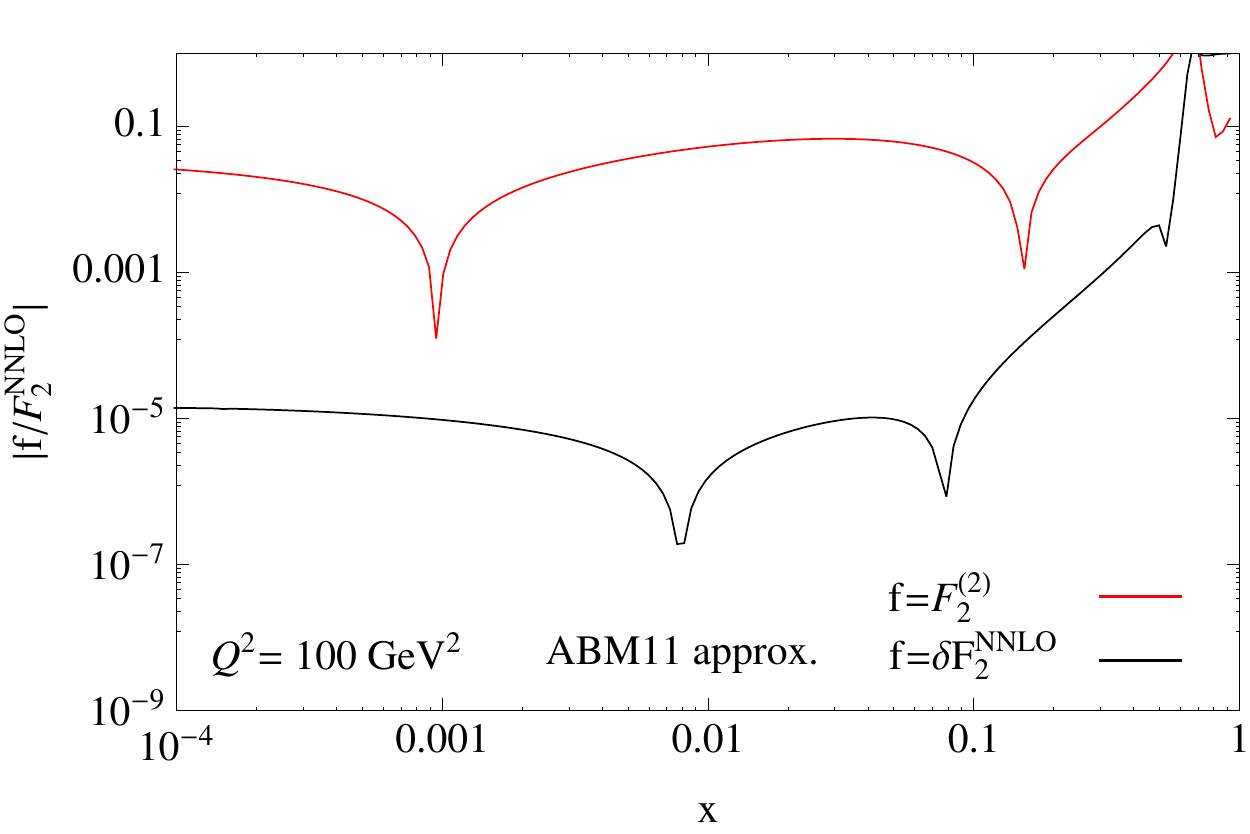}
  \includegraphics[width=.5\textwidth]{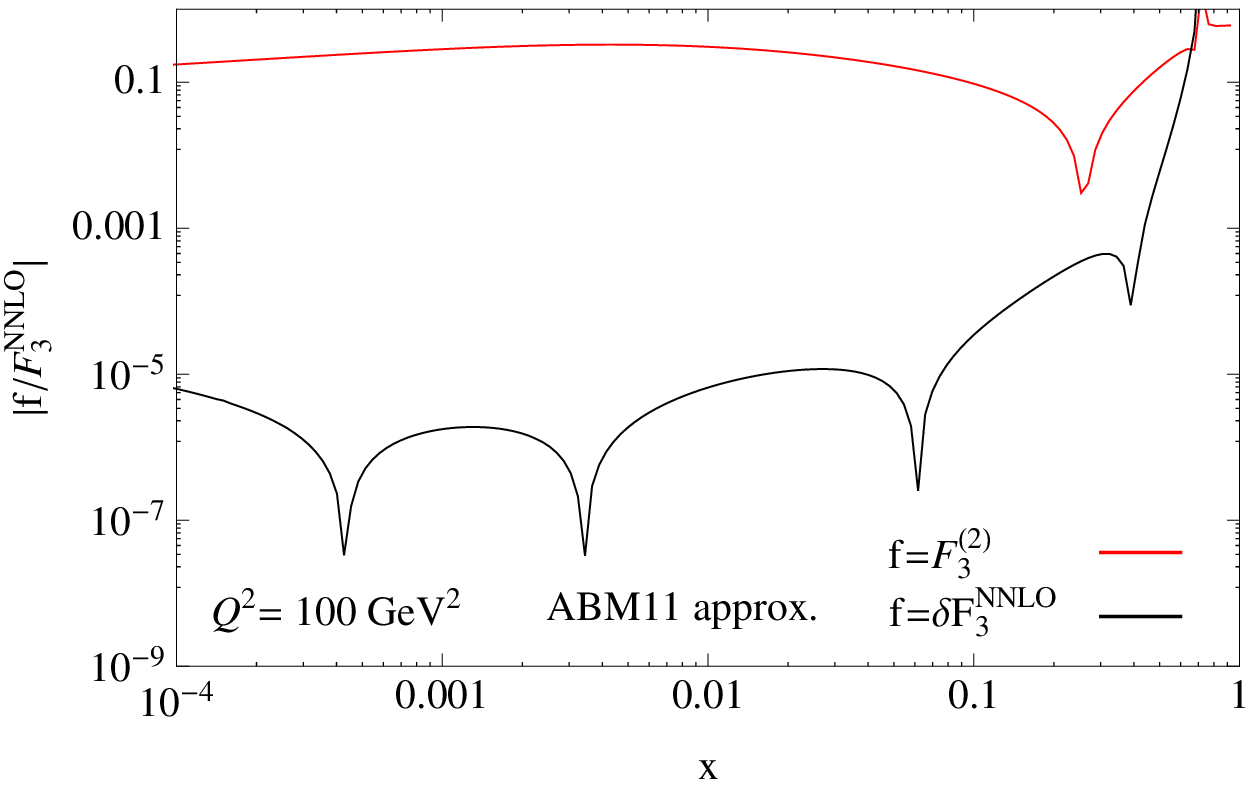}
  \caption{The relative numerical precision of the NNLO heavy quark
           contributions structure functions for charm production
	   (black) in comparison with the ratio of the $O(\alpha_s^2)$ 
           correction to the structure function (red).}
  \label{fig:precisions}
\end{figure}

Using the expressions derived in the previous chapter, a {\tt FORTRAN}
program was developed to calculate the 2-loop charm contribution to
the structure functions $F_2$ and $F_3$. The code is based on
earlier work on the exact 1-loop contributions \cite{Blumlein:2011zu}. It
works in $N$-space using the analytic continuation of the $N$-space
representation to complex values of $N$. The Mellin inversion into the
physical $x$-space is performed using a single complex contour
integral picking up the residues of all poles on the real axis. Since
the necessary points on the contour can be held fixed for different
values of $x$ at a given value of $Q^2$ for all PDFs, the calculation is
naturally very fast. For the analytic continuation of the harmonic
sums the {\tt ANCONT} implementations of Mellin transforms \citeANCONT{} are
used. The numerical accuracy of the implementation is checked by
calculating test values of $F_{2,c}$ and $F_{3,c}$ for different values
of $x$. For this purpose we used shape-fits to the ABM11 PDF sets at
$Q^2=100~GeV^2$ given by \,:
\begin{align}
  \begin{array}{ll}
    g(x) = 2.37 x^{-0.3}(1-x)^{12}\comma &
    s(x) = \bar{s}(x) = 0.108 x^{-0.29} (1-x)^{10}\comma \\
    d(x) = (0.145 x^{-0.27}+1.6 x^{0.6}) (1-x)^{4.5}\comma &
    \bar{d}(x) = 0.14 x^{-0.275} (1-x)^{7}\comma \\
    u(x) = (0.16 x^{-0.26}+3.5 x^{0.7}) (1-x)^{3.7}\comma &
    \bar{u}(x) = 0.14 x^{-0.275} (1-x)^{9}\period
  \end{array}
\end{align}
The relative numerical uncertainties of the 2-loop contributions are
below $10^{-3}$ for a wide range of $x$ values. So in total the numeric
uncertainties are at the order of $10^{-5}$ and at least 3 orders of
magnitude smaller than the 2-loop corrections, see Figures
\ref{fig:precisions}.
\section{Conclusions}

\vspace{1mm}
\noindent
The $O(\alpha_s^2)$ QCD corrections to the heavy flavor contributions of the deep-inelastic structure
functions in charged-current scattering have been calculated in the region $Q^2 \gg m^2$ using
the method of Ref.~\cite{Buza:1995ie}. We completed the set of Wilson coefficients and corrected
previous results in Ref.~\cite{Buza:1997mg}, presenting a detailed outline of the differences found.
The Wilson coefficients obey representations in terms of harmonic sums in Mellin-$N$ space and 
weighted harmonic polylogarithms in $x$-space, respectively, in both cases to weight 
{\sf w = 4}. Numerical studies were performed for the structure
functions $F_{i,c}^{W^{\pm}}(x,Q^2),~i = 1,2,3$ in the kinematic region available at HERA comparing 
the corrections from LO to NNLO. The NNLO results come out close to those at NLO in a wide range of 
$Q^2$. Numerical implementations both in Mellin-$N$ and $x$-space were performed at high accuracy.
The Wilson coefficients in $x$-space may all be expressed in terms of Nielsen integrals. The 
corresponding {\tt FORTRAN} codes are available on request.  
\begin{appendix}
\section{Relative Signs in Wilson Coefficients
\label{sec:SIGNS}}

\vspace*{1mm}
\noindent
Since the sign in front of the OME in the gluonic heavy flavor Wilson coefficient of 
Eq.~(\ref{eq237}) contradicts the asymptotic representation given in Eq.~(A.17) of 
\cite{Buza:1997mg}, a recalculation of the full gluonic $O(\alpha_s)$ correction was performed 
which will be presented in the following. We confirm the result given in \cite{Gluck:1996ve,Kretzer99}. 
As the minus sign was confirmed in this analysis, further changes in signs in the relations 
(A.18) and (A.19) of \cite{Buza:1997mg} are anticipated.  The reasoning follows the idea of 
calculating leading logarithms in the Altarelli-Parisi picture of scaling violations 
\cite{Altarelli:1977zs}.

\begin{figure}
  \centering
  \parbox[t]{.4\textwidth}{\centering
  \includegraphics[width=.3\textwidth]{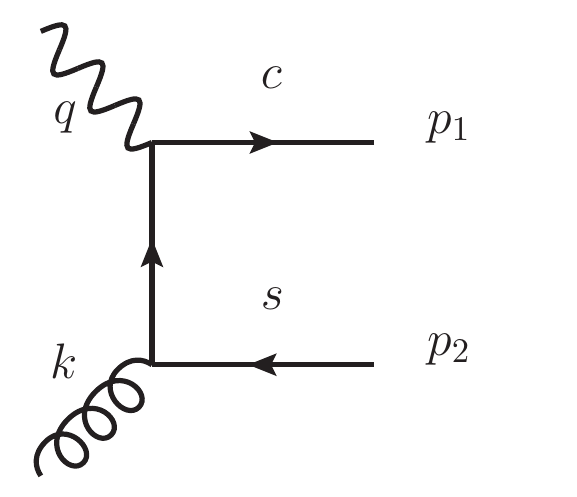}\\
  (a)}
  \parbox[t]{.4\textwidth}{\centering
  \includegraphics[width=.3\textwidth]{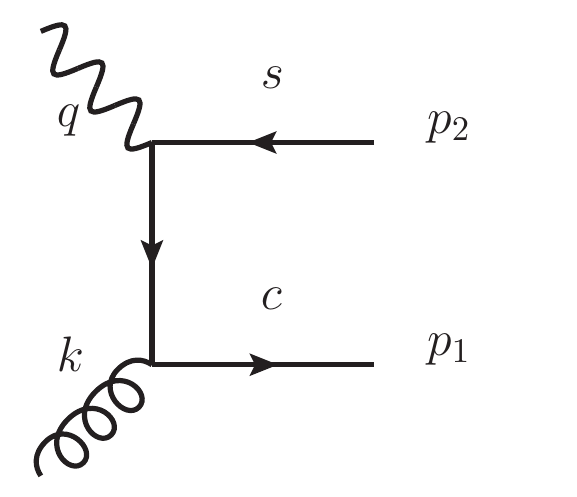}\\
  (b)}
  \caption{Graphs contributing to $H_{g}^{(2)}$}
  \label{fig:gluonDiags}
\end{figure}

The heavy flavor Wilson coefficient $H_{g}$ is obtained from the diagrams in Figure 
\ref{fig:gluonDiags}
with the matrix element
\begin{eqnarray}
   M^{\mu}_{a} &=&
   \bar{u}(p_1)
   i\gamma^\mu(1-\gamma_5)
   \frac{i(\slashed{p}_1 - \slashed{q})}{(p_1-q)^2}
   \gamma^\rho
   i g_s t^a
   v(p_2)
   \ep_{\rho}^{a}(k)
\nonumber\\ &&
  +
   \bar{u}(p_1)
   \gamma^\rho
   i g_s t^a
   \frac{i(\slashed{p}_1 - \slashed{k} + m)}{(p_1-k)^2 - m^2}
   i \gamma^\mu
   (1-\gamma_5)
   v(p_2)
   \ep_{\rho}^{a}(k)
\end{eqnarray}
contributing to the hadronic tensor.  For the implementation of $\gamma_5$, the prescription of 
\cite{Larin:1993tq} was used, which amounts to the replacement
\begin{align}
  \gamma_\mu \gamma_5
  =
  \frac{i}{6}\ep_{\mu\nu\rho\sigma}
  \gamma_{\nu}
  \gamma_{\rho}
  \gamma_{\sigma}
\comma
\end{align}
in the matrix element, where products of Levi-Civita symbols are
evaluated by the determinant
\begin{align}
\label{eq:LeviCivitaContraction}
  \ep_{\alpha\beta\gamma\delta}\ep_{\mu\nu\rho\sigma}
  =
  \left\vert
  \begin{array}{cccc}
    g_{\alpha\mu}&
    g_{\alpha\nu}&
    g_{\alpha\rho}&
    g_{\alpha\sigma}\\
    g_{\beta\mu}&
    g_{\beta\nu}&
    g_{\beta\rho}&
    g_{\beta\sigma}\\
    g_{\gamma\mu}&
    g_{\gamma\nu}&
    g_{\gamma\rho}&
    g_{\gamma\sigma}\\
    g_{\delta\mu}&
    g_{\delta\nu}&
    g_{\delta\rho}&
    g_{\delta\sigma}
  \end{array}
  \right\vert
\comma
\end{align}
and Lorentz contractions are performed in $D$ dimensions.  Since
$O(\alpha_s)$ is the leading order of the gluon channel, no finite
renormalization is needed.  The Lorentz-structure of the squared
matrix element is projected onto the (unrenormalized) partonic
versions of the structure functions $\hat{\mathcal{F}}_i$, $i=1,2,3$
via the projectors\,:
\begin{align}
  \hat{P}_{1}
  ={}&
  \frac{1}{2+\ep}
  \frac{x}{Q^2}
  \left(
         4 x P_\mu P_\nu
       + 2 P_\mu q_\nu
       + 2 P_\nu q_\mu
       - \frac{Q^2}{x} g_{\mu\nu}
  \right)
\comma
\N\\
  \hat{P}_{2} 
  ={}& 
    2x \left( \frac{q_\mu q_\nu}{Q^2} - \frac{g_{\mu\nu}}{2+\ep} \right)
  + 4\frac{x^2}{Q^2} \frac{3+\ep}{2+\ep}
    \left(  2x P_\mu P_\nu +  P_\mu q_\nu + P_\nu q_\mu \right)
\comma
\N\\
  \hat{P}_{3}
  ={}&
  -\frac{4 x}{Q^2}
   \frac{1}{(1+\ep)(2+\ep)}
   i\ep_{\mu\nu\rho\sigma}P^\rho q^\sigma
\period
\end{align}
The two particle phase space leads to one-dimensional integrals
which, after a partial fraction decomposition, can be solved in terms
of ${}_2 F_1$ functions, e.g. 
\begin{align}
    \int_{0}^{1} dy\;
    y^{\eph}
    (1-y)^{\eph}
    \frac{1}{(p_1-k)^2-m^2}
  =
   -\frac{1}{s+Q^2}
    B\left(1+\eph,1+\eph\right)\;
    \Ftwon{1,1+\eph}{2+\ep}{\frac{s-m^2}{s}}
\comma
\end{align}
with
\begin{align}
 y=\frac{1}{2}[1+\cos \sphericalangle(p_1,q)],
 \quad
 (p_1-k)^2-m^2 =-(s+Q^2)\left(1-\frac{s-m^2}{s}(1-y)\right)
\period
\end{align}
This particular example is the source for the mass logarithms\,:
\begin{align}
 \Ftwon{1,1 + \eph}{2 + \ep}{\frac{s-m^2}{s}}
  = -\frac{s}{s-m^2} \ln\left(\frac{m^2}{s}\right) + O(\ep)
\comma
\end{align}
and thus contributes to the OME in the asymptotic expansions.

The $t$-channel exchange of the light $s$-quark in the first diagram
of Figure~\ref{fig:gluonDiags} introduces a collinear singularity, which
has to be removed via mass factorization as described in Eq.~(2.38) of
\cite{Zijlstra:1992qd}. In the present case it proceeds via\,:
\begin{align}
\label{eq:MassFac}
  \hat{\mathcal{F}}_{i} = \Gamma_{qg}^{(1)} + H_{i,g}^{(1)}
\comma
\end{align}
with the \MSbar{} transition function
\begin{align}
  \Gamma_{qg}^{(1)}
  ={}&
  S_\ep
  \frac{1}{2\ep}
  P_{qg}^{(0)},
\quad
 P_{qg}^{(0)}(z)
 =
 8 T_F [z^2+(1-z)^2]
\period
\end{align}
In contrast to the electromagnetic case, the factor $2n_f$ in (2.38)
of \cite{Zijlstra:1992qd} is omitted, since the above calculation is performed
for only one incoming light flavor, and only for one of the two graphs
in Figure~\ref{fig:gluonDiags} the quark propagator is massless and thus
develops a collinear singularity.  The results of this calculation
agree with those in \cite{Kretzer99, Gluck:1996ve, Gottschalk:1980rv}.

In order to gain further confidence in the emergence of a minus sign
in the asymptotic representation, as well as to understand how this
observation relates to the pure singlet Wilson coefficients at 2-loop
order,  the calculation of the leading logarithmic contributions is
performed using the method also applied by Altarelli and Parisi
\cite{Altarelli:1977zs}, cf.\ also \cite{Dokshitzer:1991wu}.

A Sudakov parameterization \cite{Sudakov:1954sw} is introduced for
the $t$-channel momentum in the diagram in
Figure~\ref{fig:gluonDiags}(a)\,:
\begin{align}
  k - p_2 = \alpha k + \beta q' + k_\perp
\comma
\end{align}
denoting the gluon momentum by $k$, and the photon momentum by $q$.
Furthermore, the vectors $k_\perp$ and $q'$ are defined via
\begin{align}
  q' = q + x k,
  \quad
  q'.k_\perp = k.k_\perp = 0
\period
\end{align}
This leads to the final state momenta
\begin{align}
  p_1 ={}& (\alpha-x) k + (\beta+1) q' + k_\perp
\comma
\\
  p_2 ={}& (1-\alpha) k - \beta q' - k_\perp
\comma
\end{align}
and the Mandelstam variables
\begin{align}
  s :={}& (q+k)^2 = 2k.q - Q^2
\comma
\N\\
  t :={}& (p_1-q)^2
\comma
\N\\
  u :={}& (p_1-k)^2 = -t  + m^2 - Q^2 - s
\period
\end{align}
With the approximation $q'^2 \approx k^2 \approx 0$ and $p.q'\approx
p.q$, the phase space integral then takes the form
\begin{align}
  \int dp_1\; dp_2\;
  \delta(p_1^2 - m^2)
  \delta(p_2^2)
  ={}&
  \int
  d\beta \;
  d\alpha \;
  dk_\perp^2 \;
  \frac{\pi}{2 k.q (1-\alpha)}
  \delta\left( \beta - \frac{k_\perp^2}{2 k.q (1-\alpha)}\right)
\N\\&\times
  \delta\left( \alpha - x + \frac{1-x}{1-\alpha}
  \frac{k_\perp^2}{2 k.q} - \frac{m^2}{2 k.q} \right)
\period
\end{align}
Using the implication from the $\delta$-distributions one finds
\begin{align}
  k_\perp^2 = (1-\alpha) t
\comma	
\end{align}
and thus defines the positive variable
\begin{align}
 r^2 := -t
\period
\end{align}
The physical region\footnote{For a collection of kinematic formulae
used here see \cite{KB1973}.} is determined from the conditions
\begin{align}
  0\leq \cos \sphericalangle(q,p_1) \leq 1
  ~\text{and}~
  0\leq \cos \sphericalangle(q,p_2) \leq 1 
\end{align}
on the angles in the target system of coordinates.  As a result one finds
\begin{align}
  2 k^2 \frac{s-m^2}{s+Q^2} \leq r^2 \leq \frac{(s-m^2)(s+Q^2)}{s}
  \period
\end{align}
There are two integrals leading to logarithmic values\,:
\begin{align}
  \int dr^2 \frac{1}{(p_1-q)^2}
  ={}&
  -\int dr^2 \frac{1}{r^2}
\N\\
  ={}&
  \ln\left(\frac{2 s k^2}{(s+Q^2)^2}\right)
  \approx
  \ln\left( \frac{k^2}{s} \right)
\comma
\\
  \int dr^2 \frac{1}{(p_1-k)^2-m^2}
  ={}&
  \int dr^2 \frac{1}{r^2 - s - Q^2}
\N\\
  ={}&
  -
  \ln\left(
    \frac{s}{m^2}
   -2
    \frac{k^2}{m^2}
    \frac{s}{s+Q^2}
    \frac{(s-m^2)}{s+Q^2}
  \right)
\N\\
  \approx&
  \ln\left(\frac{m^2}{s}\right)
\period
\end{align}
Since the incoming gluon is massless, i.e.\ $k^2=0$, the first
logarithm represents a collinear singularity, which was earlier
regulated in $D=4+\ep$ dimensions and removed via mass factorization
in Eq.~(\ref{eq:MassFac}).  The second logarithm indeed constitutes the
leading mass dependence of the process.  Picking out this logarithmic
part, one finds
\begin{align}
  H_{g,1}^{W,(1),\text{LL}}
  ={}&
  H_{g,2}^{W,(1),\text{LL}}
  =
  -\frac{1}{2} P_{qg}^{(0)}(N) \ln\left( \frac{m^2}{Q^2} \right)
\comma
\\
  H_{g,3}^{W,(1),\text{LL}}
  ={}&
   \frac{1}{2} P_{qg}^{(0)}(N) \ln\left( \frac{m^2}{Q^2} \right)
\period
\end{align}
The splitting functions derive from the fermion traces after applying
the above approximations and canceling against denominators.

\begin{figure}[htpb]
  \centering
  \includegraphics[width=.3\textwidth]{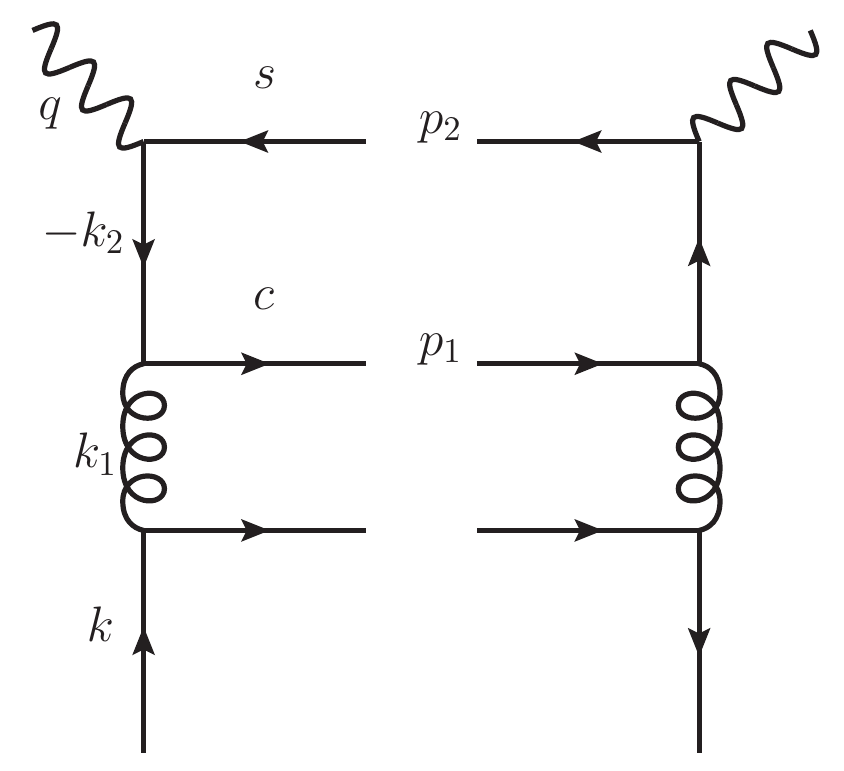}
  \caption{The leading logarithmic 2-loop PS-contribution
    $H_q^{\text{PS},(2),\text{LL}}$ can be built from the leading
    logarithmic 1-loop gluonic contribution by adding a splitting of a
    quark into a gluon.}
   \label{fig:2loopLadder}
\end{figure}

In order to obtain the 2-loop pure singlet contribution in leading
logarithmic approximation, one has to include another ladder rung
formed by a light quark line, as depicted in
Figure~\ref{fig:2loopLadder}.

Then the Sudakov parameters are introduced as above\,:
\begin{align}
  k_1 ={}& \alpha_1 k + \beta_1 q' + k_{\perp 1}
\comma
\\
  k_2 ={}& \alpha_2 k + \beta_2 q' + k_{\perp 2}
\period
\end{align}

The three-particle phase space can be treated similarly as before,
assuming a strict hierarchy $k^2 \ll \vert k_{\perp 1}^2 \vert \ll
\vert k_{\perp 2}^2 \vert \ll Q^2$.  The $\delta$-distributions
introduced by the phase space integral then take the forms\,:
\begin{align}
  \delta((k-k_1)^2)
  ={}&
  \frac{1}{2 k.q(1-\alpha_1)}
  \delta\left(\beta_1 - \frac{k_{\perp 1}^2}{2k.q (1-\alpha_1)}\right)
\comma
\N\\
  \delta((k_2-k_1)^2-m^2)
  ={}&
  \frac{1}{2k.q (\alpha_1-\alpha_2)}
  \delta\left(\beta_2 - \frac{k_{\perp 2}^2 - m^2}
    {2 k.q (\alpha_1 - \alpha_2)}\right)
\comma
\N\\
  \delta((k_2+q)^2)
  ={}&
  \frac{1}{2k.q}
  \delta\left(\alpha_2-x+\frac{(\alpha_1-x)}{(\alpha_1-\alpha_2)}
    \frac{(k_{\perp}^2-m^2)} {2k.q}\right)
\period
\end{align}
This again leads to the definition of positive squares of momenta\,:
\begin{align}
  r_1^2 ={}& -\frac{k_{\perp 1}^2}{1 - \alpha_1}
\comma
\N\\
  r_2^2 ={}& -\frac{k_{\perp 2}^2}{\alpha_1 - \alpha_2}
\period
\end{align}
Like in the case of purely massless ladder rungs \cite{Altarelli:1977zs}, see
also \cite{Dokshitzer:1991wu, Cheng:1985bj}, the integral becomes nested in both
the momentum and the Sudakov variables $\alpha_1,\alpha_2$,
\begin{align}
\label{eq:H3qPSLL}
  H_{3,q}^{W,\text{PS},(2)}
  ={}&
  \frac{1}{8}
  \int_{k^2}^{\frac{(s-m^2)(s+Q^2)}{s}}
  \frac{dr_2^2}{r_2^2-s-Q^2}
  \int_{k^2}^{\vert k_{\perp 2}^2 \vert}
  \frac{d\vert k_{\perp 1}^2\vert }{-\vert{k_{\perp 1}^2} \vert}
\N\\&
  \int_{0}^{1} \frac{d\alpha_2}{\alpha_2}
  \delta\left(1-\frac{x}{\alpha_2}\right)
  \int_{\alpha_2}^{1} \frac{d\alpha_1}{\alpha_1}
  P_{gq}^{(0)}\left(\frac{\alpha_1}{\alpha_2}\right)
  P_{qg}^{(0)}(\alpha_1)
\comma
\end{align}
where the following splitting function occurs\,:
\begin{align}
 P_{gq}^{(0)}(x)
 =
 4 C_F \frac{1+(1-x)^2}{x}
\period
\end{align}
With the variable substitution $R^2=s+Q^2-r_2^2$, the integrals over
the squared momenta can be performed\,:
\begin{align}
  \int_{k^2}^{\frac{(s-m^2)(s+Q^2)}{s}}
  \frac{dr_2^2}{r_2^2-s-Q^2}
  \int_{k^2}^{\vert k_{\perp 2}^2 \vert}
  \frac{d\vert k_{\perp 1}^2\vert }{-\vert{k_{\perp 1}^2} \vert}
  ={}&
  \int_{m^2\frac{s+Q^2}{s}}^{s+Q^2-k^2}
  \frac{dR^2}{R^2}
  \int_{k^2}^{R^2\frac{\alpha_1-\alpha_2}{\alpha_1}-m^2}
  \frac{d\vert k_{\perp 1}^2\vert }{-\vert{k_{\perp 1}^2} \vert}
\N\\
  \approx&
  \int_{m^2\frac{s+Q^2}{s}}^{s+Q^2}
  \frac{dR^2}{R^2}
  \ln\left( \frac{R^2}{k^2}\frac{\alpha_1-\alpha_2}{\alpha_1} \right)
\N\\
  =&
  \frac{1}{2}\ln^2\left(\frac{m^2}{Q^2}\right) + O(\ln(m^2/Q^2))
\period
\end{align}
Here the reference scale in the mass-logarithm was chosen to be $Q^2$.
In Mellin space the convolutions of the splitting functions in
(\ref{eq:H3qPSLL}) factorize, and one finds to $O(\ln^2(m^2/Q^2))$ the
relation
\begin{align}
  H_{3,q}^{W,\text{PS},(2)}
  =
  \frac{1}{16}
  P_{qg}^{(0)}(N)
  P_{gq}^{(0)}(N)
  \ln^2\left( \frac{m^2}{Q^2} \right)
  =
  -\frac{1}{2}
  A_{Qq}^{\text{PS},(2)}
\comma
\end{align}
which fixes the respective sign.
The additional ladder rung has the effect of introducing another
splitting function independently from the boson-quark coupling.
Hence the minus sign from the one-loop heavy flavor Wilson
coefficient in leading logarithmic approximation is simply translated
to the 2-loop pure-singlet contribution.  As in the gluonic heavy
flavor Wilson coefficient at the 1-loop order, the result above
disagrees with the asymptotic representation given in \cite{Buza:1997mg}.
This confirms the results of the derivation of the asymptotic representations at 2-loop order
given in Section 2, which captures the signs in a rigorous way.
\section{Asymptotic Expansion of the Wilson Coefficients}

\vspace*{1mm}
\noindent
In this appendix we present the asymptotic expansions of the different Wilson coefficients.
In Mellin-space codes these expressions may serve as numerical starting values for large $N \in 
\mathbb{C}$ outside the singularities being located at the integers left of an integer $N_0$.
All other values in the analytic region can be obtained by the shift relations of the analytic
continuations of the harmonic sums, 
cf.~\cite{Blumlein:2000hw,Blumlein:2005jg,Blumlein:2009ta,Blumlein:2009fz}.

As examples we show the asymptotic expansions of $L_{2,q}^{W^{+}+W^{-},\text{NS},(2)}$ and
$H_{2,q}^{W^++W^-,\text{NS},(2)}$. One obtains
\begin{align}
   L_{2,q}^{W^{+}+W^{-},\text{NS},(2)}
   ={}&
   C_F T_F \Biggl\{
      \Biggl[
        -\frac{80}{9} \ln(\bar{N})
        +\frac{16}{3} \zeta_2
        +\frac{2}{3}
        -\frac{88}{9 N}
        +\frac{356}{27 N^2}
        -\frac{16}{N^3}
        +\frac{478}{27 N^4}
        -\frac{304}{15 N^5}
\N\\&
        +\frac{13124}{567 N^6}
        -\frac{544}{21 N^7}
        +\frac{767}{27 N^8}
      \Biggr] \ln \big(\frac{m^2}{Q^2}\big)
     +\Biggl[
        -\frac{8 \ln(\bar{N})}{3}
        +2
        -\frac{4}{3 N}
        +\frac{14}{9 N^2}
\N\\&
        -\frac{4}{3 N^3}
        +\frac{59}{45 N^4}
        -\frac{4}{3 N^5}
        +\frac{254}{189 N^6}
        -\frac{4}{3 N^7}
        +\frac{119}{90 N^8}
      \Biggr] \ln ^2\big(\frac{m^2}{Q^2}\big)
     +\Biggl[
        -\frac{58}{9}
\N\\&
        -\frac{4}{3 N}
        +\frac{14}{9 N^2}
        -\frac{4}{3 N^3}
        +\frac{59}{45 N^4}
        -\frac{4}{3 N^5}
        +\frac{254}{189 N^6}
        -\frac{4}{3 N^7}
        +\frac{119}{90 N^8}
      \Biggr] \ln(\bar{N})^2
\N\\&
     +\Biggl[
        -\frac{718}{27}
        -\frac{214}{9 N}
        +\frac{455}{27 N^2}
        -\frac{182}{9 N^3}
        +\frac{2893}{135 N^4}
        -\frac{1079}{45 N^5}
        +\frac{152897}{5670 N^6}
        -\frac{28121}{945 N^7}
\N\\&
        +\frac{521287}{16200 N^8}
      \Biggr] \ln(\bar{N})
     -\frac{8}{9} \ln(\bar{N})^3
     +\frac{8}{3} \zeta_2 \ln(\bar{N})
     +\Biggl[
         \frac{70}{3}
        +\frac{4}{3 N}
        -\frac{14}{9 N^2}
        +\frac{4}{3 N^3}
\N\\&
        -\frac{59}{45 N^4}
        +\frac{4}{3 N^5}
        -\frac{254}{189 N^6}
        +\frac{4}{3 N^7}
        -\frac{119}{90 N^8}
      \Biggr] \zeta_2
     -\frac{16 \zeta_3}{9}
     +\frac{265}{9}
     -\frac{1931}{27 N}
     +\frac{4816}{81 N^2}
\N\\&
     -\frac{139}{2 N^3}
     +\frac{285637}{3240 N^4}
     -\frac{72146}{675 N^5}
     +\frac{371921}{2835 N^6}
     -\frac{6318899}{39690 N^7}
     +\frac{257172521}{1360800 N^8}
   \Biggr\}
\N\\&
  +O\left(\ln^2\left(\bar{N}\right) \frac{1}{N^{9}}\right)
\comma
\end{align}
where $\bar{N} = N \exp(\gamma_E)$ and $\gamma_E$ denotes the Euler-Mascheroni number. The asymptotic 
representation for $H_{2,q}^{W^{+}+W^{-},\text{NS},(2)}$ reads~:
\begin{align}
   H_{2,q}^{W^{+}+W^{-},\text{NS},(2)}
   ={}&
   C_F^2
   \Biggl\{
      2 \ln(\bar{N})^4
     +\Biggl[
         6
        +\frac{4}{N}
        -\frac{14}{3 N^2}
        +\frac{4}{N^3}
        -\frac{59}{15 N^4}
        +\frac{4}{N^5}
        -\frac{254}{63 N^6}
        +\frac{4}{N^7}
\N\\&
        -\frac{119}{30 N^8}
      \Biggr] \ln(\bar{N})^3
     +\Biggl[
        -\frac{27}{2}
        +\frac{39}{N}
        -\frac{33}{2 N^2}
        +\frac{65}{3 N^3}
        -\frac{683}{30 N^4}
        +\frac{797}{30 N^5}
\N\\&
        -\frac{13099}{420 N^6}
        +\frac{1063}{30 N^7}
        -\frac{140287}{3600 N^8}
      \Biggr] \ln(\bar{N})^2
     -4 \zeta_2 \ln(\bar{N})^2
     +\Biggl[
        -\frac{51}{2}
        -\frac{49}{2 N}
\N\\&
        +\frac{287}{4 N^2}
        -\frac{1561}{18 N^3}
        +\frac{332}{5 N^4}
        -\frac{1353}{25 N^5}
        -\frac{32719}{3780 N^6}
        +\frac{533443}{4410 N^7}
        -\frac{642431}{7200 N^8}
      \Biggr] \ln(\bar{N})
\N\\&
     +\Biggl[
        -18
        +\frac{28}{N}
        -\frac{106}{3 N^2}
        +\frac{36}{N^3}
        -\frac{541}{15 N^4}
        +\frac{36}{N^5}
        -\frac{2266}{63 N^6}
        +\frac{36}{N^7}
        -\frac{1081}{30 N^8}
      \Biggr] \zeta_2 \ln(\bar{N})
\N\\&
     +24 \zeta_3 \ln(\bar{N})
     +\frac{4 \zeta_2^2}{5}
     +\big(
         \frac{111}{2}
        -\frac{47}{N}
        +\frac{213}{2 N^2}
        -\frac{233}{N^3}
        +\frac{24067}{45 N^4}
        -\frac{42331}{30 N^5}
\N\\&
        +\frac{4665499}{1260 N^6}
        -\frac{6845971}{630 N^7}
        +\frac{2391890629}{75600 N^8}
      \big) \zeta_2
     +\big(
        -66
        +\frac{60}{N}
        -\frac{74}{N^2}
        +\frac{72}{N^3}
\N\\&
        -\frac{359}{5 N^4}
        +\frac{72}{N^5}
        -\frac{1514}{21 N^6}
        +\frac{72}{N^7}
        -\frac{719}{10 N^8}
      \big) \zeta_3
     +\frac{331}{8}
     -\frac{431}{4 N}
     +\frac{42}{N^2}
     +\frac{12047}{72 N^3}
\N\\&
     -\frac{864961}{1440 N^4}
     +\frac{26702713}{13500 N^5}
     -\frac{144440651}{25200 N^6}
     +\frac{16295446411}{926100 N^7}
     -\frac{82008282247}{1587600 N^8}
   \Biggr\}
\N\\&
  +C_F T_F \Biggl\{
     -\frac{8}{9} \ln(\bar{N})^3
     +\Biggl[
        -\frac{58}{9}
        -\frac{4}{3 N}
        +\frac{14}{9 N^2}
        -\frac{4}{3 N^3}
        +\frac{59}{45 N^4}
        -\frac{4}{3 N^5}
        +\frac{254}{189 N^6}
\N\\&
        -\frac{4}{3 N^7}
        +\frac{119}{90 N^8}
      \Biggr] \ln(\bar{N})^2
     +\Biggl[
        -\frac{718}{27}
        -\frac{214}{9 N}
        +\frac{455}{27 N^2}
        -\frac{182}{9 N^3}
        +\frac{2893}{135 N^4}
        -\frac{1079}{45 N^5}
\N\\&
        +\frac{152897}{5670 N^6}
        -\frac{28121}{945 N^7}
        +\frac{521287}{16200 N^8}
      \Biggr] \ln(\bar{N})
     + \ln^2\big(\frac{m^2}{Q^2}\big) \Biggl[
        -\frac{8 \ln(\bar{N})}{3}
        +2
        -\frac{4}{3 N}
\N\\&
        +\frac{14}{9 N^2}
        -\frac{4}{3 N^3}
        +\frac{59}{45 N^4}
        -\frac{4}{3 N^5}
        +\frac{254}{189 N^6}
        -\frac{4}{3 N^7}
        +\frac{119}{90 N^8}
      \Biggr]
\N\\&
     +\ln\big(\frac{m^2}{Q^2}\big) \Biggl[
        -\frac{80 \ln(\bar{N})}{9}
        +\frac{16 \zeta_2}{3}
        +\frac{2}{3}
        -\frac{88}{9 N}
        +\frac{356}{27 N^2}
        -\frac{16}{N^3}
        +\frac{478}{27 N^4}
        -\frac{304}{15 N^5}
\N\\&
        +\frac{13124}{567 N^6}
        -\frac{544}{21 N^7}
        +\frac{767}{27 N^8}
      \Biggr]
     +\Biggl[
         \frac{70}{3}
        +\frac{4}{3 N}
        -\frac{14}{9 N^2}
        +\frac{4}{3 N^3}
        -\frac{59}{45 N^4}
        +\frac{4}{3 N^5}
\N\\&
        -\frac{254}{189 N^6}
        +\frac{4}{3 N^7}
        -\frac{119}{90 N^8}
      \Biggr] \zeta_2
     +\frac{8}{3} \zeta_2 \ln(\bar{N})
     -\frac{16 \zeta_3}{9}
     +\frac{265}{9}
     -\frac{1931}{27 N}
     +\frac{4816}{81 N^2}
\N\\&
     -\frac{139}{2 N^3}
     +\frac{285637}{3240 N^4}
     -\frac{72146}{675 N^5}
     +\frac{371921}{2835 N^6}
     -\frac{6318899}{39690 N^7}
     +\frac{257172521}{1360800 N^8}
   \Biggr\}
\N\\&
  +n_f C_F T_F \Biggl\{
     -\frac{8}{9} \ln(\bar{N})^3
     +\Biggl[
        -\frac{58}{9}
        -\frac{4}{3 N}
        +\frac{14}{9 N^2}
        -\frac{4}{3 N^3}
        +\frac{59}{45 N^4}
        -\frac{4}{3 N^5}
\N\\&
        +\frac{254}{189 N^6}
        -\frac{4}{3 N^7}
        +\frac{119}{90 N^8}
      \Biggr] \ln(\bar{N})^2
     +\Biggl[
        -\frac{494}{27}
        -\frac{214}{9 N}
        +\frac{455}{27 N^2}
        -\frac{182}{9 N^3}
        +\frac{2893}{135 N^4}
\N\\&
        -\frac{1079}{45 N^5}
        +\frac{152897}{5670 N^6}
        -\frac{28121}{945 N^7}
        +\frac{521287}{16200 N^8}
      \Biggr] \ln(\bar{N})
     +\frac{8}{3} \zeta_2 \ln(\bar{N})
     +\Biggl[
         \frac{170}{9}
        +\frac{4}{3 N}
\N\\&
        -\frac{14}{9 N^2}
        +\frac{4}{3 N^3}
        -\frac{59}{45 N^4}
        +\frac{4}{3 N^5}
        -\frac{254}{189 N^6}
        +\frac{4}{3 N^7}
        -\frac{119}{90 N^8}
      \Biggr] \zeta_2
     +\frac{8 \zeta_3}{9}
     +\frac{457}{18}
\N\\&
     -\frac{1699}{27 N}
     +\frac{3668}{81 N^2}
     -\frac{859}{18 N^3}
     +\frac{21229}{360 N^4}
     -\frac{46946}{675 N^5}
     +\frac{711653}{8505 N^6}
     -\frac{3962699}{39690 N^7}
\N\\&
     +\frac{158737961}{1360800 N^8}
   \Biggr\}
  +C_F C_A \big(
      \frac{22 \ln(\bar{N})^3}{9}
     +\big(
         \frac{367}{18}
        +\frac{11}{3 N}
        -\frac{77}{18 N^2}
        +\frac{11}{3 N^3}
        -\frac{649}{180 N^4}
\N\\&
        +\frac{11}{3 N^5}
        -\frac{1397}{378 N^6}
        +\frac{11}{3 N^7}
        -\frac{1309}{360 N^8}
      \big) \ln(\bar{N})^2
     -4 \zeta_2 \ln(\bar{N})^2
     +\big(
         \frac{3155}{54}
        +\frac{1333}{18 N}
\N\\&
        -\frac{8365}{108 N^2}
        +\frac{161}{2 N^3}
        -\frac{44219}{540 N^4}
        +\frac{75433}{900 N^5}
        -\frac{1451011}{22680 N^6}
        +\frac{531973}{26460 N^7}
\N\\&
        -\frac{21246299}{453600 N^8}
      \big) \ln(\bar{N})
     +\big(
        -\frac{22}{3}
        -\frac{20}{N}
        +\frac{74}{3 N^2}
        -\frac{24}{N^3}
        +\frac{359}{15 N^4}
        -\frac{24}{N^5}
        +\frac{1514}{63 N^6}
\N\\&
        -\frac{24}{N^7}
        +\frac{719}{30 N^8}
      \big) \zeta_2 \ln(\bar{N})
     -40 \zeta_3 \ln(\bar{N})
     +\frac{51}{5} \zeta_2^2
     +\big(
        -\frac{1139}{18}
        +\frac{13}{3 N}
        -\frac{805}{18 N^2}
\N\\&
        +\frac{326}{3 N^3}
        -\frac{11719}{45 N^4}
        +\frac{20953}{30 N^5}
        -\frac{6970693}{3780 N^6}
        +\frac{244171}{45 N^7}
        -\frac{1195407011}{75600 N^8}
      \big) \zeta_2
     +\big(
         \frac{464}{9}
\N\\&
        -\frac{44}{N}
        +\frac{160}{3 N^2}
        -\frac{50}{N^3}
        +\frac{149}{3 N^4}
        -\frac{50}{N^5}
        +\frac{3160}{63 N^6}
        -\frac{50}{N^7}
        +\frac{299}{6 N^8}
      \big) \zeta_3
     -\frac{5465}{72}
     +\frac{17579}{108 N}
\N\\&
     -\frac{30811}{324 N^2}
     +\frac{181}{72 N^3}
     +\frac{19531}{96 N^4}
     -\frac{11910503}{13500 N^5}
     +\frac{233947499}{85050 N^6}
     -\frac{48217963501}{5556600 N^7}
\N\\&
     +\frac{979578746771}{38102400 N^8}
   \big)
+O\left(\ln^3\left(\bar{N}\right) \frac{1}{N^{9}}\right)
\period
\end{align}

\end{appendix}

\section{\label{sec:xspaceRes}The heavy flavor Wilson coefficients in $x$-space}

\vspace*{1mm}
\noindent
The $x$-space representations of the heavy flavor Wilson coefficients
can be expressed in terms of harmonic polylogarithms to 2--loop order.
Also here  arguments $x$ will not be written
explicitly. They read~:
\begin{align}
   \LtwoWpWqNStwo
   ={}&
   C_F T_F \Biggl\{
      \frac{265}{9} \delta(1-x)
     +\Biggl[
          \frac{16 H_{0,1}}{3 (1-x)}
         +\frac{8 H_0{}^2}{1-x}
         +\frac{16 H_1 H_0}{3 (1-x)}
         +\frac{268 H_0}{9 (1-x)}
         +\frac{8 H_1{}^2}{3 (1-x)}
\N\\&
         +\frac{116 H_1}{9 (1-x)}
         -\frac{32 \zeta_2}{3 (1-x)}
         +\frac{718}{27 (1-x)}
      \Biggr]_+
     -\frac{8}{3}  (x+1) H_{0,1}
     -(4 x+4) H_0{}^2
\N\\&
     -\frac{8}{9}  (34 x+19) H_0
     -\frac{8}{3}  (x+1 ) H_1 H_0
     -\frac{4}{3}  (x+1) H_1{}^2
     -\frac{8}{9}  (17 x+8) H_1
\N\\&
     +\frac{16}{3} (x+1) \zeta_2
     -\frac{1244 x}{27}
     -\frac{272}{27}
   \Biggr\}
  +C_F T_F \ln\Biggl(\frac{m^2}{Q^2}\Biggr) \Biggl\{
      \frac{2}{3} \delta(1-x)
\N\\&
     +\Biggl[\frac{16 H_0}{3 (1-x)}+\frac{80}{9 (1-x)}\Biggr]_+
     -\frac{8}{3} (x+1) H_0
     -\frac{88 x}{9}
     +\frac{8}{9}
   \Biggr\}
\N\\&
  +C_F T_F \ln^2\Biggl(\frac{m^2}{Q^2}\Biggr) \Biggl\{
      2 \delta(1-x)
     +\Biggl[\frac{8}{3 (1-x)}\Biggr]_+
     -\frac{4 x}{3}
     -\frac{4}{3}
   \Biggr\},
\end{align}

\begin{align}
  \HtwoWpWqNStwo
  ={}&
  C_F^2 \Biggl\{
    \Biggl(\frac{64 \zeta_2^2}{5}+8 \zeta_2-72 \zeta_3+\frac{331}{8}\Biggr) \delta(1-x)
   +\Biggl[
     -\frac{8 H_0{}^3}{3 (1-x)}
     -\frac{12 H_1 H_0{}^2}{1-x}
\N\\&
     -\frac{3 H_0{}^2}{1-x}
     -\frac{32 H_1{}^2 H_0}{1-x}
     +\frac{48 \zeta_2 H_0}{1-x}
     -\frac{36 H_1 H_0}{1-x}
     +\frac{48 H_{0,-1} H_0}{1-x}
     -\frac{24 H_{0,1} H_0}{1-x}
\N\\&
     +\frac{61 H_0}{1-x}
     -\frac{8 H_1{}^3}{1-x}
     -\frac{18 H_1{}^2}{1-x}
     +\frac{24 \zeta_2}{1-x}
     +\frac{64 \zeta_3}{1-x}
     +\frac{16 \zeta_2 H_1}{1-x}
     +\frac{27 H_1}{1-x}
     +\frac{16 H_1 H_{0,1}}{1-x}
\N\\&
     +\frac{12 H_{0,1}}{1-x}
     -\frac{96 H_{0,0,-1}}{1-x}
     +\frac{24 H_{0,0,1}}{1-x}
     -\frac{24 H_{0,1,1}}{1-x}
     +\frac{51}{2 (1-x)}
    \Biggr]_+ 
\N\\&
   +\Biggl(x+5-\frac{4}{x+1}\Biggr) H_0{}^3
   +\Biggl(40 x-16+\frac{40}{x+1}\Biggr) H_{-1} H_0{}^2
   +(10-14 x) H_1 H_0{}^2
\N\\&
   +\Biggl(-56 x+8-\frac{32}{x+1}\Biggr) H_{-1}{}^2 H_0
   +16(x+1) H_1{}^2 H_0
   +\Biggl(\frac{72 x^3}{5}-2 x+12\Biggr) H_0{}^2
\N\\&
   +\Biggl(
       \frac{144 x^2}{5}
      -\frac{502 x}{5}
      -\frac{132}{5}
      -\frac{16}{x+1}
      -\frac{16}{5 x}
    \Biggr) H_0
   +\Biggl(-24 x-40+\frac{16}{x+1}\Biggr) \zeta_2 H_0
\N\\&
   +\Biggl(-\frac{144 x^3}{5}+40 x+72+\frac{16}{5 x^2}\Biggr) H_{-1} H_0
   +32 (x+1) H_1 H_0
\N\\&
   +\Biggl(-80 x-\frac{32}{x+1}\Biggr) H_{0,-1} H_0
   +(56 x+8) H_{0,1} H_0
   +4 (x+1) H_1{}^3
   +\frac{144 x^2}{5}
\N\\&
   +(18 x+14) H_1{}^2
   -\frac{461 x}{5}
   +\Biggl(-\frac{144 x^3}{5}-8 x-32\Biggr) \zeta_2
\N\\&
   +\Biggl(72 x-64+\frac{56}{x+1}\Biggr) \zeta_3
   +\Biggl(-72 x-\frac{64}{x+1}+24\Biggr) \zeta_2 H_{-1}
\N\\&
   +(16-68 x) H_1
   +(32 x-16) \zeta_2 H_1
   +\Biggl(\frac{144 x^3}{5}-40 x-72-\frac{16}{5 x^2}\Biggr) H_{0,-1}
\N\\&
   +\Biggl(112 x-16+\frac{64}{x+1}\Biggr) H_{-1} H_{0,-1}
   +16 x H_{0,1}
   +\Biggl(16 x-16+\frac{32}{x+1}\Biggr) H_{-1} H_{0,1}
\N\\&
   -8(x+1) H_1 H_{0,1}
   +\Biggl(-112 x+16-\frac{64}{x+1}\Biggr) H_{0,-1,-1}
\N\\&
   +\Biggl(-16 x+16-\frac{32}{x+1}\Biggr) H_{0,-1,1}
   +\Biggl(80 x+32-\frac{16}{x+1}\Biggr) H_{0,0,-1}
\N\\&
   +\Biggl(-60 x+4-\frac{16}{x+1}\Biggr) H_{0,0,1}
   +\Biggl(-16 x+16-\frac{32}{x+1}\Biggr) H_{0,1,-1}
\N\\&
   +16 (x+1) H_{0,1,1}
   +\frac{16}{5 x}
   -\frac{124}{5}
  \Biggr\}
 +C_A C_F \Biggl\{
    \Biggl(
      -\frac{32 \zeta_2^2}{5}
      -4 \zeta_2
\N\\&
      +54 \zeta_3
      -\frac{5465}{72}
    \Biggr) \delta(1-x)
   +\Biggl[
      -\frac{2 H_0{}^3}{1-x}
      -\frac{8 H_1 H_0{}^2}{1-x}
      -\frac{55 H_0{}^2}{3 (1-x)}
      +\frac{4 H_1{}^2 H_0}{1-x}
\N\\&
      +\frac{8 \zeta_2 H_0}{1-x}
      -\frac{44 H_1 H_0}{3 (1-x)}
      -\frac{24 H_{0,-1} H_0}{1-x}
      +\frac{16 H_{0,1} H_0}{1-x}
      -\frac{239 H_0}{3 (1-x)}
      -\frac{22 H_1{}^2}{3 (1-x)}
\N\\&
      +\frac{88 \zeta_2}{3 (1-x)}
      +\frac{4 \zeta_3}{1-x}
      +\frac{24 \zeta_2 H_1}{1-x}
      -\frac{367 H_1}{9 (1-x)}
      -\frac{16 H_1 H_{0,1}}{1-x}
      -\frac{44 H_{0,1}}{3 (1-x)}
\N\\&
      +\frac{48 H_{0,0,-1}}{1-x}
      -\frac{24 H_{0,0,1}}{1-x}
      +\frac{24 H_{0,1,1}}{1-x}
      -\frac{3155}{54 (1-x)}
    \Biggr]_+ 
   +\Biggl(2 x+\frac{2}{x+1}\Biggr) H_0{}^3
\N\\&
   +\Biggl(-\frac{36 x^3}{5}+\frac{115 x}{6}+\frac{55}{6}\Biggr) H_0{}^2
   +\Biggl(-20 x+8-\frac{20}{x+1}\Biggr) H_{-1} H_0{}^2
\N\\&
   +(14 x+2) H_1 H_0{}^2
   +\Biggl(28 x-4+\frac{16}{x+1}\Biggr) H_{-1}{}^2 H_0
   -2(x+1) H_1{}^2 H_0
\N\\&
   +\Biggl(-\frac{72 x^2}{5}+\frac{1693 x}{15}+\frac{8}{x+1}+\frac{583}{15}+\frac{8}{5 x}\Biggr) H_0
   +\Biggl(-8 x-\frac{8}{x+1}\Biggr) \zeta_2 H_0
\N\\&
   +\Biggl(\frac{72 x^3}{5}-20 x-36-\frac{8}{5 x^2}\Biggr) H_{-1} H_0
   +\frac{22}{3}(x+1) H_1 H_0
\N\\&
   +\Biggl(40 x+\frac{16}{x+1}\Biggr) H_{0,-1} H_0
   -(28 x+4) H_{0,1} H_0
   -\frac{72 x^2}{5}
   +\frac{11}{3}(x+1) H_1{}^2
\N\\&
   +\frac{17626 x}{135}
   +\Biggl(\frac{72 x^3}{5}-\frac{104 x}{3}-\frac{44}{3}\Biggr) \zeta_2
   +\Biggl(-56 x+12-\frac{28}{x+1}\Biggr) \zeta_3
\N\\&
   +\Biggl(36 x-12+\frac{32}{x+1}\Biggr) \zeta_2 H_{-1}
   +\frac{4}{9}(167 x+14) H_1
   -(32 x+8) \zeta_2 H_1
\N\\&
   +\Biggl(-\frac{72 x^3}{5}+20 x+36+\frac{8}{5 x^2}\Biggr) H_{0,-1}
   +\Biggl(-56 x+8-\frac{32}{x+1}\Biggr) H_{-1} H_{0,-1}
\N\\&
   +\frac{22}{3}(x+1) H_{0,1}
   +\Biggl(-8 x+8-\frac{16}{x+1}\Biggr) H_{-1} H_{0,1}
   +8 (x+1) H_1 H_{0,1}
\N\\&
   +\Biggl(56 x-8+\frac{32}{x+1}\Biggr) H_{0,-1,-1}
   +\Biggl(8 x-8+\frac{16}{x+1}\Biggr) H_{0,-1,1}
\N\\&
   +\Biggl(-40 x-16+\frac{8}{x+1}\Biggr) H_{0,0,-1}
   +\Biggl(36 x+4+\frac{8}{x+1}\Biggr) H_{0,0,1}
\N\\&
   +\Biggl(8 x-8+\frac{16}{x+1}\Biggr) H_{0,1,-1}
   -12 (x+1) H_{0,1,1}
   -\frac{8}{5 x}
   +\frac{3709}{135}
  \Biggr\}
\N\\&
  +\ln^2\Biggl(\frac{m^2}{Q^2}\Biggr) C_F T_F \Biggl(
     +\Biggl[\frac{8}{3 (1-x)}\Biggr]_+
     +2 \delta(1-x)
     -\frac{4 x}{3}
     -\frac{4}{3}
   \Biggr)
\N\\&
  +\ln\Biggl(\frac{m^2}{Q^2}\Biggr) C_F T_F \Biggl(
     -\frac{88 x}{9}
     +\frac{2}{3} \delta(1-x)
     -\frac{8}{3} (x+1) H_0
     +\frac{8}{9}
\N\\&
     +\Biggl[\frac{16 H_0}{3 (1-x)}+\frac{80}{9 (1-x)}\Biggr]_+ 
   \Biggr)
  +C_F T_F \Biggl(
     (-4 x-4) H_0{}^2
    -\frac{8}{9} (34 x+19) H_0
\N\\&
    -\frac{8}{3} (x+1) H_1 H_0
    -\frac{4}{3} (x+1) H_1{}^2
    -\frac{1244 x}{27}
    +\frac{16}{3} (x+1) \zeta_2
    +\frac{265}{9} \delta(1-x)
\N\\&
    -\frac{8}{9} (17x+8) H_1
    -\frac{8}{3} (x+1) H_{0,1}
    +\Biggl[
       +\frac{8 H_0{}^2}{1-x}
       +\frac{16 H_1 H_0}{3 (1-x)}
       +\frac{268 H_0}{9 (1-x)}
\N\\&
       +\frac{8 H_1{}^2}{3 (1-x)}
       -\frac{32 \zeta_2}{3 (1-x)}
       +\frac{116 H_1}{9 (1-x)}
       +\frac{16 H_{0,1}}{3 (1-x)}
       +\frac{718}{27 (1-x)}
     \Biggr]_+ 
    -\frac{272}{27}
   \Biggr)
\N\\&
  +n_f C_F T_F \Biggl(
     -\frac{10}{3} (x+1) H_0{}^2
     -\frac{4}{3}(19 x+13) H_0
     -\frac{8}{3} (x+1) H_1 H_0
\N\\&
     -\frac{4}{3} (x+1) H_1{}^2
     -\frac{976 x}{27}
     +\frac{16}{3} (x+1) \zeta_2
     +\frac{457}{18} \delta(1-x)
     -\frac{8}{9} (17x+8) H_1
\N\\&
     -\frac{8}{3} (x+1) H_{0,1}
     -\frac{316}{27}
     +\Biggl[
         \frac{20 H_0{}^2}{3 (1-x)}
        +\frac{16 H_1 H_0}{3 (1-x)}
        +\frac{76 H_0}{3 (1-x)}
        +\frac{8 H_1{}^2}{3 (1-x)}
\N\\&
        -\frac{32 \zeta_2}{3 (1-x)}
        +\frac{116 H_1}{9 (1-x)}
        +\frac{16 H_{0,1}}{3 (1-x)}
        +\frac{494}{27 (1-x)}
      \Biggr]_+ 
   \Biggr),
\end{align}

\begin{align}
\HtwoWmWqNStwo
  ={}&
  \HtwoWpWqNStwo
  + C_F (C_F - C_A/2) \Biggl\{
     \Biggl(
       -\frac{144 x^3}{5}
       +96 x^2
       +\frac{16}{5 x^2}
\N\\&
       +64 x
       +64
     \Biggr) H_{0,-1}
    +\Biggl(32 x-32+\frac{64}{x+1}\Biggr) H_0 H_{0,-1}
\N\\&
    +\Biggl(-224 x+32-\frac{128}{x+1}\Biggr) H_{-1} H_{0,-1}
    +\Biggl(-32 x+32-\frac{64}{x+1}\Biggr) H_{-1} H_{0,1}
\N\\&
    +16 (x+1) H_{0,1}
    +\Biggl(224 x-32+\frac{128}{x+1}\Biggr) H_{0,-1,-1}
\N\\&
    +\Biggl(32 x-32+\frac{64}{x+1}\Biggr) H_{0,-1,1}
    +\Biggl(96 x+\frac{32}{x+1}\Biggr) H_{0,0,-1}
\N\\&
    +\Biggl(16 x-16+\frac{32}{x+1}\Biggr) H_{0,0,1}
    +\Biggl(32 x-32+\frac{64}{x+1}\Biggr) H_{0,1,-1}
\N\\&
    +\Biggl(-\frac{144 x^2}{5}+\frac{292 x}{5}+\frac{32}{x+1}-\frac{28}{5}+\frac{16}{5 x}\Biggr) H_0
    +\Biggl(
       -\frac{72 x^3}{5}
       +48 x^2
\N\\&
       +32 x
       +8
     \Biggr) H_0{}^2
    +\Biggl(\frac{144 x^3}{5}-96 x^2-\frac{16}{5 x^2}-64 x-64\Biggr) H_{-1} H_0
\N\\&
    +\Biggl(-16 x+16-\frac{32}{x+1}\Biggr) \zeta_2 H_0
    +\Biggl(144 x-48+\frac{128}{x+1}\Biggr) \zeta_2 H_{-1}
\N\\&
    +\Biggl(4 x-4+\frac{8}{x+1}\Biggr) H_0{}^3
    +\Biggl(-80 x+32-\frac{80}{x+1}\Biggr) H_{-1} H_0{}^2
\N\\&
    +\Biggl(112 x-16+\frac{64}{x+1}\Biggr) H_{-1}{}^2 H_0
    -32(x-1) H_1
    +\Biggl(
        \frac{144 x^3}{5}
       -96 x^2
\N\\&
       -56 x
       -8
     \Biggr) \zeta_2
    +\Biggl(-136 x+40-\frac{112}{x+1}\Biggr) \zeta_3
    -\frac{144 x^2}{5}
    -\frac{164 x}{5}
    -\frac{16}{5 x}
    +\frac{324}{5}
   \Biggr\}
\comma
\end{align}

\begin{align}
  \HtwoWqPStwo
  ={}&
   C_F T_F \ln^2\Biggl(\frac{m^2}{Q^2}\Biggr) \Biggl\{
    (-4 x-4) H_0+\frac{8 x^2}{3}+2 x-\frac{8}{3 x}-2
   \Biggr\}
\N\\&
  +C_F T_F \ln\Biggl(\frac{m^2}{Q^2}\Biggr) \Biggl\{
      \Biggl(-\frac{32 x^2}{3}-20 x-4\Biggr) H_0
     +(4 x+4) H_0{}^2
     +\frac{224 x^2}{9}
     -24 x
     +8
\N\\&
     -\frac{80}{9 x}
   \Biggr\}
  +C_F T_F \Biggl\{
      \Biggl(\frac{32 x^2}{3}+32 x+\frac{32}{3 x}+32\Biggr) H_{0,-1}
     +\Biggl(-16 x^2+12 x-\frac{16}{x}-12\Biggr) H_{0,1}
\N\\&
     +(24 x+24) H_0 H_{0,1}
     +(-16 x-16) H_{0,0,1}
     +(16 x+16) H_{0,1,1}
\N\\&
     +\Biggl(-\frac{56 x^2}{3}+35 x-1\Biggr) H_0{}^2
     +\Biggl(-\frac{160 x^2}{3}-\frac{220 x}{3}+\frac{308}{3}\Biggr) H_0
\N\\&
     +\Biggl(-\frac{32 x^2}{3}-32 x-\frac{32}{3 x}-32\Biggr) H_{-1} H_0
     +\Biggl(-16 x^2-12 x+\frac{16}{x}+12\Biggr) H_1 H_0
\N\\&
     +\Biggl(-\frac{16 x^2}{3}-4 x+\frac{16}{3 x}+4\Biggr) H_1{}^2
     +\Biggl(\frac{64 x^2}{9}-\frac{160 x}{3}-\frac{208}{9 x}+\frac{208}{3}\Biggr) H_1
\N\\&
     +(-32 x-32) \zeta_2 H_0
     +(6 x+6) H_0{}^3
     +\Biggl(32 x^2-32 x-\frac{32}{3 x}\Biggr) \zeta_2
     +\frac{1696 x^2}{27}
\N\\&
     -\frac{1030 x}{9}
     +\frac{464}{27 x}
     +\frac{310}{9}
   \Biggr\},
\end{align}

\begin{align}
  \LtwoWgtwo
  ={}&
  \ln\Biggl(\frac{m^2}{Q^2}\Biggr) T_F^2 \Biggl\{
     \Biggl(-\frac{32 x^2}{3}+\frac{32 x}{3}-\frac{16}{3}\Biggr) H_0
    +\Biggl(-\frac{32 x^2}{3}+\frac{32 x}{3}-\frac{16}{3}\Biggr) H_1
\N\\&
    -\frac{128 x^2}{3}
    +\frac{128 x}{3}
    -\frac{16}{3}
  \Biggr\},
\end{align}

\begin{align}
  \HtwoWgtwo
  ={}&
   \ln^2\Biggl(\frac{m^2}{Q^2}\Biggr) \Biggl\{
      T_F^2 \Biggl(-\frac{16 x^2}{3}+\frac{16 x}{3}-\frac{8}{3}\Biggr)
     +C_F T_F (
         4 x-1
        +\left(-8 x^2+4 x-2\right) H_0
\N\\&
        +\left(-8 x^2+8 x-4\right) H_1)
     +C_A T_F \Biggl(
         \frac{62 x^2}{3}
        -16 x
        -2
        -\frac{8}{3 x}
        +(-16 x-4) H_0
\N\\&
        +\left(8 x^2-8 x+4\right) H_1
      \Biggr)
   \Biggr\}
  +\ln\Biggl(\frac{m^2}{Q^2}\Biggr) \Biggl\{
      T_F^2 \Biggl(
        -\frac{128 x^2}{3}
        +\frac{128 x}{3}
        -\frac{16}{3}
\N\\&
        +\Biggl(-\frac{32 x^2}{3}+\frac{32 x}{3}-\frac{16}{3}\Biggr) H_0
        +\Biggl(-\frac{32 x^2}{3}+\frac{32 x}{3}-\frac{16}{3}\Biggr) H_1
      \Biggr)
\N\\&
     +C_A T_F \Biggl(
         \frac{872 x^2}{9}
        -100 x
        +8
        -\frac{80}{9 x}
        +16 x \zeta_2
        +(8 x+4) H_0{}^2
        +\left(8 x^2-8 x+4\right) H_1{}^2
\N\\&
        +\Biggl(-\frac{176 x^2}{3}-32 x-4\Biggr) H_0
        +\left(16 x^2+16 x+8\right) H_{-1} H_0
        +\left(16 x^2-16 x\right) H_1
\N\\&
        +\left(-16 x^2-16 x-8\right) H_{0,-1}
      \Biggr)
     +C_F T_F \Biggl(
        -8 x^2
        +34 x
        -18
        +\left(-16 x^2+8 x-4\right) H_0{}^2
\N\\&
        +\left(-16 x^2+16 x-8\right) H_1{}^2
        +\left(32 x^2-24 x+12\right) \zeta_2
        +\left(-40 x^2+24 x-4\right) H_0
\N\\&
        +\left(-40 x^2+48 x-14\right) H_1
        +\left(-32 x^2+32 x-16\right) H_0 H_1
        +(4-8 x) H_{0,1}
      \Biggr)
   \Biggr\}
\N\\&
  +C_A T_F \Biggl\{
      \Biggl(\frac{52 x}{3}+6\Biggr) H_0{}^3
     +\Biggl(-\frac{365 x^2}{3}+180 x-1\Biggr) H_0{}^2
     +\left(28 x^2+20 x+10\right) H_{-1} H_0{}^2
\N\\&
     +\left(-16 x^2+24 x-12\right) H_1 H_0{}^2
     +\left(-8 x^2+8 x+4\right) H_{-1}{}^2 H_0
     +\left(-28 x^2+28 x-14\right) H_1{}^2 H_0
\N\\&
     +\Biggl(-\frac{1660 x^2}{3}+\frac{1082 x}{3}+\frac{320}{3}\Biggr) H_0
     +\left(32 x^2-128 x-16\right) \zeta_2 H_0
\N\\&
     +\Biggl(\frac{184 x^2}{3}+8 x-48-\frac{32}{3 x}\Biggr) H_{-1} H_0
     +\Biggl(-222 x^2+192 x-2+\frac{16}{x}\Biggr) H_1 H_0
\N\\&
     +\left(-24 x^2-40 x-20\right) H_{0,-1} H_0
     +(80 x+32) H_{0,1} H_0
     +\Biggl(-\frac{4 x^2}{3}+\frac{4 x}{3}-\frac{2}{3}\Biggr) H_1{}^3
\N\\&
     -\frac{5810 x^2}{27}
     +\frac{1202 x}{9}
     +\frac{472}{9}
     +\frac{464}{27 x}
     +\Biggl(-\frac{229 x^2}{3}+68 x-5+\frac{16}{3 x}\Biggr) H_1{}^2
\N\\&
     +\Biggl(294 x^2-280 x+16-\frac{32}{3 x}\Biggr) \zeta_2
     +\left(48 x^2-40 x+24\right) \zeta_3
     +\left(-40 x^2-24 x-12\right) \zeta_2 H_{-1}
\N\\&
     +\Biggl(-\frac{3068 x^2}{9}+\frac{884 x}{3}+\frac{118}{3}-\frac{208}{9 x}\Biggr) H_1
     +\left(16 x^2-32 x+16\right) \zeta_2 H_1
\N\\&
     +\Biggl(-\frac{184 x^2}{3}-8 x+48+\frac{32}{3 x}\Biggr) H_{0,-1}
     +\left(16 x^2-16 x-8\right) H_{-1} H_{0,-1}
\N\\&
     +\Biggl(-72 x^2+96 x-14-\frac{16}{x}\Biggr) H_{0,1}
     +\left(32 x^2+32 x+16\right) H_{-1} H_{0,1}
\N\\&
     +\left(32 x^2-32 x+16\right) H_1 H_{0,1}
     +\left(-16 x^2+16 x+8\right) H_{0,-1,-1}
\N\\&
     +\left(-32 x^2-32 x-16\right) H_{0,-1,1}
     +\left(-8 x^2+40 x+20\right) H_{0,0,-1}
     +(-80 x-24) H_{0,0,1}
\N\\&
     +\left(-32 x^2-32 x-16\right) H_{0,1,-1}
     +\left(-40 x^2+104 x-4\right) H_{0,1,1}
   \Biggr\}
\N\\&
  +C_F T_F \Biggl\{
      \left(-12 x^2+6 x-3\right) H_0{}^3
     +\Biggl(-\frac{96 x^3}{5}-62 x^2+\frac{26 x}{3}-\frac{7}{2}\Biggr) H_0{}^2
\N\\&
     +\left(16 x^2+32 x+16\right) H_{-1} H_0{}^2
     +\left(-20 x^2+4 x-2\right) H_1 H_0{}^2
\N\\&
     +\left(-32 x^2-64 x-32\right) H_{-1}{}^2 H_0
     +\left(-32 x^2+32 x-16\right) H_1{}^2 H_0
\N\\&
     +\Biggl(-\frac{552 x^2}{5}+\frac{181 x}{5}-\frac{592}{15}-\frac{16}{15 x}\Biggr) H_0
     +\left(96 x^2-64 x+32\right) \zeta_2 H_0
\N\\&
     +\Biggl(\frac{192 x^3}{5}+\frac{128 x}{3}+96+\frac{16}{15 x^2}\Biggr) H_{-1} H_0
     +\left(-124 x^2+136 x-50\right) H_1 H_0
\N\\&
     +\left(32 x^2-64 x+32\right) H_{0,-1} H_0
     +\left(-24 x^2+48 x-24\right) H_{0,1} H_0
\N\\&
     +\Biggl(-\frac{44 x^2}{3}+\frac{44 x}{3}-\frac{22}{3}\Biggr) H_1{}^3
     +\frac{128 x^2}{5}
     +\frac{273 x}{5}
     +\frac{16}{15 x}
     -\frac{1099}{15}
\N\\&
     +\left(-78 x^2+84 x-24\right) H_1{}^2
     +\Biggl(\frac{192 x^3}{5}+156 x^2-\frac{280 x}{3}+28\Biggr) \zeta_2
     +\left(152 x^2+8 x+60\right) \zeta_3
\N\\&
     +\left(-32 x^2-64 x-32\right) \zeta_2 H_{-1}
     +\left(-72 x^2+106 x-28\right) H_1
     +\left(64 x^2-32 x+16\right) \zeta_2 H_1
\N\\&
     +\Biggl(-\frac{192 x^3}{5}-\frac{128 x}{3}-96-\frac{16}{15 x^2}\Biggr) H_{0,-1}
     +\left(64 x^2+128 x+64\right) H_{-1} H_{0,-1}
\N\\&
     +\left(22-32 x^2\right) H_{0,1}
     +\left(-32 x^2+32 x-16\right) H_1 H_{0,1}
     +\left(-64 x^2-128 x-64\right) H_{0,-1,-1}
\N\\&
     +\left(-96 x^2+64 x-96\right) H_{0,0,-1}
     +\left(-8 x^2-40 x+20\right) H_{0,0,1}
\N\\&
     +\left(40 x^2-56 x+28\right) H_{0,1,1}
   \Biggr\},
\end{align}

\begin{align}
\LthreeWpWqNStwo
   ={}&
   C_F T_F \ln^2\Biggl(\frac{m^2}{Q^2}\Biggr) \Biggl\{
      2 \delta(1-x)
     +\Biggl[\frac{8}{3 (1-x)}\Biggr]_+
     -\frac{4 x}{3}
     -\frac{4}{3}
   \Biggr\}
\N\\&
  +C_F T_F \ln\Biggl(\frac{m^2}{Q^2}\Biggr) \Biggl\{
      \frac{2}{3} \delta(1-x)
     +\Biggl[\frac{16 H_0}{3 (1-x)}+\frac{80}{9 (1-x)}\Biggr]_+
     +\Biggl(-\frac{8 x}{3}-\frac{8}{3}\Biggr) H_0
\N\\&
     -\frac{88 x}{9}
     +\frac{8}{9}
   \Biggr\}
  +C_F T_F \Biggl\{
      \Biggl[
         +\frac{16 H_{0,1}}{3 (1-x)}
         +\frac{8 H_0{}^2}{1-x}
         +\frac{16 H_1 H_0}{3 (1-x)}
         +\frac{268 H_0}{9 (1-x)}
         +\frac{8 H_1{}^2}{3 (1-x)}
\N\\&
         +\frac{116 H_1}{9 (1-x)}
         -\frac{32 \zeta_2}{3 (1-x)}
         +\frac{718}{27 (1-x)}
      \Biggr]_+ 
     -\frac{8}{3} (x+1) H_{0,1}
     +\frac{265}{9} \delta(1-x)
\N\\&
     +(-4 x-4) H_0{}^2
     +\Biggl(-\frac{224 x}{9}-\frac{104}{9}\Biggr) H_0
     +\Biggl(-\frac{8 x}{3}-\frac{8}{3}\Biggr) H_1 H_0
     -\frac{872 x}{27}
     -\frac{188}{27}
\N\\&
     +\Biggl(-\frac{4 x}{3}-\frac{4}{3}\Biggr) H_1{}^2
     +\Biggl(-\frac{112 x}{9}-\frac{40}{9}\Biggr) H_1
     +\Biggl(\frac{16 x}{3}+\frac{16}{3}\Biggr) \zeta_2
   \Biggr\},
\end{align}

\begin{align}
\HthreeWpWqNStwo
  ={}&
  \HtwoWpWqNStwo
 +C_F^2 \Biggl\{
     \Biggl(4 x-4+\frac{8}{x+1}\Biggr) H_0{}^3
    +\Biggl(
       -\frac{72 x^3}{5}
       +8 x^2
\N\\&
       +28 x
       +12
     \Biggr) H_0{}^2
    +\Biggl(-56 x+40-\frac{80}{x+1}\Biggr) H_{-1} H_0{}^2
    +(24 x-8) H_1 H_0{}^2
\N\\&
    +\Biggl(64 x-32+\frac{64}{x+1}\Biggr) H_{-1}{}^2 H_0
    +\Biggl(
       -\frac{144 x^2}{5}
       +\frac{192 x}{5}
       -\frac{68}{5}
\N\\&
       +\frac{32}{x+1}
       +\frac{16}{5 x}
     \Biggr) H_0
    +\Biggl(-16 x+16-\frac{32}{x+1}\Biggr) \zeta_2 H_0
    +\Biggl(
        \frac{144 x^3}{5}
       -16 x^2
       -48 x
\N\\&
       -80
       -\frac{16}{x}
       -\frac{16}{5 x^2}
     \Biggr) H_{-1} H_0
    -8 (x+1) H_1 H_0
    +\Biggl(80 x-48+\frac{64}{x+1}\Biggr) H_{0,-1} H_0
\N\\&
    +(16-48 x) H_{0,1} H_0
    -4 (x+1) H_1{}^2
    +\Biggl(\frac{144 x^3}{5}-16 x^2-44 x-12\Biggr) \zeta_2
\N\\&
    +\Biggl(-136 x+72-\frac{112}{x+1}\Biggr) \zeta_3
    +\Biggl(96 x-64+\frac{128}{x+1}\Biggr) \zeta_2 H_{-1}
    +(46 x-10) H_1
\N\\&
    +(16-48 x) \zeta_2 H_1
    +\Biggl(-\frac{144 x^3}{5}+16 x^2+48 x+80+\frac{16}{x}+\frac{16}{5 x^2}\Biggr) H_{0,-1}
\N\\&
    +\Biggl(-128 x+64-\frac{128}{x+1}\Biggr) H_{-1} H_{0,-1}
    +12 (x+1) H_{0,1}
\N\\&
    +\Biggl(-32 x+32-\frac{64}{x+1}\Biggr) H_{-1} H_{0,1}
    +\Biggl(128 x-64+\frac{128}{x+1}\Biggr) H_{0,-1,-1}
\N\\&
    +\Biggl(32 x-32+\frac{64}{x+1}\Biggr) H_{0,-1,1}
    +\Biggl(-48 x+16+\frac{32}{x+1}\Biggr) H_{0,0,-1}
\N\\&
    +\Biggl(64 x-32+\frac{32}{x+1}\Biggr) H_{0,0,1}
    +\Biggl(32 x-32+\frac{64}{x+1}\Biggr) H_{0,1,-1}
    -\frac{144 x^2}{5}
\N\\&
    +\frac{561 x}{5}
    -\frac{231}{5}
    -\frac{16}{5 x}
  \Biggr\}
\N\\&
 +C_A C_F \Biggl\{
     \Biggl(-2 x+2-\frac{4}{x+1}\Biggr) H_0{}^3
    +\Biggl(\frac{36 x^3}{5}-4 x^2-16 x-8\Biggr) H_0{}^2
\N\\&
    +\Biggl(28 x-20+\frac{40}{x+1}\Biggr) H_{-1} H_0{}^2
    +(4-12 x) H_1 H_0{}^2
\N\\&
    +\Biggl(-32 x+16-\frac{32}{x+1}\Biggr) H_{-1}{}^2 H_0
    +\Biggl(\frac{72 x^2}{5}-\frac{478 x}{15}+\frac{2}{15}-\frac{16}{x+1}-\frac{8}{5 x}\Biggr) H_0
\N\\&
    +\Biggl(8 x-8+\frac{16}{x+1}\Biggr) \zeta_2 H_0
    +\Biggl(-\frac{72 x^3}{5}+8 x^2+24 x+40+\frac{8}{x}+\frac{8}{5 x^2}\Biggr) H_{-1} H_0
\N\\&
    +\Biggl(-40 x+24-\frac{32}{x+1}\Biggr) H_{0,-1} H_0
    +(24 x-8) H_{0,1} H_0
    +\Biggl(
       -\frac{72 x^3}{5}
       +8 x^2
\N\\&
       +28 x
       +12
     \Biggr) \zeta_2
    +\Biggl(68 x-36+\frac{56}{x+1}\Biggr) \zeta_3
    +\Biggl(-48 x+32-\frac{64}{x+1}\Biggr) \zeta_2 H_{-1}
\N\\&
    -\frac{2}{3}(47 x-1) H_1
    +(24 x-8) \zeta_2 H_1
    +\Biggl(
        \frac{72 x^3}{5}
       -8 x^2
       -24 x
       -40
\N\\&
       -\frac{8}{x}
       -\frac{8}{5 x^2}
     \Biggr) H_{0,-1}
    +\Biggl(64 x-32+\frac{64}{x+1}\Biggr) H_{-1} H_{0,-1}
    -8 (x+1) H_{0,1}
\N\\&
    +\Biggl(16 x-16+\frac{32}{x+1}\Biggr) H_{-1} H_{0,1}
    +\Biggl(-64 x+32-\frac{64}{x+1}\Biggr) H_{0,-1,-1}
\N\\&
    +\Biggl(-16 x+16-\frac{32}{x+1}\Biggr) H_{0,-1,1}
    +\Biggl(24 x-8-\frac{16}{x+1}\Biggr) H_{0,0,-1}
\N\\&
    +\Biggl(-32 x+16-\frac{16}{x+1}\Biggr) H_{0,0,1}
    +\Biggl(-16 x+16-\frac{32}{x+1}\Biggr) H_{0,1,-1}
    +\frac{72 x^2}{5}
\N\\&
    -\frac{3517 x}{45}
    +\frac{8}{5 x}
    +\frac{647}{45}
  \Biggr\}
  +C_F T_F \Biggl\{
      \frac{124 x}{9}
     +\frac{28}{9}
     +\frac{16}{3} (x+1) H_0
     +\frac{8}{3} (x+1) H_1
   \Biggr\}
\N\\&
  +n_f C_F T_F \Biggl\{
      \frac{124 x}{9}
     +\frac{28}{9}
     +\frac{16}{3} (x+1) H_0
     +\frac{8}{3} (x+1) H_1
   \Biggr\}
\comma
\end{align}

\begin{align}
\HthreeWmWqNStwo
  ={}&
  \HthreeWpWqNStwo
 +C_F (C_F - C_A/2) \Biggl\{
     \Biggl(-16 x^2-\frac{16}{x}\Biggr) H_{0,-1}
\N\\&
    +\Biggl(-32 x+32-\frac{64}{x+1}\Biggr) H_0 H_{0,-1}
    +\Biggl(32 x-96+\frac{128}{x+1}\Biggr) H_{-1} H_{0,-1}
\N\\&
    -16 (x+1) H_{0,1}
    +\Biggl(32 x-32+\frac{64}{x+1}\Biggr) H_{-1} H_{0,1}
\N\\&
    +\Biggl(-32 x+96-\frac{128}{x+1}\Biggr) H_{0,-1,-1}
    +\Biggl(-32 x+32-\frac{64}{x+1}\Biggr) H_{0,-1,1}
\N\\&
    +\Biggl(32-\frac{32}{x+1}\Biggr) H_{0,0,-1}
    +\Biggl(-16 x+16-\frac{32}{x+1}\Biggr) H_{0,0,1}
\N\\&
    +\Biggl(-32 x+32-\frac{64}{x+1}\Biggr) H_{0,1,-1}
    +\left(-8 x^2-8 x-16\right) H_0{}^2
\N\\&
    +\Biggl(16 x^2+\frac{16}{x}\Biggr) H_{-1} H_0
    +\Biggl(16 x-16+\frac{32}{x+1}\Biggr) \zeta_2 H_0
\N\\&
    +\Biggl(-48 x+80-\frac{128}{x+1}\Biggr) \zeta_2 H_{-1}
    +\Biggl(-4 x+4-\frac{8}{x+1}\Biggr) H_0{}^3
\N\\&
    +\Biggl(32 x-48+\frac{80}{x+1}\Biggr) H_{-1} H_0{}^2
    +\Biggl(-16 x+48-\frac{64}{x+1}\Biggr) H_{-1}{}^2 H_0
\N\\&
    +\Biggl(60 x+28-\frac{32}{x+1}\Biggr) H_0
    +32 (x-1) H_1
    +\left(16 x^2+8 x+24\right) \zeta_2
\N\\&
    +\Biggl(40 x-72+\frac{112}{x+1}\Biggr) \zeta_3
    -60 x
    +60
  \Biggr\}
\comma
\end{align}

\begin{align}
   \HthreeWqPStwo
   ={}&
   C_F T_F \ln^2\Biggl(\frac{m^2}{Q^2}\Biggr) \Biggl\{
      (4 x+4) H_0
     -\frac{8 x^2}{3}
     -2 x+\frac{8}{3 x}+2
   \Biggr\}
\N\\&
  +C_F T_F \ln\Biggl(\frac{m^2}{Q^2}\Biggr) \Biggl\{
      \Biggl(\frac{32 x^2}{3}+20 x+4\Biggr) H_0
     +(-4 x-4) H_0{}^2
     -\frac{224 x^2}{9}
     +24 x
\N\\&
     -8
     +\frac{80}{9 x}
   \Biggr\}
  +C_F T_F \Biggl\{
      \Biggl(-\frac{16 x^2}{3}-4 x+\frac{16}{3 x}+4\Biggr) H_{0,1}
     +(-8 x-8) H_0 H_{0,1}
\N\\&
     +(16 x+16) H_{0,0,1}
     +\Biggl(-\frac{8 x^2}{3}-5 x-1\Biggr) H_0{}^2
     +\Biggl(\frac{224 x^2}{9}+\frac{44 x}{3}+\frac{28}{3}\Biggr) H_0
\N\\&
     +\Biggl(\frac{16 x^2}{3}+4 x-\frac{16}{3 x}-4\Biggr) H_1 H_0
     +\Biggl(\frac{2 x}{3}+\frac{2}{3}\Biggr) H_0{}^3
     -\frac{800 x^2}{27}
     +(-16 x-16) \zeta_3
\N\\&
     +\frac{62 x}{3}
     +\frac{224}{27 x}
     +\frac{2}{3}
   \Biggr\},
\end{align}

\begin{align}
\HthreeWgtwo
  ={}&
  \ln^2\Biggl(\frac{m^2}{Q^2}\Biggr) \Biggl\{
     T_F^2 \Biggl(\frac{16 x^2}{3}-\frac{16 x}{3}+\frac{8}{3}\Biggr)
    +C_A T_F \Biggl(
       -\frac{62 x^2}{3}
       +16 x
       +(16 x+4) H_0
\N\\&
       +\left(-8 x^2+8 x-4\right) H_1
       +2
       +\frac{8}{3 x}
     \Biggr)
    +C_F T_F \Bigl(
       -4 x
       +\left(8 x^2-4 x+2\right) H_0
\N\\&
       +\left(8 x^2-8 x+4\right) H_1
       +1
     \Bigr)
  \Biggr\}
 +\ln\Biggl(\frac{m^2}{Q^2}\Biggr) \Biggl\{
     C_A T_F \Biggl(
       -16 \zeta_2 x
       +(-8 x-4) H_0{}^2
\N\\&
       +\left(-8 x^2+8 x-4\right) H_1{}^2
       +\Biggl(\frac{176 x^2}{3}+32 x+4\Biggr) H_0
       +\left(-16 x^2-16 x-8\right) H_{-1} H_0
\N\\&
       +\left(16 x-16 x^2\right) H_1
       +\left(16 x^2+16 x+8\right) H_{0,-1}
       -\frac{872 x^2}{9}
       +100 x
       -8
       +\frac{80}{9 x}
     \Biggr)
\N\\&
    +C_F T_F (
        { 20 x^2 -46 x}
       +\left(16 x^2-8 x+4\right) H_0{}^2
       +\left(16 x^2-16 x+8\right) H_1{}^2
\N\\&
       +\left(-32 x^2+24 x-12\right) \zeta_2
       +\left(40 x^2{-32 x+8}\right) H_0
       +\left(40 x^2-48 x+14\right) H_1
\N\\&
       +\left(32 x^2-32 x+16\right) H_0 H_1
       +(8 x-4) H_{0,1}
       {-18}
     )
  \Biggr\}
 +C_A T_F \Biggl\{
     \Biggl(\frac{4 x}{3}+\frac{2}{3}\Biggr) H_0{}^3
\N\\&
    +\Biggl(-\frac{23 x^2}{3}-4 x-1\Biggr) H_0{}^2
    +\left(-4 x^2-4 x-2\right) H_{-1} H_0{}^2
    +\left(8 x^2+8 x+4\right) H_{-1}{}^2 H_0
\N\\&
    +\left(4 x^2-4 x+2\right) H_1{}^2 H_0
    +\Biggl(\frac{800 x^2}{9}+\frac{86 x}{3}+\frac{28}{3}\Biggr) H_0
    +\left(-8 x^2-8 x\right) H_{-1} H_0
\N\\&
    +\Biggl(\frac{130 x^2}{3}-32 x-6-\frac{16}{3 x}\Biggr) H_1 H_0
    +\left(8 x^2+8 x+4\right) H_{0,-1} H_0
    +(-32 x-8) H_{0,1} H_0
\N\\&
    +\Biggl(-\frac{4 x^2}{3}+\frac{4 x}{3}-\frac{2}{3}\Biggr) H_1{}^3
    -\frac{3176 x^2}{27}
    +\left(-5 x^2+4 x+1\right) H_1{}^2
    +\frac{314 x}{3}
    +\left(2 x^2-8 x\right) \zeta_2
\N\\&
    +(-56 x-16) \zeta_3
    +\left(8 x^2+8 x+4\right) \zeta_2 H_{-1}
    +\left(-8 x^2+8 x+2\right) H_1
    +\left(8 x^2+8 x\right) H_{0,-1}
\N\\&
    +\left(-16 x^2-16 x-8\right) H_{-1} H_{0,-1}
    +\Biggl(-\frac{136 x^2}{3}+32 x+6+\frac{16}{3 x}\Biggr) H_{0,1}
\N\\&
    +\left(16 x^2+16 x+8\right) H_{0,-1,-1}
    +\left(-8 x^2-8 x-4\right) H_{0,0,-1}
    +(64 x+16) H_{0,0,1}
\N\\&
    +\left(-8 x^2+8 x-4\right) H_{0,1,1}
    +\frac{224}{27 x}
    +\frac{2}{3}
  \Biggr\}
 +C_F T_F \Biggl\{
     \Biggl(-\frac{4 x^2}{3}+\frac{2 x}{3}-\frac{1}{3}\Biggr) H_0{}^3
\N\\&
    +\Biggl(-10 x^2+6 x+\frac{1}{2}\Biggr) H_0{}^2
    +\left(-4 x^2+4 x-2\right) H_1 H_0{}^2
    +\left(24 x^2+9 x+8\right) H_0
\N\\&
    +\left(-20 x^2+24 x-2\right) H_1 H_0
    +\left(8 x^2-16 x+8\right) H_{0,1} H_0
    +\Biggl(\frac{4 x^2}{3}-\frac{4 x}{3}+\frac{2}{3}\Biggr) H_1{}^3
\N\\&
    -40 x^2+\left(6 x^2-4 x-2\right) H_1{}^2
    +41 x+\left(-12 x^2+24 x+4\right) \zeta_2
    +\left(-8 x^2-8 x+4\right) \zeta_3
\N\\&
    +\left(24 x^2-26 x\right) H_1
    +\left(32 x^2-48 x-2\right) H_{0,1}
    +\left(-8 x^2+24 x-12\right) H_{0,0,1}
\N\\&
    +\left(8 x^2-8 x+4\right) H_{0,1,1}
    -13
  \Biggr\}
\period
\end{align}
The $+$-distribution in the above relations is defined by
\begin{eqnarray}
\int_0^1 dx \left[f(x)\right]_+~g(x) = \int_0^1 dx f(x) [g(x) - g(1)]~.
\end{eqnarray}

\vspace*{5mm}\noindent
{\bf Acknowledgment.}~We would like to thank A. Behring, A. De Freitas and C. Schneider for 
discussions. This work has been supported in part by DFG Sonderforschungsbereich Transregio 9, 
Computergest\"utzte Theoretische Teilchenphysik, by the Austrian Science Fund (FWF) grant 
P20347-N18, by the EU Network grants {\sf LHCPHENOnet} PITN-GA-2010-264564 and {\sf 
HIGGSTOOLS} PITN-GA-2012-316704.


\begin{thebibliography}{99}
%
\bibitem{Blumlein:1996sc}
  J.~Bl\"umlein and S.~Riemersma.
  In~: Proceedings of the International Workshop {\sf Future physics at HERA}, pp. 82--85
  [hep-ph/9609394].
%
\bibitem{Blumlein:1987xk}
  J.~Bl\"umlein, M.~Klein, T.~Naumann and T.~Riemann.
  In~: Proceedings of the {\sf HERA Workshop}, ed.~R.D. Peccei, pp. 67--106,~ 
  PHE-88-01.
%
\bibitem{Blumlein:1989pd}
  J.~Bl\"umlein, M.~Klein, G.~Ingelman and R.~R\"uckl.
  {\it Z.\ Phys.\ C}, {\bf 45} (1990) 501--513.
%
\bibitem{Buza:1995ie}
M.~Buza, Y.~Matiounine, J.~Smith, R.~Migneron, and W.~L. van Neerven.
\newblock {\em Nucl. Phys. B}, {\bf 472} (1996) 611--658.
\newblock (arXiv:hep-ph/9601302).
%
\bibitem{Furmanski:1981cw}
  W.~Furmanski and R.~Petronzio.
  {\it Z.\ Phys.\ C}, {\bf 11} (1982) 293--314.
%
\bibitem{vanNeerven:1991nn}
W.~L. van Neerven and E.~B. Zijlstra.
\newblock {\em Phys. Lett. B}, {\bf 272} (1991) 127--133.
%
\bibitem{Zijlstra:1991qc}
E.~B. Zijlstra and W.~L. van Neerven.
\newblock {\em Phys. Lett. B}, {\bf 273} (1991) 476--482.
%
\bibitem{Zijlstra:1992kj}
E.~B. Zijlstra and W.~L. van Neerven.
\newblock {\em Phys. Lett. B}, {\bf 297} (1992) 377--384.
%
\bibitem{Zijlstra:1992qd}
E.~B. Zijlstra and W.~L. van Neerven.
\newblock {\em Nucl. Phys. B}, {\bf 383} (1992) 525--574.
%
\bibitem{Moch:1999eb}
S.~Moch and J.~A.~M. Vermaseren.
\newblock {\em Nucl. Phys. B}, {\bf 573} (2000) 853--907.
\newblock (arXiv:hep-ph/9912355).
%
\bibitem{Buza:1996xr}
M.~Buza, Y.~Matiounine, J.~Smith, and W.~L. van Neerven.
\newblock {\em Nucl. Phys. B}, {\bf 485} (1997) 420--456.
\newblock (arXiv:hep-ph/9608342).
%
\bibitem{Buza:1996wv}
M.~Buza, Y.~Matiounine, J.~Smith, and W.~L. van Neerven.
\newblock {\em Eur. Phys. J. C}, {\bf 1} (1998) 301--320.
%
\bibitem{Bierenbaum:2007dm}
I.~Bierenbaum, J.~Bl\"umlein, and S.~Klein.
\newblock {\em Phys. Lett. B}, {\bf 648} (2007) 195--200.
\newblock (arXiv:hep-ph/0702265).
%
\bibitem{Blumlein:2006mh}
  J.~Bl\"umlein, A.~De Freitas, W.~L.~van Neerven and S.~Klein.
  {\em Nucl.\ Phys.\ B} {\bf 755}, (2006) 272--285
  [hep-ph/0608024].
%
\bibitem{Bierenbaum:2007qe}
I.~Bierenbaum, J.~Bl\"umlein, and S.~Klein.
\newblock {\em Nucl. Phys. B}, {\bf 780} (2007) 40--75.
\newblock (arXiv:hep-ph/0703285).
%
\bibitem{Bierenbaum:2008yu}
I.~Bierenbaum, J.~Bl\"umlein, S.~Klein, and C.~Schneider.
\newblock {\em Nucl. Phys. B}, {\bf 803} (2008) 1--41.
\newblock (arXiv:0803.0273 [hep-ph]).
%
\bibitem{Bierenbaum:2009zt}
I.~Bierenbaum, J.~Bl\"umlein, and S.~Klein.
\newblock {\em Phys. Lett. B}, {\bf 672} (2009) 401--406.
\newblock (arXiv:0901.0669 [hep-ph]).
%
\bibitem{Bierenbaum:2009mv}
I.~Bierenbaum, J.~Bl\"umlein, and S.~Klein.
\newblock {\em Nucl. Phys. B}, {\bf 820} (2009) 417--482.
\newblock (arXiv:0904.3563 [hep-ph]).
%
\bibitem{Ablinger:2010ty}
  J.~Ablinger, J.~Bl\"umlein, S.~Klein, C.~Schneider and F.~Wi\ss{}brock.
  {\em Nucl.\ Phys.\ B}, {\bf 844} (2011) 26--54
  [arXiv:1008.3347 [hep-ph]].
%
\bibitem{Ablinger:2012qm}
  J.~Ablinger, J.~Bl\"umlein, A.~Hasselhuhn, S.~Klein, C.~Schneider and F.~Wi\ss{}brock.
  {\em Nucl.\ Phys.\ B}, {\bf 864} (2012) 52--84
  [arXiv:1206.2252 [hep-ph]].
%
\bibitem{Blumlein:2012vq}
  J.~Bl\"umlein, A.~Hasselhuhn, S.~Klein and C.~Schneider.
  {\em Nucl.\ Phys.\ B}, {\bf 866} (2013) 196--211
  [arXiv:1205.4184 [hep-ph]].
%
\bibitem{Behring:2013dga}
  A.~Behring et al. 
  arXiv:1312.0124 [hep-ph].
%
\bibitem{Buza:1997mg}
M.~Buza and W.~L. van Neerven.
\newblock {\em Nucl. Phys. B}, {\bf 500} (1997) 301--324.
\newblock (arXiv:hep-ph/9702242).
%
\bibitem{Gottschalk:1980rv}
T.~Gottschalk.
\newblock {\em Phys. Rev. D}, {\bf 23} (1981) 56--74.
%
\bibitem{Gluck:1996ve}
M.~Gl\"uck, S.~Kretzer, and E.~Reya.
\newblock {\em Phys. Lett. B}, {\bf 380} (1996) 171--176.
\newblock [Erratum-ibid. 405 (1997) 391], (arXiv:hep-ph/9603304).
%
\bibitem{Blumlein:2011zu}
J.~Bl\"umlein, A.~Hasselhuhn, P.~Kovacikova, and S.~Moch.
\newblock {\em Phys. Lett. B}, {\bf 700} (2011) 294--304.
\newblock (arXiv:1104.3449 [hep-ph]).
%
\bibitem{Alekhin:2013nda}
  S.~Alekhin, J.~Bl\"umlein and S.~Moch,
  arXiv:1310.3059 [hep-ph].
%
\bibitem{Alekhin:2012ig}
  S.~Alekhin, J.~Bl\"umlein and S.~Moch.
  {\em Phys.\ Rev.\ D}, {\bf 86} (2012) 054009
  [arXiv:1202.2281 [hep-ph]].
%
\bibitem{Schmitz1997}
N.~Schmitz.
\newblock {\em {Neutrinophysik}}.
\newblock (Teubner, Stuttgart, 1997).
%
\bibitem{Cabibbo:1963}
  N.~Cabibbo,
  {\it Phys.\ Rev.\ Lett.},  {\bf 10} (1963) 531.
%
\bibitem{Kobayashi:1973fv}
  M.~Kobayashi and T.~Maskawa,
  {\it Prog.\ Theor.\ Phys.},\  {\bf 49} (1973) 652.
%
\bibitem{Buras:1979yt}
A.~J. Buras.
\newblock {\em Rev. Mod. Phys.}, {\bf 52} (1980) 199--276.
%
\bibitem{Carlson1914}
F.~D. Carlson.
\newblock {\em {Sur une classe de s\'eries de Taylor}}.
\newblock PhD thesis, Uppsala University, 1914.
\newblock {see also \url{http://en.wikipedia.org/wiki/Carlson's_theorem}}.
%
\bibitem{Titchmarsh39}
E.~C. Titchmarsh.
\newblock {\em {The Theory of Functions}}.
\newblock (Oxford University Press, 1939), 2nd ed. edition.
%
\bibitem{Ablinger:2013cf}
J.~Ablinger, J.~Bl\"umlein, and C.~Schneider.
\newblock {\em J. Math. Phys.}, {\bf 54} (2013) 082301.
\newblock (arXiv:1302.0378 [math-ph]).
%
\bibitem{Politzer:1974fr}
H.~D. Politzer.
\newblock {\em Phys. Rept.}, {\bf 14} (1974) 129--180.
%
\bibitem{Blumlein:1996tp}
J.~Bl\"umlein and N.~Kochelev.
\newblock {\em Phys. Lett. B}, {\bf 381} (1996) 296--304.
\newblock (arXiv:hep-ph/9603397).
%
\bibitem{Blumlein:1998nv}
J.~Bl\"umlein and A.~Tkabladze.
\newblock {\em Nucl. Phys. B}, {\bf 553} (1999) 427--464.
\newblock (arXiv:hep-ph/9812478).
%
\bibitem{Vermaseren:2005qc}
J.~A.~M. Vermaseren, A.~Vogt, and S.~Moch.
\newblock {\em Nucl. Phys. B}, {\bf 724} (2005) 3--182.
\newblock (arXiv:hep-ph/0504242).
%
\bibitem{Blumlein:2009tj}
J.~Bl\"umlein, M.~Kauers, S.~Klein, and C.~Schneider.
\newblock {\em Comput. Phys. Commun.}, {\bf 180} (2009) 2143--2165.
\newblock (arXiv:0902.4091 [hep-ph]).
%
\bibitem{Moch:2007rq}
S.~Moch, M.~Rogal, and A.~Vogt.
\newblock {\em Nucl. Phys. B}, {\bf 790} (2008) 317--335.
\newblock (arXiv:0708.3731 [hep-ph]).
%
\bibitem{Vermaseren:1998uu}
  J.~A.~M.~Vermaseren.
  {\em Int.\ J.\ Mod.\ Phys.\ A}, {\bf 14} (1999) 2037--2076
  [hep-ph/9806280].
%
\bibitem{Blumlein:1998if}
  J.~Bl\"umlein and S.~Kurth,
  {\em Phys.\ Rev.\ D}, {\bf 60} (1999) 014018
  [hep-ph/9810241].
%
\bibitem{Ablinger:2010kw}
J.~Ablinger,
\newblock Diploma thesis, Johannes Kepler Universit\"at, Linz, Austria, 2010.
\newblock (arXiv:1011.1176 [math-ph]).
%
\bibitem{Ablinger:2013hcp}
J.~Ablinger.
\newblock PhD thesis, Johannes Kepler Universit\"at, Linz, Austria, 2012.
\newblock (arXiv:1305.0687 [math-ph]).
%
\bibitem{Ablinger:2011te}
J.~Ablinger, J.~Bl\"umlein, and C.~Schneider.
\newblock {\em J. Math. Phys.}, {\bf 52} (2011) 102301--1--52.
\newblock (arXiv:1105.6063 [math-ph]).
%
\bibitem{Remiddi:1999ew}
  E.~Remiddi and J.~A.~M.~Vermaseren,
  {\em Int.\ J.\ Mod.\ Phys.\ A} {\bf 15} (2000) 725--754
  [hep-ph/9905237].
%
\bibitem{NIELSEN}
N.~Nielsen.  {\sf Der Eulersche Dilogarithmus und seine Verallgemeinerungen},
{\it Nova Acta Leopold.}, {\bf XC} Nr. 3 (1909) 125--211.
%
\bibitem{Blumlein:2000hw}
J.~Bl\"umlein.
\newblock {\em Comput. Phys. Commun.}, {\bf 133} (2000) 76--104.
\newblock (arXiv:hep-ph/0003100).
%
\bibitem{Alekhin:2009ni}
  S.~Alekhin, J.~Bl\"umlein, S.~Klein and S.~Moch.
  {\em Phys.\ Rev.\ D} {\bf 81} (2010) 014032
  [arXiv:0908.2766 [hep-ph]].
%
\bibitem{Blumlein:2005jg}
J.~Bl\"umlein and S.-O. Moch.
\newblock {\em Phys. Lett. B}, {\bf 614} (2005) 53--61.
\newblock (arXiv:hep-ph/0503188).
%
\bibitem{Blumlein:2009ta}
J.~Bl\"umlein.
\newblock {\em Comput. Phys. Commun.}, {\bf 180} (2009) 2218--2249.
\newblock (arXiv:0901.3106 [hep-ph]).
%
\bibitem{Blumlein:2009fz}
J.~Bl\"umlein.
\newblock In: A.~Carey, D.~Ellwood, S.~Paycha, and S.~Rosenberg, Eds., {\em
  {Proceedings of the Workshop Motives, Quantum Field Theory, and
  Pseudodifferential Operators, Boston, 2008}}. Clay Mathematics Proceedings,
  (2010) 167--186.
\newblock (arXiv:0901.0837 [math-ph]).
%
\bibitem{Kretzer99}
S.~Kretzer.
\newblock {\em {Heavy Quark Production and Fragmentation Processes in
  Next-to-Leading Order QCD}}.
\newblock PhD thesis, Universit\"at Dortmund, Germany, 1999.
%
\bibitem{Altarelli:1977zs}
G.~Altarelli and G.~Parisi.
\newblock {\em Nucl. Phys. B}, {\bf 126} (1977) 298--318.
%
\bibitem{Larin:1993tq}
S.~A. Larin.
\newblock {\em Phys. Lett. B}, {\bf 303} (1993) 113--118.
%
\bibitem{Dokshitzer:1991wu}
Yu.~L. Dokshitzer, V.~A. Khoze, A.~H. Mueller, and S.~I. Troian.
\newblock {\em {Basics of perturbative QCD}}.
\newblock (Editions Fronti\`eres, 1991).
%
\bibitem{Sudakov:1954sw}
V.~V. Sudakov.
\newblock {\em Sov. Phys. JETP}, {\bf 3} (1956) 65--71.
%
\bibitem{KB1973}
E.~Byckling and K.~Kajantie.
\newblock {\em {Particle Kinematics}}.
\newblock (John Wiley \& Sons, New York, 1973).
%
\bibitem{Cheng:1985bj}
T.-P. Cheng and L.-F. Li.
\newblock {\em {Gauge theory of elementary particle physics}}.
\newblock (Oxford University Press, Oxford, 1984).
\end{thebibliography}
\end{document}